\RequirePackage{fix-cm}
\documentclass[twocolumn]{svjour3}          
\smartqed  
\usepackage{graphicx}
\usepackage{color}
\usepackage{amsmath}
\usepackage{amsfonts}
\usepackage[export]{adjustbox}
\usepackage{pgfplots}
\usepackage{float}
\usepackage[square,numbers]{natbib}
\setcitestyle{numbers}
\usepackage[normalem]{ulem}
\usepackage{algorithm}
\usepackage[noend]{algpseudocode}
\algnewcommand\And{\textbf{and}}

\begin{document}

\title{Automatic Flare Spot Artifact Detection and Removal in Photographs}

\author{Patricia Vitoria \and Coloma Ballester 
}

\institute{Patricia Vitoria \at
                Universitat Pompeu Fabra \\
                C/ Roc Boronat 138 \\
                08018 Barcelona, Spain\\
              Tel.: +34 93 542 1449 \\
              \email{patricia.vitoria@upf.edu}          
           \and
           Coloma Ballester \at
            Universitat Pompeu Fabra \\ 
              Tel.: +34 93 542 2711 \\
              \email{coloma.ballester@upf.edu}
}

\date{Received: date / Accepted: date}

\maketitle

\begin{abstract}
Flare spot is one type of flare artifact caused by a number of conditions, frequently provoked by one or more high-luminance sources within or close to the camera field of view. 
When light rays coming from a high-luminance source reach the front element of a camera, it can produce intra-reflections within camera elements that emerge  at the film plane forming non-image information or flare on the captured image.
Even though preventive mechanisms are used, artifacts can appear. In this paper, we propose a robust computational method to automatically detect and remove flare spot artifacts. 
Our contribution is threefold: firstly, we propose a characterization which is based on intrinsic properties that a flare spot is likely to satisfy; secondly, we define a new confidence measure able to select flare spots among the candidates; and, finally, a method to accurately determine the flare region is given. Then, the detected artifacts are removed by using exemplar-based inpainting.
We show that our algorithm achieve top-tier quantitative and qualitative performance. 

\end{abstract}

\section{Introduction}
\label{introduction}

\begin{figure}
    \centering
    \begin{tabular}{cc}
        \vspace*{0.15cm}
    \includegraphics[angle=270,width=3.7cm]{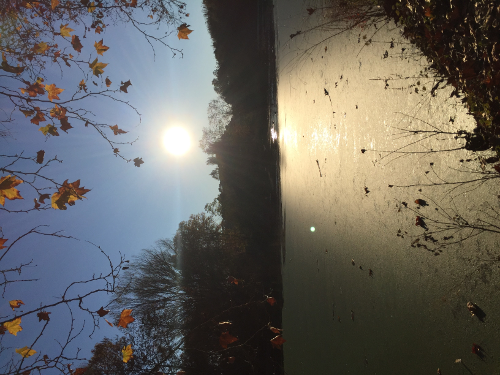} &
    \includegraphics[angle=270,width=3.7cm]{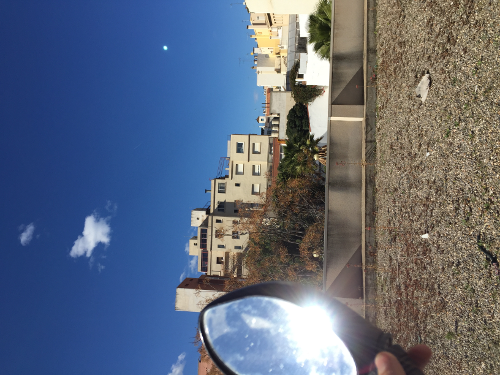} \\
    (a) & (b) \\
    \vspace*{0.15cm}
    \includegraphics[angle=270,width=3.7cm]{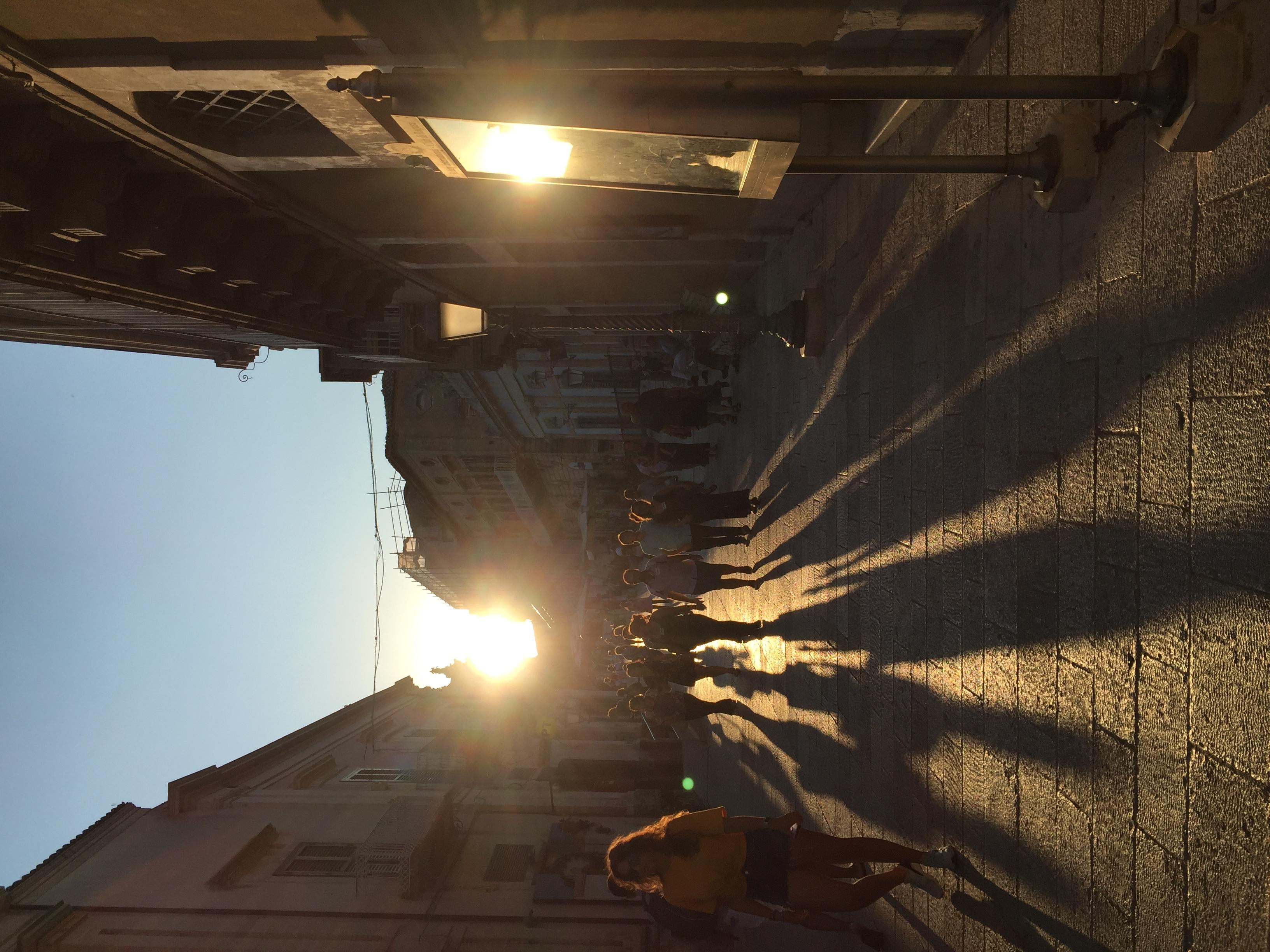} &
    \includegraphics[angle=270,width=3.7cm]{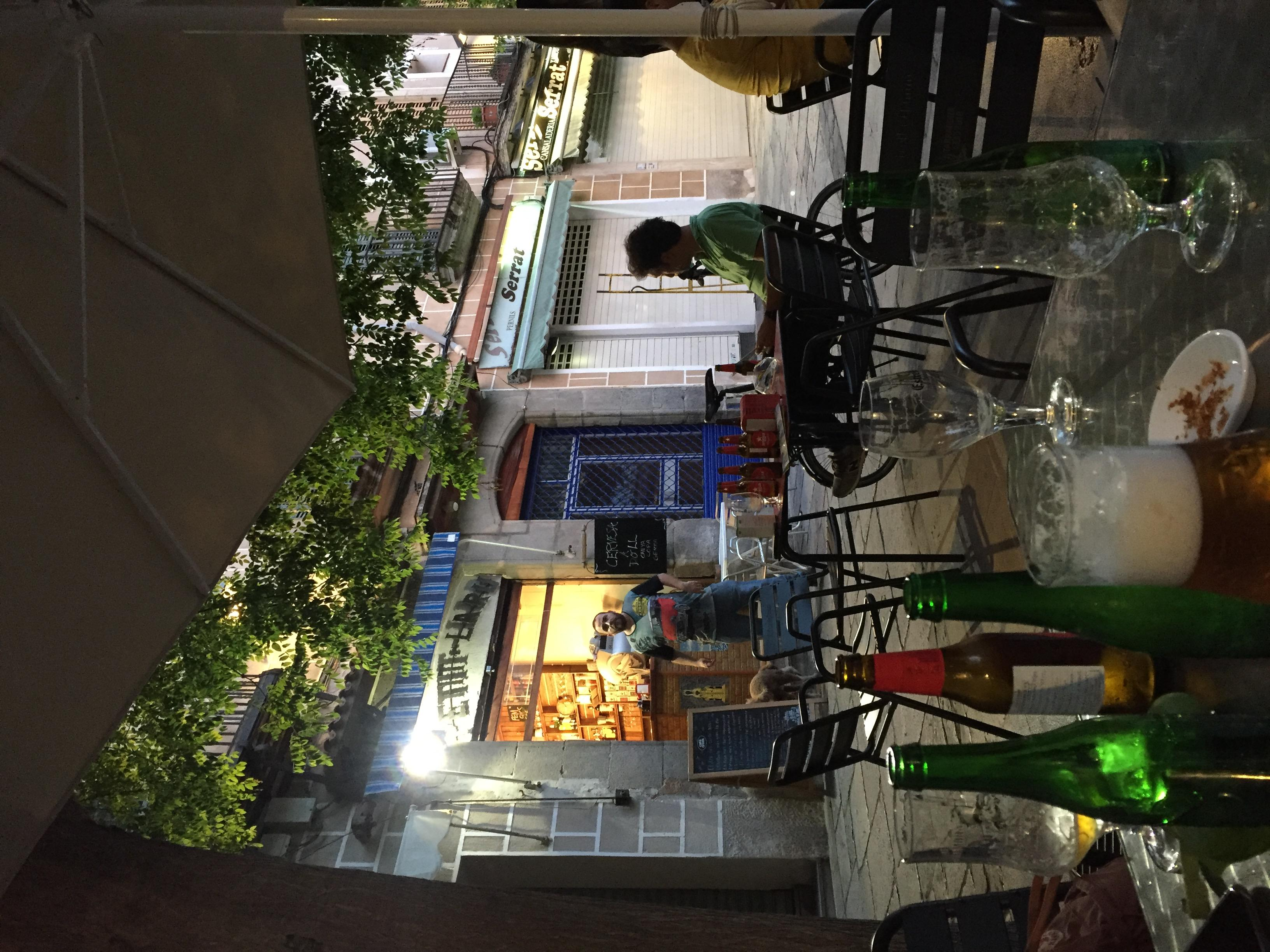} \\
    (c) & (d) \\
    \end{tabular}
    \caption{Four images with flare spot produced by: (a) the sun, (b) a reflection on a mirror, (c) the sun and the reflection on a mirror and (d) a light bulb (bottom right)}
    \label{fig:LensFlareExamples}
\end{figure}

\begin{figure}[H]
    \centering
    \includegraphics[angle=270,width=3cm]{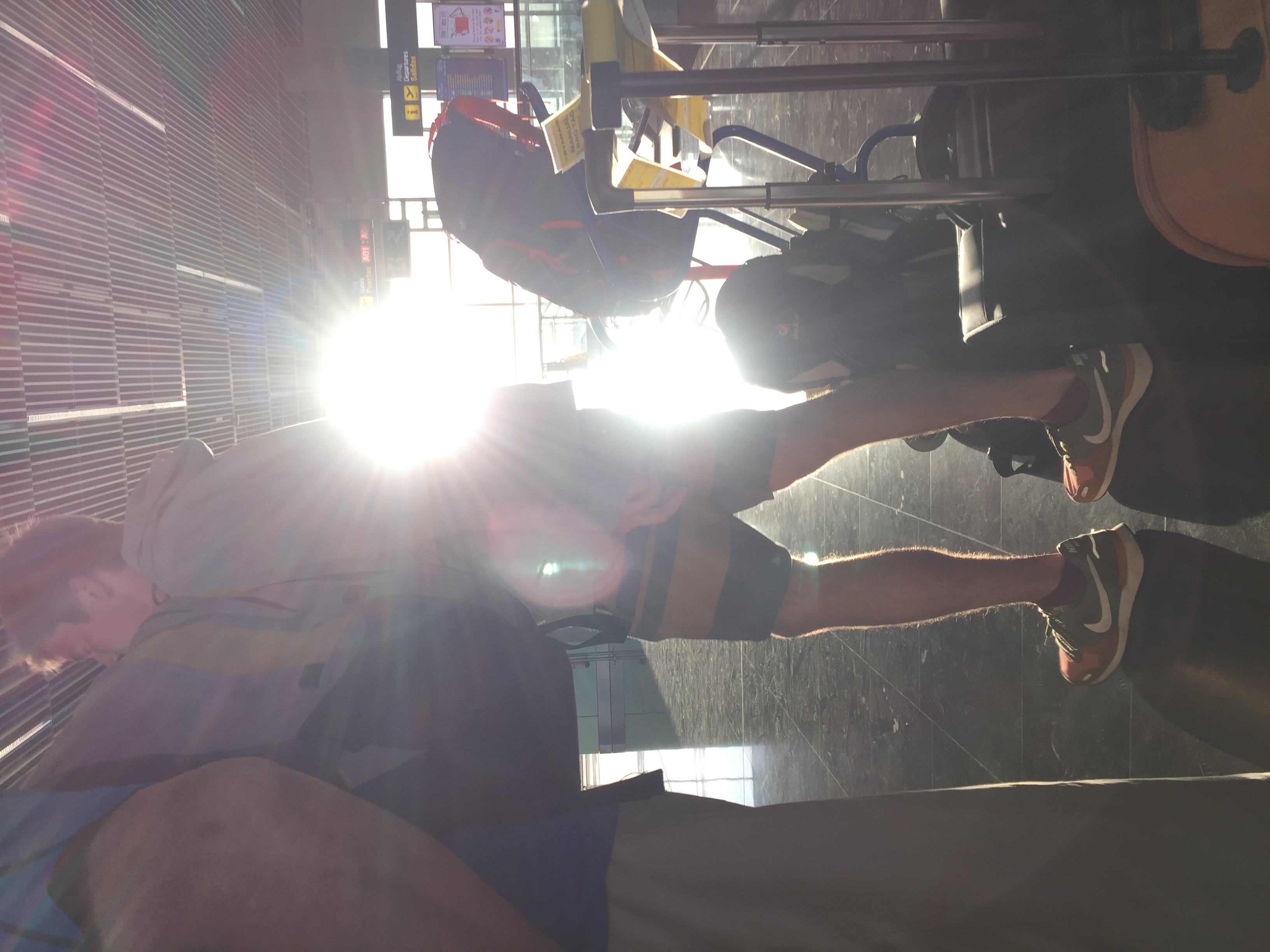}
    \caption{Image corrupted with several types of lens flare artifacts such as veiling glare, red artifact and flare spot.}
    \label{fig:SeveralArtifacts}
\end{figure}

When light rays coming from bright sources of light reach a camera, they can reflect and scatter in between its optical elements such as surfaces of a lens,  mechanical surfaces adjacent to or within to the camera and lenses like, for example, lens protection filters. This situation potentially degrades the image quality and creates unwanted artifacts in photographs. 
Better known as flare, this phenomenon can impact images in a number of ways: it can drastically reduce image contrast by introducing haze 
(also called veiling glare), it can add circular or semi-circular halos or “ghosts” (named as ghosting or flare spot; some examples are shown in Figure \ref{fig:LensFlareExamples}) and semi-transparent odd-shaped objects of different colors, to mention but a few of the many kinds of possible flare artifacts. Some examples of flare artifacts (veiling glare, red artifact, flare spot) appear in the image displayed in Figure \ref{fig:SeveralArtifacts}.

 Although there are applications where the flare artifacts are not undesirable $-$ e.g., some film directors use flare as an element of realism or even as an extra character in their movies $-$, in other contexts flare identification is a pivotal problem, for instance, in order to avoid false detections. This is the case in medical diagnosis assistance $-$ e.g., in slit lamp retina images where flare anomalies corrupt the retinal content \citep{prokopetc2017slim} $-$, in order to distinguish from aerial objects following a collision track in aerial images \citep{nussberger2016robust}, or in the context of surveillance camera systems, to mention but a few.  Moreover, there are situations where in unique shooting circumstances (e.g., studio photographs, special events,  etc), the flare artifacts are unwanted or annoying and significantly degrade the quality of the image. Thus, an automatic and powerful flare detection and removal mechanism is essential. Other applications of such a method are video or image editing for 3D chroma keying purposes, or in general in situations where a strong light source is close to the object to be photographed and it is impossible to avoid the light source to be inside the field of view (such as, for instance, in the scenes of Figure~\ref{fig:LensFlareExamples}a and c).

The use of preventative measures during the image capture such as special coating covers, lens hoods or anti-reflective coating, does not always eliminate entirely the flare artifacts~\citep{nussberger2016robust}. For instance, anti-reflection coatings have been largely studied in thin-film optics. Current cameras, where 20 or more elements may be included, would be unusable without anti-reflection coatings.
In fact, nowadays most of the lenses are designed with special multi-coating technologies to reduce flare.
Unfortunately, this fact does not completely prevent the reflections in the presence of a light source with 4+ orders of magnitude brighter than other scene elements. Additionally, most of the commercial cameras are equipped with lens protector filters that also increase the likelihood of flare artifacts.

In this paper, we tackle the problem of one type of flare artifacts, namely, flare spot or ghosting. Figure \ref{fig:LensFlareExamples} displays four examples where one or several flare spot artifacts appear. This paper proposes an automatic method to detect and remove them. Flare spot can appear in images captured by cameras equipped with multi-coated lenses and a lens protection filter, elements which are wide-spread incorporated in current cameras.  The case of veiling glare is very different in nature and is out of the scope of the present work as other approaches would be more convenient.
Flare spot can arise in daytime as well as night-time photo captures. It appears when a bright light source is within or close to the camera field of view.  
The very bright light source is often the sun but can also be bright studio lights or specular reflection surfaces such as, e.g., the case shown in Figure \ref{fig:LensFlareExamples}b and \ref{fig:LensFlareExamples}c. Sunlight reflected by water or silica sand are also a common source of flare.

\subsection{Contributions}

We propose an automatic method for flare spot detection and removal. Specific technical contributions are 

\noindent
$-$ A computational characterization of the main properties that identify flare spot artifacts in photographs taken by a camera. This characterization grounds on geometrical, morphological, luminous intensity and chromatic properties. Additionally, it takes into account the mechanical architecture of optical lens systems together with the nature of the lens protector filters.

\noindent 
$-$ An efficient algorithm to  estimate a set of candidates that fulfill those properties. Then, we define a confidence measure, which can be interpreted as the likelihood to be a flare spot and which is used to select the correct flare spots among those candidates by maximizing it. Again, this confidence measure grounds on geometrical, luminous intensity and chromatic properties.

\noindent
$-$ Once the flare is detected, a flare mask that delimits the  flare spot region is estimated. As a final step, the region is reconstructed via exemplar-based inpainting of the area affected.

Let us note again that our goal is not to identify all the many types of flare artifacts since each artifact have a very different origin. Instead, we want to characterize, detect and remove those artifacts present in widely used cameras that are equipped with multi-coated lenses with a lens protection filter.

The remainder of the paper is organized as follows. In Section~\ref{sec:relwork}, we review the related work. Section~\ref{sec:method} details our method for flare spot detection and removal. In Section~\ref{sec:results}, we present both quantitative and qualitative assessments of all parts of the proposed method. Finally, Section~\ref{sec:conclusions} concludes the paper.

\section{Related Research}\label{sec:relwork}

Over the years, a number of approaches have been proposed to measure, compensate or reduce flare artifacts due to reflection or  scattering of light in lens and camera systems. Fresnel equations  describe the amount of reflection when light impacts on an interface between two media having different refractive indices.  
Flare is more visible for large aperture, wider field of view, shorter wavelenght and near the center of the image. Systematic laboratory evaluations of flare characteristics of various lenses have been carried out mainly by \citep{martin1972glare,matsuda1972flare,steenstrup1980automatic,ideki1982veiling}.

On the other hand, several metrics have been defined in order to measure the amount of flare light that might be present on photographs.

Another method is the \emph{Black spot's method} where a small area of zero luminance is surrounded by a large bright field. The illuminance of the image in the black spot ($E_S$) and in the field ($E_F$) are measured and the flare is quantified with the \emph{Flare index} as a percentage where $G=100 \times\frac{E_S}{E_F}$ \citep{ray2002oapplied}. On the other hand, \cite{raskar2008glare} discussed that glare is inherently a 4D ray-space phenomenon and proposed a technique that, considering glare as high frequency noise, reduces it by outlier rejection.

Most of the methods to reduce all sorts of flare artifacts focus on improvements in the optical elements of the system. For example, most of the actual cameras use anti-reflective coating to reduce the reflections from lens surfaces. It was found that a quite satisfactory reduction in flare could be achieved by replacing a circular polarizer by a neutral density filter such as a sheet of absorbing glass or plastic. Specular reflectance from the filter is eliminated by anti-reflection coatings front and back. Flare light then passes through the filter twice, while signal light passes through only once \cite{macleod2010thin}. 
Also, other approaches apply a slight modification to the camera in order to reduce flare \cite{raskar2008glare,boynton2003liquid}. For example \cite{raskar2008glare} place a printed film transparency mask on the top of the sensor to prevent glare-producing light from reaching the sensor pixels and achieve glare reduction. The authors of \cite{boynton2003liquid} constructed a fluid-filled camera to reduce internal reflections due to lens surfaces. 
In any case, the mentioned approaches aim to reduce the artifacts without removing them completely.

In the particular case of flare spot, the stray light rays from inside and outside the system field of view that reflect between lens surfaces and optical elements focus to a region near the image plane.  They represent a large concentration of energy which produces a flare spot which can hide information as well as create false information in the image plane \cite{evans1988analysis}.
The more lens elements there are the more ghosts will appear in images. For a lens with $n$ surfaces, the maximum number of possible ghost artifacts is equal to $\frac{n(n-1)}{2}$ \citep{kojima1980computer,ray2002oapplied}. 

The modifications of the optical elements explained above are frequently not enough to avoid flare spot artifact occurrence. Moreover, sometimes the photographs are already taken and contain them. Our work fits in those contexts. In order to completely remove flare spot artifacts from an image by leveraging the use of image processing techniques, an algorithm generally goes through two phases: first of all, the detection of the flare artifacts inside the image and second, the reconstruction of the flare region by filling-in the missing information in a plausible way. However, some of the state-of-the-art algorithms about flare spot focus only in the first phase and omit the reconstruction of the region. For example,  \cite{nussberger2016robust} tracks aerial objects in the airspace while, at the same time, trying to avoid false detection due to flare spots present in the aerial images. Due to their purpose, the detection stage is enough in order to avoid the identification of a flare spot as an aerial object. Accordingly, we include in this related work review the literature on flare spot detection. The methods found can be classified into two groups, depending on the provided input information: semi-automatic \cite{Psotny09,nussberger2016robust,wu2005bayesian} and automatic \cite{chabertautomated,prokopetc2017slim}. 

In  semi-automatic algorithms \cite{nussberger2016robust,Psotny09,wu2005bayesian}, additional information besides the input image is needed. In \cite{nussberger2016robust} the detection of the artifacts within aerial images is based on the (known) position of the sun with respect to the observer. They use the date, time, position and altitude of the observer to predict the flare direction within the image. The authors impose consistent motion of the candidates together with the flare direction over multiple time steps. Once the direction is known the position, size and shape of the flares are extracted. From the detected flares an associated mask is generated to discard it in their aerial object detection aim. The recovery of the missing pixels within the mask is outside of the scope of their work. In the case of \cite{Psotny09}, the aim is on the removal and to this goal the user has to select either the area where the flare spot is present or specify the color of the flare for each case. In \cite{wu2005bayesian}, an algorithm for shadow extraction is also applied to flare spot detection and removal. Their method uses hints from the user that roughly indicates about region properties such as ``regions with flare", ``regions without flare", ``uncertain regions" or ``excluded regions". 

The aim of automatic flare spot detection algorithms is to detect the flare spot without any user interaction. At the time this paper was written, to the best of our knowledge only one technical report has been found regarding that topic applied to photography \cite{chabertautomated}. The author uses a blob detector from OpenCV that detects and filters blobs based on several parameters such as area,  circularity, convexity and elongation of the blob. Once all the flare spots are detected, the algorithm creates an inpainting mask and recovers it using the exemplar-based inpainting method by \cite{criminisi2004region}. However, that algorithm only identifies blob-like structures with those particular attributes and moreover does not generalize to all types of flare spot. This fact entails a a high percentage of false positives and a low  detection precision. The recent work \cite{prokopetc2017slim} aims to detect and correct light reflections in retinal images. It detects the flare using the specular-free image concept \cite{TanIkeuchiTIP05} and a probability map. Finally, they perform an image blending on a spatially aligned images. 

Let us notice, that, even being flare spot artifacts one of the most common artifacts present in nowadays photographs, there is not a lot of research done on that field. By contrast, more emphasis has been placed on other type of lens artifacts such as veiling glare \cite{wu2005bayesian,raskar2008glare,talvala2007veiling}, camera flash artifacts \cite{petschnigg2004digital}, or dirty camera lenses \cite{gu2009removing,hara2009removal}. In the next section we propose and detail an automatic method for flare spot detection and removal that can be applied automatically to a single photograph without any previous knowledge of the camera specifications.

\section{Proposed Method}\label{sec:method}

Our method consists of three main blocks: light source detection, flare spot detection and flare spot removal. The first one aims at detecting the presence and location of high light sources in the image which in turn can be used for a faster flare spot detection. 
Our flare spot detection is then based on hypothesizing a number of properties that a flare spot is likely to satisfy. They take into account both geometric and color properties as well as the mechanical architecture of optical lens systems together with the nature of the lens protector filters. As a result, a set of candidates is identified. Then, we propose a confidence measure that selects the flare spot. Lastly, after accurately computing the flare spot region, exemplar-based inpainting is applied in that region. A pseudocode of our algorithm can be found in Algorithm \ref{OurPseudocode}. 

\begin{algorithm*}
\caption{Lens Flare Detection and Removal Algorithm Pseudo-code}\label{OurPseudocode}
\begin{algorithmic}[1] 
\Function{FlareSpotRemoval}{u} \Comment{see Section \ref{sec:method}}
\State $\{\mathbf{x}_{S_i}\}_{i=1}^s= \Call{FindLightSources}{u}$  \Comment{Find all light sources present in the image  (see Section \ref{sec:sourcelight})}
\State $\{\mathbf{x}_k(j)\}_j = \Call{BlobDetector}{u}$ \Comment{Find brigh blobs in the image (see Section \ref{sec:hypothesis1})}
\State $\mathbf{X}_{fs} = \{\}$ \Comment{Set used to save all the detected flare spot points}
\ForAll {$i=1,\dots,s$} 
\Comment{For each light source}
\State MaxConfMeasure = 0
\ForAll {$j$} \Comment{For each blob}
\State $(\lambda_1, \lambda_2) = \Call{Eingenvalues}{ \mathbf{x}_k(j)}$
\If{$\lambda_1 < 0 || \lambda_2 >  4\lambda_1$} \Comment{Omit too elongated blobs (see Equation (\ref{eq:nonelongated}) in Section \ref{sec:hypothesis4})}
\State Continue (with $j+1$);
\EndIf
\If{$\Call{area}{cc(B_{\delta}(\mathbf{x}_k(j));\mathbf{x}_k(j)} \geq \frac{area(S_i)}{100}$ } \Comment{Omit this keypoint  (see Equation (\ref{eq:CC10}) in Section \ref{sec:hypothesis2})}
\State Continue (with $j+1$);
\EndIf
\If{ \Call{relative brightness}{ $\mathbf{x}_k(j)$} $<\alpha$} \Comment{Omit not high exposured  keypoints (see Section \ref{sec:hypothesis3})}
\State Continue  (with $j+1$);
\EndIf
\If{$e^{-E(\mathbf{x}_k(j))} > MaxConfMeasure$} \Comment{Compute confidence measure (see Section \ref{sec:hypothesis5})}
\State $MaxConfMeasure = e^{-E(\mathbf{x}_k(j))}$ \Comment{Keep the keypoints with maximum confidence}

\State $\mathbf{x}^i_{fs} = \mathbf{x}_k(j)$

\EndIf
\EndFor
$\mathbf{X}_{fs} \gets \mathbf{X}_{fs} \cup \{\mathbf{x}^i_{fs}\}$ \Comment{Add the  new flare spot coordinates}
\EndFor
\State $M = \Call{MaskComputation}{\{\mathbf{X}_{fs}\}}$ \Comment{Compute the mask having into account all the flares (see Section \ref{sec:MaskCreation})}
\State $u_{new} = \Call{ExemplarBasedInpainting}{u,M}$ \Comment{Flare spot removal via Exemplar-based inpainting \ref{sec:removal}}
\EndFunction
\end{algorithmic}
\end{algorithm*}


\subsection{Light Source Detector}\label{sec:sourcelight}

Flare spot artifacts often emerge when light rays coming from a bright source of light (for instance, the sun) directly reach the camera and the light source is captured in the scene. This situation creates an almost circular artifact, the flare spot, on the image. In the case of modern cameras which are equipped with multi-coated lenses and a lens protection filter, the flare spot artifact is roughly located at a symmetric point in the opposite side of the light source \cite{kojima1980computer,evans1988analysis}. More precisely, a digital camera focusing to the infinite works as a retro-reflector, then if the reflected light from the camera towards the original bright light source is reflected by the lens protector filter, the flare appears at a symmetric point. 
Let $u:\Omega\to\mathbb{R}^3$ be a given color image, where $\Omega\subset\mathbb{R}^2$ is bounded and denotes the image domain.
We can roughly locate the flare spot in the opposite side of the light source in the image domain and close to the line that passes through the light source and the principal point of the image. The principal point is the point where the principal axis, which is the line that passes through the optical centre of the camera and is perpendicular to the image plane, meets the image plane. The optical center of the camera (and the principal point) can be identified by camera calibration \cite{ray2002oapplied,hartley2003multiple}. In this paper and aiming at a real-time algorithm, we approximate it by the center of the image domain $\Omega$, denoted by $\mathbf{x}_c$ below. Thus, finding out the two-dimensional position of each bright light source, denoted here by $\mathbf{x}_{S_i}$, for $i=1,\dots,s$, for a certain positive integer $s$, inside of the image domain, is a key step for an efficient detection of flare spots. 

The very bright light sources captured in the image are likely to correspond to the brightest regions of the image. 
To identify them, let $u^{La^*b^*}=(u^L, u^{a^*}, u^{b^*}):\Omega\to\mathbb{R}^3$ be the image $u$ in the CIELab color space. The value $u^L(\mathbf{x})$ of the brightness or luminance channel $u^L$ at a point $\mathbf{x}\in\Omega$ indicates its brightness perception in a range between $0$ and $100$, where $0$ provides the black, which corresponds to minimum brightness, and $100$ the brightest white. Hence, in order to find the brightest regions of the image we consider the upper level set of level $\iota$ of $u^L$, denoted by $X _\iota u^L$, for a real value of $\iota$ close to $100$. Let us recall that an upper level set $X _\iota u^L$ of $u^L$ is defined as:
\begin{equation}\label{eq:levelset}
    X_\iota u^L := [u^L \geq \iota] = \{\mathbf{x}\in\Omega : u^L(\mathbf{x}) \geq \iota\}. 
\end{equation}

In general, the set $X_\iota u^L$ will be made of a finite union of connected components. Let us recall that a subset $C$ of a set $X$ is called a connected component if $C$ is a maximal connected subset of $X$. In our case, once the brightest set, given by $X_\iota u^L$, is identified, we compute its decomposition in connected components. Before computing the connected components, a morphological opening filter is applied to $X_\iota u^L$ in order to remove small noise that might affect our results - in all experiments, we use an opening with a disc of radius $1.5$ as structuring element -. For the discrete computation of the connected components, we use the discrete 8-connectivity notion. Then, among those connected components of $X_\iota u^L$, we select the one
with biggest area. This resulting connected region corresponds to what we call the main light source, denoted here by $S_1$. Sometimes, more than one bright light source is present in the image. Also, some  specular reflection surfaces or materials might also act as light sources. For instance, sunlight reflected by water or silica sand are a common source of flare. An example is shown in Figure \ref{fig:InpaintingResultschabertOurs2}. Now, in order to consider other possible light sources we keep also the connected  components of $X_\iota u^L$ with area bigger or equal than $0.8 \times$area$(S_1)$. 
For each one of the potential light sources $S_i$ ($i=1,\dots,s$), we find the approximate center by computing the center of mass (or centroid) of $S_i$, denoted by $\mathbf{x}_{S_i}$ and given by
\begin{equation*}
    \mathbf{x}_{S_i} = \frac{1}{\text{area}(S_i)} \sum_{\mathbf{x}\in S_i} \mathbf{x},
\end{equation*}
where $\mathbf{x}=(x,y)$. In the next Section we propose a method to detect flare spot artifacts and we apply it for each one of the light sources  $S_i$. To simplify notations, we will forget about the sub-index and denote it by $S$.

\subsection{Flare Spot Detector}\label{sec:FSdetector}

In order to detect a flare spot, which will be denoted by $\mathbf{x}_{fs}$ in the following, we start by hypothesizing the properties that it is likely to satisfy. From the previous step, we know the position of each potential light source $S$, which is represented by its centroid $\mathbf{x}_S$. As previously mentioned, the flare spot  is located in the opposite side of the image domain $\Omega$  and close to the line that connects $\mathbf{x}_S$ with the center of $\Omega$, denoted here by $\mathbf{x}_c$.
We claim that flare spot candidates meet the following properties:
\begin{itemize}
    \item  To be a bright blob of the image $u$. A blob can be described as a region of the image that is either brighter or darker than the background and is surrounded by a smoothly curved boundary. In our case, we hypothesize that the flare spot is a region brighter than the background. A blob will be represented by a representative point, also called keypoint, that we will denote by  $\mathbf{x}_{k}$.
    \item The blob associated to a certain keypoint $\mathbf{x}_{k}$ should not be too elongated as flare spots tend to have slightly round or elliptical shapes.
    \item Overexposure or high luminous energy in the area of the flare spot candidate in comparison with the remaining parts of the image. Indeed, stray light rays that strike the image plane focus to regions representing a large concentration of energy which produces the flare spot \cite{evans1988analysis}.  
    \item Limited area of the flare spot region candidate; it is bounded by a threshold depending on the light source region.
    \item To be close to the straight line passing through $\mathbf{x}_S$ and the image center $\mathbf{x}_c$. Moreover, the distance from the flare spot candidate to the center $\mathbf{x}_c$ should be similar to the distance between the light source $\mathbf{x}_S$ and the center $\mathbf{x}_c$. 
    \item In general, the color value at that candidate in the CIELab color space should have a high $L$ value and a negative $a^*$. This property includes in our flare spot's characterization the most usual chromaticity cases in  amateur photography. There are several reasons for that, including: on the one hand, due to Rayleigh formula, blue light is scattered nearly six times as strongly as red light; so filters are used; on the other hand, multiple coating layers are used to increase the transmittance of multi-element lenses \cite{macleod2010thin}, and the particular material of the layers affects the flare spot appearance.
    \end{itemize}
In the following we detail how the candidates fulfilling the previous first four hypotheses are computed. Then, we propose a confidence measure which is based on the last two hypothesis and which is able to finally select the correct flare spot among those candidates.


\subsubsection{Blob detection}\label{sec:hypothesis1}

First, we claim 
that a flare spot is a blob brighter than the background. To estimate it, we use the definition of blobs and keypoints introduced by David Lowe \cite{lowe1999object,lowe2004distinctive} in his proposal of the SIFT method (see also \cite{lindeberg1994scale,lindeberg1998feature}). A \emph{keypoint} is a blob-like structure whose center
is an extremum in the scale-space of the Laplacian of the convolution of Gaussians with the image. More precisely, a point $\mathbf{x}_{k}\in\Omega\subset\mathbb{R}$ is a keypoint of $u$ at the scale $\sigma_{k}$ if
$(\mathbf{x}_{k},\sigma_{k})$ is a space-scale local extrema of the space Laplacian of
\begin{equation*}
   L(\mathbf{x},\sigma)=G(\mathbf{x};\sigma) * u^g(\mathbf{x})
\end{equation*}
where $*$ denotes the convolution operator in $\mathbb{R}^2$, $u^g$ is a gray version of $u$ (for instance, the brightness information $u^L$)  and  $G(\mathbf{x};\sigma)$ 
denotes the isotropic Gaussian function of standard deviation $\sigma$ and (space) integral equal to one. Let us notice that the subscript $k$ in the notation $\mathbf{x}_k(j)$ (and in $(\mathbf{x}_k(j),\sigma_k(j))$) stands for {\it{keypoint}}. The space Laplacian of $L$, usually denoted by $\Delta L$ or $\nabla^2 L$, is the common second order differential operator $\Delta L=\frac{\partial^2 L}{\partial x^2}+\frac{\partial^2 L}{\partial y^2}$.
For efficiency purposes, David Lowe \cite{lowe2004distinctive} proposed to compute the keypoints as the scale-space extrema of the so-called \emph{difference-of-Gaussian (DoG)}. The DOG, denoted by $D(\mathbf{x},\sigma)$, is defined as the difference of two nearby scales of $L(\mathbf{x},\sigma)$, separated by an appropriate constant multiplicative factor $k$, as 
\begin{equation}\label{eq:DoG}
D(\mathbf{x},\sigma) = L(\mathbf{x},k\sigma)- L(\mathbf{x},\sigma).
\end{equation}
$D(\mathbf{x},\sigma)$ provides a close approximation to $\sigma^2\Delta L(\mathbf{x},\sigma)$, the normalized Laplacian of $L$. 
We refer to~\cite{reyotero14} for more details and a rigorous analysis including the stability of the keypoint computation. The seminal paper introducing SIFT~\cite{lowe1999object} sparked an explosion of local keypoints detectors and descriptors seeking discrimination and invariance to specific groups of image transformations~\cite{tuytelaars2008local}. 
Mikolajczyk \cite{mikolajczyk2002detection} found that the maxima and minima of the operator $\sigma^2\Delta L(\mathbf{x},\sigma)$ produce the most stable image keypoints compared to a range of other possible image operators, such as the gradients, Hessian, or Harris corner detector. The SIFT keypoints have been used in countless image processing and computer vision applications, such as image registration~\cite{hartley2003multiple,snavely2006photo}, camera calibration~\cite{von2010towards}, 3D reconstruction~\cite{agarwal2011building}, object recognition~\cite{fergus2003object,bay2006interactive,zhang2007local,zinemanas2017visual}.

\begin{figure}
    \centering
     \begin{tikzpicture}
        \node[anchor=south west,inner sep=0] at (0,0) {
         \includegraphics[angle=270,width=3.5cm]{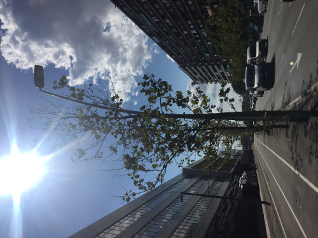}
        \includegraphics[angle=270,width=3.5cm]{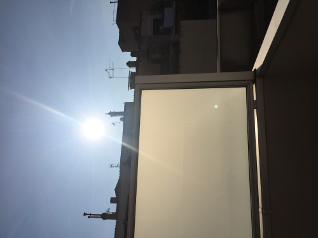}
         };
        \draw[red] (1.75,0) rectangle (3.5,2.3);
        \draw[red] (5.35,0) rectangle (7.1,2.3);
    \end{tikzpicture}
       \vspace{0.2cm}
    \includegraphics[ angle=270,width=3.5cm]{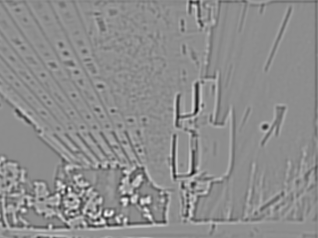}
    \includegraphics[ angle=270,width=3.5cm]{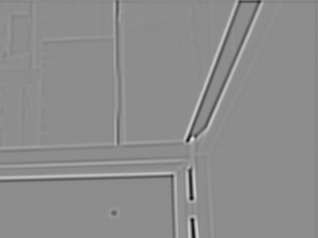}
       \vspace{0.2cm}
    \includegraphics[ angle=270,width=3.5cm]{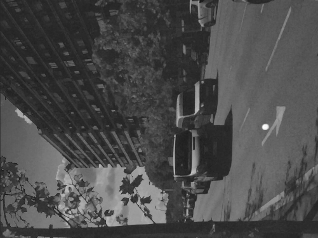}
    \includegraphics[ angle=270,width=3.5cm]{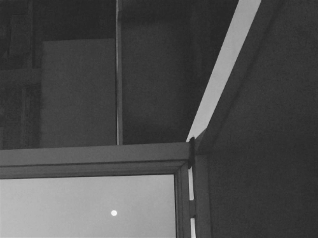}
    \caption{First row: two different original images of our dataset of $405$ images which have been captured by cameras with different technical specifications. Second row: zoom of the portion in the red box on the corresponding difference-of-Gaussians $D(\approx\sigma^2\Delta L)$ (given by Equation ~\eqref{eq:DoG}). Third row: zoom of the portion in the red box showing the values of $E_3(\mathbf{x})$ (given by Equation \eqref{eq:E3} and used in our confidence measure \eqref{eq:detected}). For visualization purposes, in this figure the values of $E_3$ have been normalized}
    \label{fig:ExampleOfLaplacians}
\end{figure}

In our case, we hypothesize that the flare spot is a region brighter than the background and thus it is among the local minima of $D$ (equivalently, of $\sigma^2\Delta L$). We illustrate this assumption on the second row of Figure \ref{fig:ExampleOfLaplacians}, which shows the values of $D$ for the two different original images displayed in the first row of Figure \ref{fig:ExampleOfLaplacians}.  Hence, varying the scales $\sigma$ in a range between a minimum $\sigma_{min}$ and a maximum $\sigma_{max}$ (see values in Table \ref{tab:parameters}, with $k=2^{1/5}$ in Equation~\eqref{eq:DoG}), we compute the set of local minima of $D$, denoted by $\{ (\mathbf{x}_k(j),\sigma_k(j))\}_j$. The corresponding set $\{ \mathbf{x}_k(j)\}_j$ is made of centers of blobs or keypoints. They constitute our first set of flare spot candidates.


\subsubsection{Bounded Area of the Flare Spot}\label{sec:hypothesis2}

As mentioned before, we claim a flare spot to have a limited area. Thus, if $\mathbf{x}_k(j)$ is one of the keypoints computed as above, we verify this property with a fast algorithm as follows. First, we compute the neighboring pixels having a similar brightness value than $\mathbf{x}_k(j)$ by considering the morphological bi-level set of $u^L$, of radius $\delta>0$, 
defined by
\begin{equation}\label{eq:CC10bis}
    B_{\delta}(\mathbf{x}_k(j))=\{\mathbf{x}\in\Omega \, | \; |u^L(\mathbf{x}_k(j))-u^L(\mathbf{x})|\leq\delta\}. 
\end{equation}
Let $cc\left(B_{\delta}(\mathbf{x}_k(j));\mathbf{x}_k(j)\right) $ be the connected component of $B_{\delta}(\mathbf{x}_k(j))$ containing the keypoint $\mathbf{x}_k(j)$. This connected component includes all the connected pixels having a brightness value similar 
to $u^L(\mathbf{x}_k(j))$. To ensure bounded area of the flare spot, we keep the keypoints whose associated connected component has an area below $1\%$ of the light source area, that is,   
\begin{equation}\label{eq:CC10}
area\left(cc\left(B_{\delta}(\mathbf{x}_k(j));\mathbf{x}_k(j)\right)\right) < \frac{area(S)}{100}.
\end{equation}


\subsubsection{Reject Elongated Blobs}\label{sec:hypothesis4}

Based on the assumption that  flare spots have an approximately rounded shape, we will further refine the candidate keypoints ensuring that the corresponding associated blob is not too elongated nor belonging to an edge. Following \cite{lowe2004distinctive}, we compute the eigenvalues $\lambda_1$ and $\lambda_2$ of the Hessian matrix of $D$ at $\mathbf{x}_k(j)$. From our assumption in Section~\ref{sec:hypothesis1}, $\mathbf{x}_k(j)$ is a minima of $D$ (equivalently, of $\sigma^2\Delta L$). Thus, $\lambda_1,\lambda_2 \geq 0$.
Assuming $\lambda_1 \leq \lambda_2$, we verify if the following conditions are satisfied

\begin{equation}\label{eq:nonelongated}
\lambda_1 > 0,\quad \lambda_2 <  4\lambda_1.    
\end{equation}
The first condition guarantees that it is indeed a strict minimum and the second one assures that the local minimum does not correspond to a considerably elongated structure.

\subsubsection{Overexposure in the Flare Spot Region using an speed up computation characterization-based strategy}\label{sec:hypothesis3}

One of the main properties of the flare spot region is the overexposure or high luminance with respect to all remaining pixels, especially for current cameras which are equipped with multi-coated lenses with a lens protection filter \cite{kojima1980computer,evans1988analysis}.
Following this characterization, those keypoints having high brightness will be kept. 

To verify it in practice, and also aiming at an efficient real-time algorithm, we take advantage of another hypothesized flare spot property; namely, that it lies close to the straight line passing through the light source $\mathbf{x}_S$ and the image center $\mathbf{x}_c$, and at a distance from  $\mathbf{x}_c$ similar to the distance between $\mathbf{x}_S$ and $\mathbf{x}_c$. Accordingly, we define a \emph{flare spot search window}  of fixed radius equal to one fifth of the image size around that hypothesized location. 

By doing so, we reduce drastically the computation time.

Now, to keep those keypoints having high brightness compared with other pixels in the flare spot search window, denoted here by $w$, we first normalize the values of $u^L$ in $w$ by
\begin{equation}
 u^L_\text{norm}(\mathbf{x})=\frac{u^L(\mathbf{x})-\min (u^L,w)}{\max (u^L,w)-\min (u^L,w)},
  \label{eq:ulnorm}
\end{equation}
where $\min (u^L,w)$ and $\max (u^L,w)$ denote the minimum and maximum value, respectively, of $u^L$ restricted to $w$.
Notice that the range of values of 
$u^L_\text{norm}$ is from 0 to 1. We select our candidates to flare spot among those with value bigger than a fixed parameter $\beta$ (see Table \ref{tab:parameters} in Section \ref{sec:results} for more details). This condition ensures high probability of overexposure around the keypoint location.

\subsubsection{A confidence measure for final flare spot detection}\label{sec:hypothesis5} 
Our detected candidates fulfill properties such as to be a bright not elongated blob. 
In this section we propose a confidence measure that allows to rank them depending on the likelihood to be a flare spot. The confidence measure is built on the following three assumptions; two of them are grounded in the location of the keypoint and the remaining one theorizes about its $La^*b^*$ value.

\paragraph{\textbf{Keypoint Location}.} 

A keypoint  $\mathbf{x}_k(j)$ is most likely to belong to a flare spot if it is close to the symmetric of $\mathbf{x}_S$. In practice, it translates in:
\begin{enumerate}
        \item The distance from the keypoint $\mathbf{x}_k(j)$ to the center $\mathbf{x}_c$ is similar to the distance from the light source $\mathbf{x}_S$ to $\mathbf{x}_c$.
    \item The distance from  $\mathbf{x}_k(j)$ to the line $l_\text{s2center}$ passing through the light source $\mathbf{x}_S$ and the center of the image $\mathbf{x}_c$ is small. 
\end{enumerate}

\paragraph{\textbf{$\mathbf{La^*b^*}$ value}.} 

Flare spot artifacts are characterized by  having a high lightness and a chromaticity  ranging from blue to green. As mentioned before, this includes in our flare spot's characterization the most usual chromaticity cases in widely used cameras. It can be computationally identified by using the information provided by the $La^*b^*$ color space.
Let us recall from the CIELab perceptual color space that the values $a^*$ and $b^*$ of a given color $(L,a^*,b^*)$ provide the chrominance information in color opponents, in particular, green-red and blue-yellow, respectively. A positive $a^*$ indicates that the color is closer to red, and negative to green. In the case of $b^*$, negative means that it is closer to blue and positive to yellow. 
Thus, in our image $u=(u^L,u^a,u^b)$, it can be translated to a large $u^L$ value, a negative $u^a$ and no restriction on the value of $u^b$.  

\paragraph{\textbf{Confidence measure}.}

We define the confidence measure by a combination of the previous three assumptions. 

\begin{definition}\label{def:confidence}
Given a point $\mathbf{x}\in\Omega$, its likelihood to be part of a flare spot, also called \emph{confidence level}, is defined by $
        \exp{\left(-E(\mathbf{x})\right)}$, 
where 
$$
E(\mathbf{x})=\frac{E_{1,\text{norm}}(\mathbf{x})+E_{2,\text{norm}}(\mathbf{x})}{2} -E_{3,\text{norm}}(\mathbf{x})
$$
and $E_{1,\text{norm}},E_{2,\text{norm}},E_{3,\text{norm}}$ denote the normalized versions of $E_1,E_2,E_3$ defined by
\begin{equation*}
    E_1(\mathbf{x}) = \big|  \|\mathbf{x}_c-\mathbf{x}_s\| - \|\mathbf{x}_c - \mathbf{x}\|\big|
\end{equation*}
$-$ i.e., it gives the difference between the Euclidean distances from the light source to the center of the image and from the center to $\mathbf{x} -$, 
\begin{equation*}
    E_2(\mathbf{x})=\frac{|m_\text{s2center}\cdot x+b_\text{s2center}-y)|}{\sqrt{m_\text{s2center}^2+1}}
\end{equation*}
$-$ it quantifies the distance from $\mathbf{x}=(x,y)$ to the line $l_\text{s2center}$, which we assume having a slope and y-intercept given by $m_\text{s2center}$ and $b_\text{s2center}$, respectively $-$, and 
\begin{equation}\label{eq:E3}
    E_3(\mathbf{x}) = u^L(\mathbf{x})-u^{a^*}(\mathbf{x}). 
\end{equation}
\end{definition}
As mentioned, $E_{1,\text{norm}},E_{2,\text{norm}},E_{3,\text{norm}}$ denote the (separately) linearly normalized versions of $E_1,E_2,E_3$ for them to take values in the range $[0,1]$.

Now, let $\{ \mathbf{x}_k(j)\}_j$ be the set of previously detected candidates. Using Definition~\ref{def:confidence}, we finally obtain the flare spot. The keypoint $\mathbf{x}_{fs}$ associated to the flare spot is given by
\begin{equation}\label{eq:detected}
    \mathbf{x}_{fs} = \arg\max_j e^{-E\left(\mathbf{x}_k(j)\right)}.
\end{equation}
That is, the candidate that maximizes the defined likelihood. In other words, the one that minimizes $E=(E_{1,\text{norm}}+E_{2,\text{norm}})/2-E_{3,\text{norm}}$. Let us finally notice that each term is normalized separately.

\subsection{Flare Spot Removal}\label{sec:removal}

The last step of the algorithm is to remove the identified flare spot artifacts present on the image. As an output of the previous stage, we have the coordinates of the detected flare spots $\{ \mathbf{x}^i_{fs}\}_i$ satisfying Equation~\eqref{eq:detected} for each of the detected light sources $S_i$, $i=1,\dots,s$. In this step we first create a mask that specifies the pixels belonging to each of the flare spot regions. Subsequently, it will be used as an input for the reconstruction of the affected areas by means of exemplar-based inpainting. As before, in order to simplify notation, we will forget about the index $i$ and simply denote the flare spot point by $\mathbf{x}_{fs}$ in order to detail our algorithm for the estimation of the binary mask $M$ providing the flare spot region $ F(\mathbf{x}_{fs})$ of $\mathbf{x}_{fs}$.

\subsubsection{Creation of the flare mask}\label{sec:MaskCreation}
 
In order to define the flare mask, we propose to identify the pixels affected by flare spot artifacts as follows:
\begin{itemize}
    \item \textbf{\textit{Step 1:}} Consider the morphological set $B_{\delta}(\mathbf{x}_{fs})$ (see Equation~\eqref{eq:CC10bis}) for a given $\delta>0$, and select the connected component  of $B_{\delta}(\mathbf{x}_{fs})$ containing $\mathbf{x}_{fs}$. For simplicity of notation, let us denote it by $C_{\delta}(\mathbf{x}_{fs})$.
\item \textbf{\textit{Step 2:}} Apply a morphological dilation of $C_{\delta}(\mathbf{x}_{fs})$ with an structuring element given by a disc of a small radius $\varepsilon>0$. Let $C_{\delta}^{\varepsilon}(\mathbf{x}_{fs})$ be this set. This ensures that the estimated flare spot region keeps all the pixels affected by it. Notice that with this simple procedure, $C_{\delta}^{\varepsilon}(\mathbf{x}_{fs})$ might include pixels not connected within $B_{\delta}(\mathbf{x}_{fs})$ to the flare spot pixel  $\mathbf{x}_{fs}$.
    \item \textbf{\textit{Step 3:}} The \emph{flare spot region}, denoted by $F(\mathbf{x}_{fs})$, is finally defined also taking into account brightness properties as
    \begin{equation*}
        F(\mathbf{x}_{fs})=C_{\delta}^{\varepsilon}(\mathbf{x}_{fs})\cap X_\alpha u^L_\text{norm}
    \end{equation*}
    where $u^L_\text{norm}$ is defined by Equation \eqref{eq:ulnorm} and $X_\alpha u^L_\text{norm}$ denotes the level set $[u^L_\text{norm} \geq \alpha]$ defined from Equation \eqref{eq:levelset}. The value of $\alpha\in [0,1]$ has been experimentally set to 0.2 (see Table~\ref{tab:parameters}).
\end{itemize}
Finally, the flare mask, $M:\Omega\to\mathbb{R}$, is defined as
\begin{equation*}
  M(\mathbf{x}) = \left\{
  \begin{array}{lr}
    1 &  \quad \text{if } \mathbf{x}\in F(\mathbf{x}_{fs})\\
    0 &  \text{otherwise}
  \end{array}
\right.
\end{equation*}
Our final mask is made of the union of the masks associated to each of the $\{ \mathbf{x}^i_{fs}\}_i$, $i=1,\dots,s$. The resulting flare mask will be used together with the original image $u$ as an input for recovering an image free of flare spot artifacts in the following section.


\subsubsection{Flare Spot Removal}

In order to reconstruct the regions of the image damaged by the flare spot  we use the method of \cite{inpainting_arias_2011} providing a variational framework for exemplar-based image inpainting and, in particular, the corresponding algorithm and implementation by  \cite{fedorov2015variational}. 

Let us briefly recall that, given an input image $u$ which is unknown on a region $O\subset\Omega\subset\mathbb{R}^2$, the problem of exemplar-based inpainting can be stated \cite{demanet2003image} as that of finding a correspondence map  $\varphi:O\rightarrow O^c$, which assigns to each point $\mathbf{x}$ in the inpainting domain $O$ a corresponding point $\varphi(\mathbf{x}) \in O^c$ in the complementary set $O^c=\Omega\setminus O$, where the image is known. 
A solution to the image inpainting problem can be obtained by joint minimization of the following energy \cite{wexler2007space,kawai2009image,inpainting_arias_2011} 
\begin{equation*}
\mathcal{E}_E(\hat{u},\varphi)= \int_{\tilde{O}} E(p_{\hat{u}}(\mathbf{x})-p_{u}(\varphi(\mathbf{x})))d\mathbf{x},
\end{equation*}
where $\hat{u}$ is unknown and set to be equal to the given $u$ in $O^c$,  $E$ is a patch error function which measures the patch similarity, $p_{\hat{u}}(\mathbf{x})$ and $p_u(\varphi(\mathbf{x}))$ denote a patch of $\hat{u}$ and $u$, respectively, centered at $\mathbf{x}$ and $\varphi(\mathbf{x})$, and  $\tilde{O}$ is the set of centers of patches that intersect the inpainting domain $O$. In our case, $O=\cup_{i=1}^s F(\mathbf{x}^i_{fs})$. In \cite{inpainting_arias_2011} four different patch similarity measures $E$ were proposed, namely, the so-called patch non-local means, patch non-local medians, patch non-local Poisson and patch non-local gradient means. The authors of \cite{fedorov2015variational} provide an algorithm and an implementation on the first three methods. In the present paper, we use the patch non-local medians method in all our results.


\section{Evaluation and Results}\label{sec:results}

This section focuses on the quantitative and qualitative evaluation of the proposed flare spot artifacts detection and removal method. Specifically, we separately analyze each of the three main steps of our algorithm: flare spot detection, flare mask computation and flare spot removal. For every step and all the experiments in this section, we compare our results with the ones obtained with the Automatic Lens Flare Removal (ALFR) \cite{chabertautomated} which we find to be a representative automatic flare spot detection method. Furthermore, we will not compare the results with \cite{nussberger2016robust,Psotny09,wu2005bayesian} as their scope is different. Indeed, the goal of \cite{nussberger2016robust} is the detection of artifacts in aerial images, \cite{Psotny09} is not an automatic method since the user has to select either the area where the flare spot is present or specify the color of the flare for each case, and \cite{wu2005bayesian} is neither automatic nor it has the same goal. Besides, for the method in \cite{chabertautomated}, the author's implementation is available. We use this implementation with the author's choice of parameters. 
In the following, we will refer to this method \cite{chabertautomated} by \emph{ALFR}.
In the case of our method, all the parameters have been fixed for all our experimental results. We refer to Table \ref{tab:parameters} for these parameter values. 

\begin{table}[ht!]
        \caption{Parameter values; they are fixed for all the experimental results}
  \begin{center}
      \label{tab:parameters}
\begin{tabular}{lll}
Section & Parameter & Value \\
\hline\noalign{\smallskip}
~\ref{sec:sourcelight} & $\iota$  & 99  \\
\noalign{\smallskip}\noalign{\smallskip}
~\ref{sec:FSdetector} 
&$\sigma_{min}$  & 3  \\
\noalign{\smallskip}\noalign{\smallskip}
&$\sigma_{max}$  & 15  \\
\noalign{\smallskip}\noalign{\smallskip}
 & $\delta$ & 10 \\
 \noalign{\smallskip}\noalign{\smallskip}
 & $\beta$ & 0.7 \\
\noalign{\smallskip}\noalign{\smallskip}
~\ref{sec:removal} &$\varepsilon$   & 5\\
\noalign{\smallskip}\noalign{\smallskip}
&$\alpha$  & 0.2   \\
\noalign{\smallskip}\hline\noalign{\smallskip}
    \end{tabular}
  \end{center}
\end{table}

\begin{figure*}
    \centering
    \begin{tikzpicture}
        \node[anchor=south west,inner sep=0] at (0,0) {
        \includegraphics[angle=270,width=2.7cm]{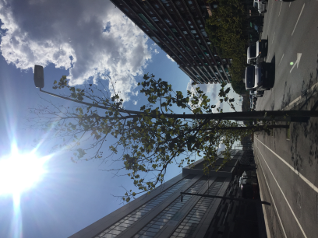}
         \includegraphics[width=3.6cm, angle=270]{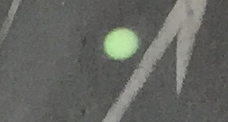}
         \includegraphics[ angle=270,width=2.7cm]{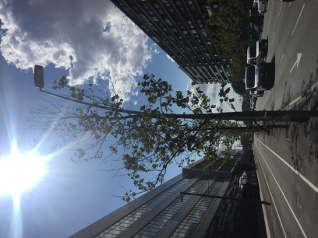}
    \includegraphics[width=3.6cm, angle=270]{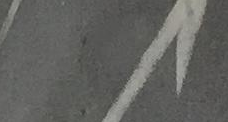}
    \includegraphics[ angle=270,width=2.7cm]{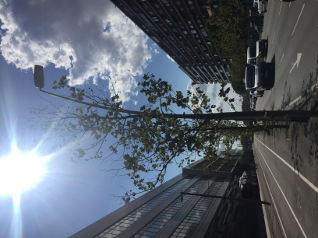}
    \includegraphics[width=3.6cm, angle=270]{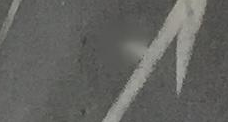}
         };
        \draw[red] (1.88,0.2) rectangle (2.05,0.45);
        \draw[red] (6.72,0.2) rectangle (6.89,0.45);
        \draw[red] (11.56,0.2) rectangle (11.73,0.45);
    \end{tikzpicture}
          \vspace{0.2cm}
     \begin{tikzpicture}
        \node[anchor=south west,inner sep=0] at (0,0) {
         \includegraphics[angle=270,width=2.7cm]{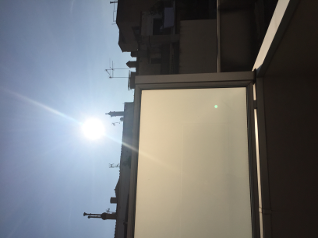}
    \includegraphics[width=3.6cm, angle=270]{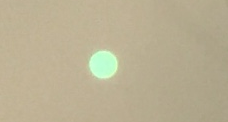}
        \includegraphics[ angle=270,width=2.7cm]{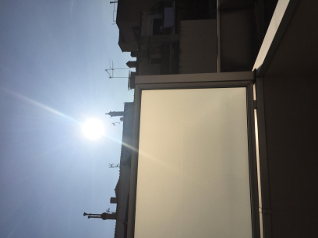}
    \includegraphics[width=3.6cm, angle=270]{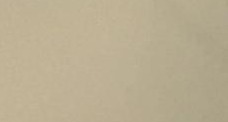}
    \includegraphics[ angle=270,width=2.7cm]{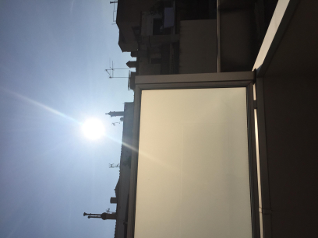}
    \includegraphics[width=3.6cm, angle=270]{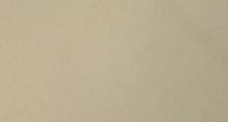}
         };
        \draw[red] (1.43,1.0) rectangle (1.6,1.25);
        \draw[red] (6.27,1.0) rectangle (6.44,1.25);
        \draw[red] (11.11,1.0) rectangle (11.28,1.25);
    \end{tikzpicture}
          \vspace{0.2cm}
     \begin{tikzpicture}
        \node[anchor=south west,inner sep=0] at (0,0) {
\includegraphics[angle=270,width=2.7cm]{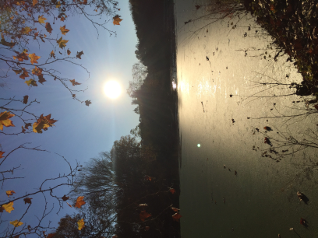}
    \includegraphics[width=3.6cm, angle=270]{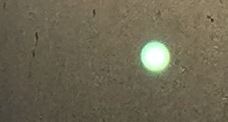}
    \includegraphics[ angle=270,width=2.7cm]{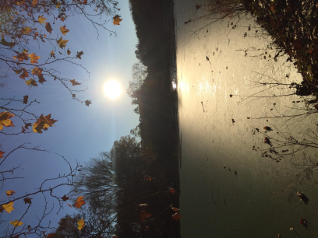}
    \includegraphics[width=3.6cm, angle=270]{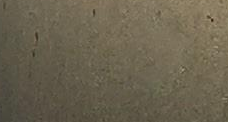}
    \includegraphics[ angle=270,width=2.7cm]{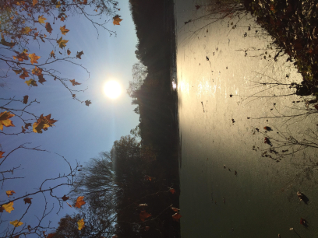}
    \includegraphics[width=3.6cm, angle=270]{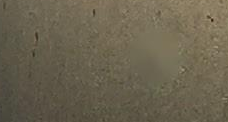}
         };
        \draw[red] (0.97,1.27) rectangle (1.14,1.52);
        \draw[red] (5.81,1.27) rectangle (5.98,1.52);
        \draw[red] (10.65,1.27) rectangle (10.82,1.52);
    \end{tikzpicture}
          \vspace{0.2cm}
       \begin{tikzpicture}
        \node[anchor=south west,inner sep=0] at (0,0) {
        \includegraphics[angle=270,width=2.7cm]{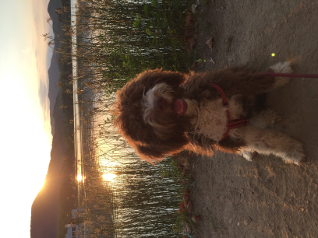}
    \includegraphics[width=3.6cm, angle=270]{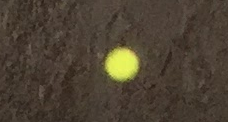}
        \includegraphics[ angle=270,width=2.7cm]{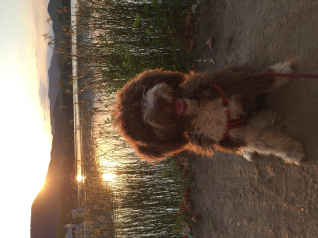}
    \includegraphics[width=3.6cm, angle=270]{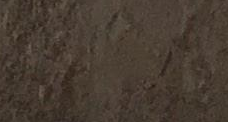}
    \includegraphics[ angle=270,width=2.7cm]{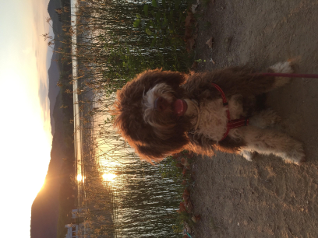}
    \includegraphics[width=3.6cm, angle=270]{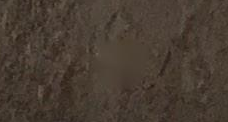}
         };
        \draw[red] (2.00,0.41) rectangle (2.17,0.66);
        \draw[red] (6.84,0.41) rectangle (7.01,0.66);
        \draw[red] (11.68,0.41) rectangle (11.85,0.66);
    \end{tikzpicture}
          \vspace{0.2cm}
       \begin{tikzpicture}
        \node[anchor=south west,inner sep=0] at (0,0) {
             \includegraphics[width=2.7cm]{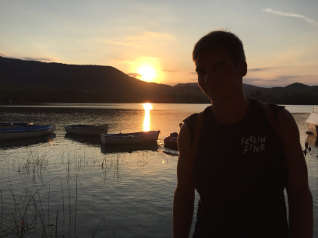}
    \includegraphics[width=1.95cm]{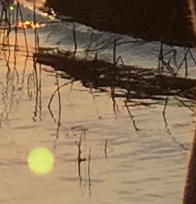}
        \includegraphics[ width=2.7cm]{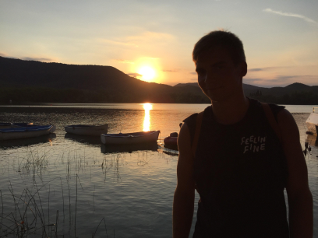}
    \includegraphics[width=1.95cm]{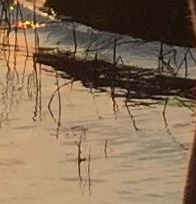}
    \includegraphics[ width=2.7cm]{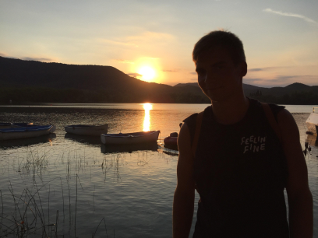}
    \includegraphics[width=1.95cm]{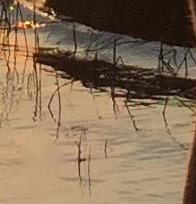}
         };
        \draw[red] (1.35,0.6) rectangle (1.52,0.79);
        \draw[red] (6.21,0.6) rectangle (6.38,0.79);
        \draw[red] (11.07,0.6) rectangle (11.24,0.79);
    \end{tikzpicture}
          \vspace{0.2cm}
           \begin{tikzpicture}
        \node[anchor=south west,inner sep=0] at (0,0) {
    \includegraphics[width=2.7cm]{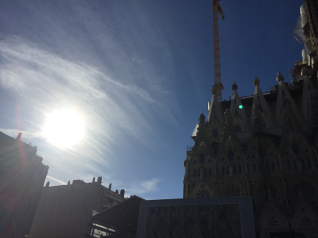}
    \includegraphics[width=1.95cm]{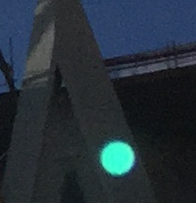}
        \includegraphics[ width=2.7cm]{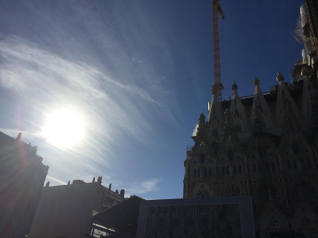}
    \includegraphics[width=1.95cm]{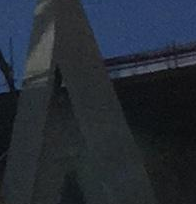}
    \includegraphics[ width=2.7cm]{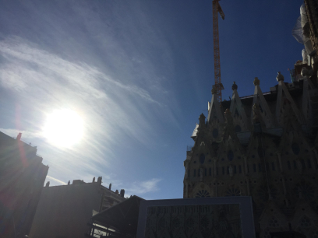}
    \includegraphics[width=1.95cm]{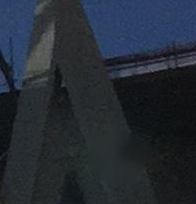}
         };
        \draw[red] (1.94,1.05) rectangle (2.11,1.24);
        \draw[red] (6.8,1.05) rectangle (6.97,1.24);
        \draw[red] (11.66,1.05) rectangle (11.83,1.24);
    \end{tikzpicture}

    \caption{Resulting images after flare spot reconstruction. Each row corresponds to one example. First and second column: original image and zoom around the flare spot artifact (indicated with a red box in the original image). Third and fourth column: resulting image using our algorithm and zoom around the reconstructed flare spot. Fifth and sixth column: resulting image using ALFR algorithm \citep{chabertautomated} and zoom around the reconstructed flare spot }
    \label{fig:InpaintingResultschabertOurs}
    
\end{figure*}

\begin{figure*}
    \centering
    \begin{tikzpicture}
        \node[anchor=south west,inner sep=0] at (0,0) {
        \includegraphics[angle=270,width=2.7cm]{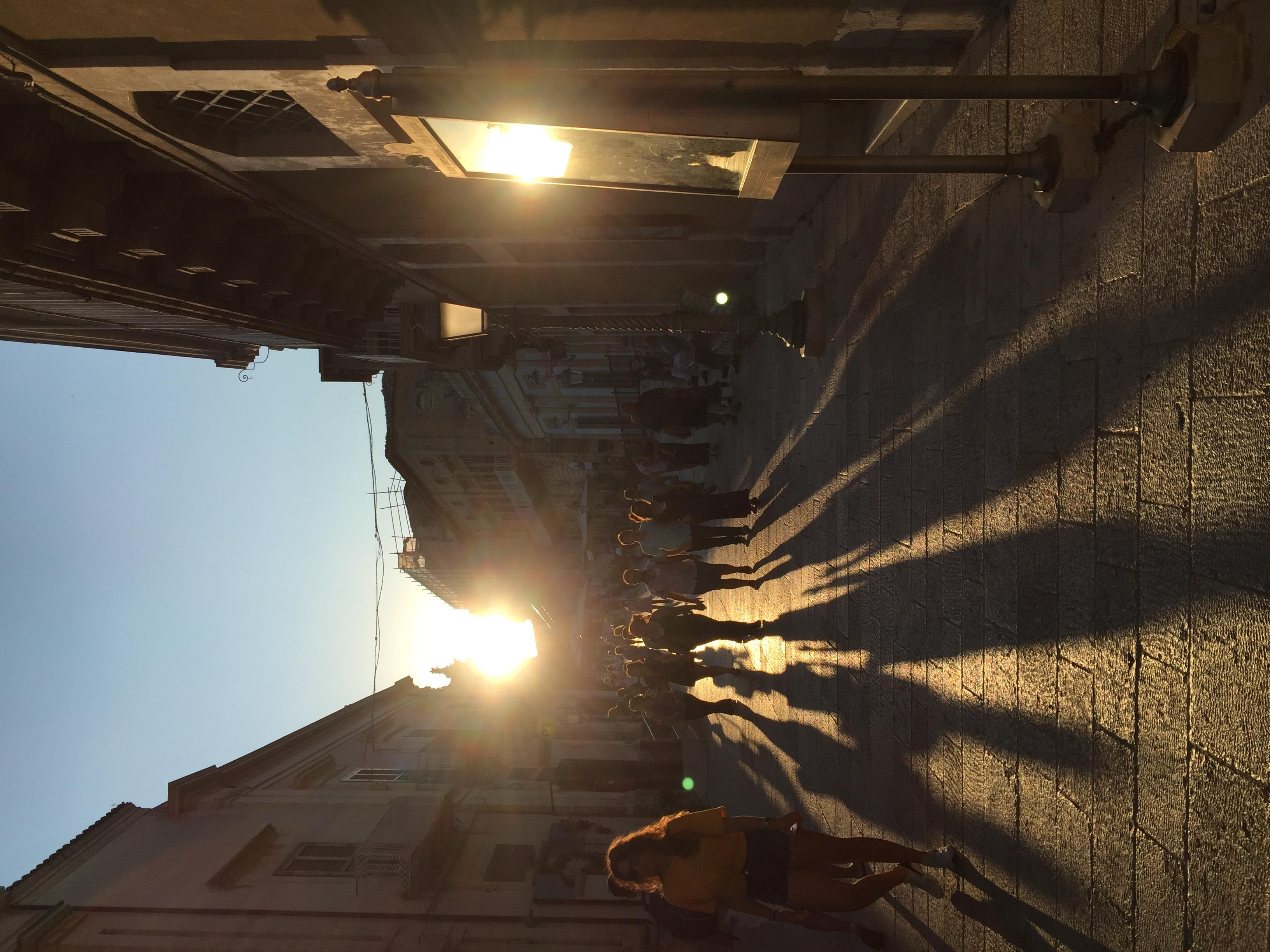}
         \includegraphics[height=3.6cm, angle=180]{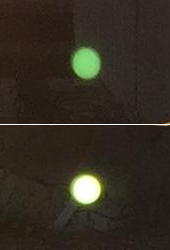}
         \includegraphics[ angle=270,width=2.7cm]{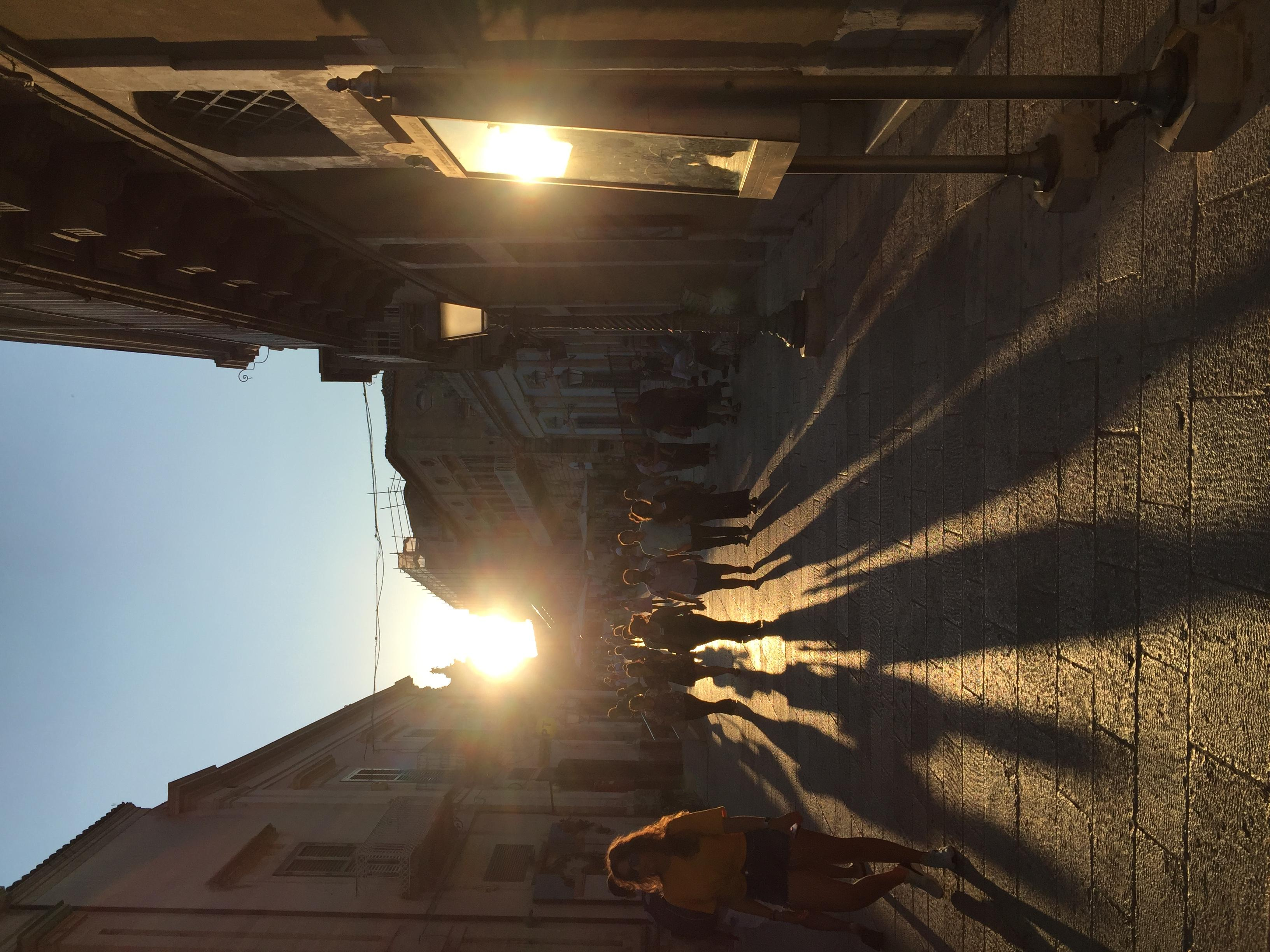}
    \includegraphics[height=3.6cm, angle=180]{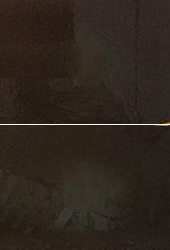}
    \includegraphics[ angle=270,width=2.7cm]{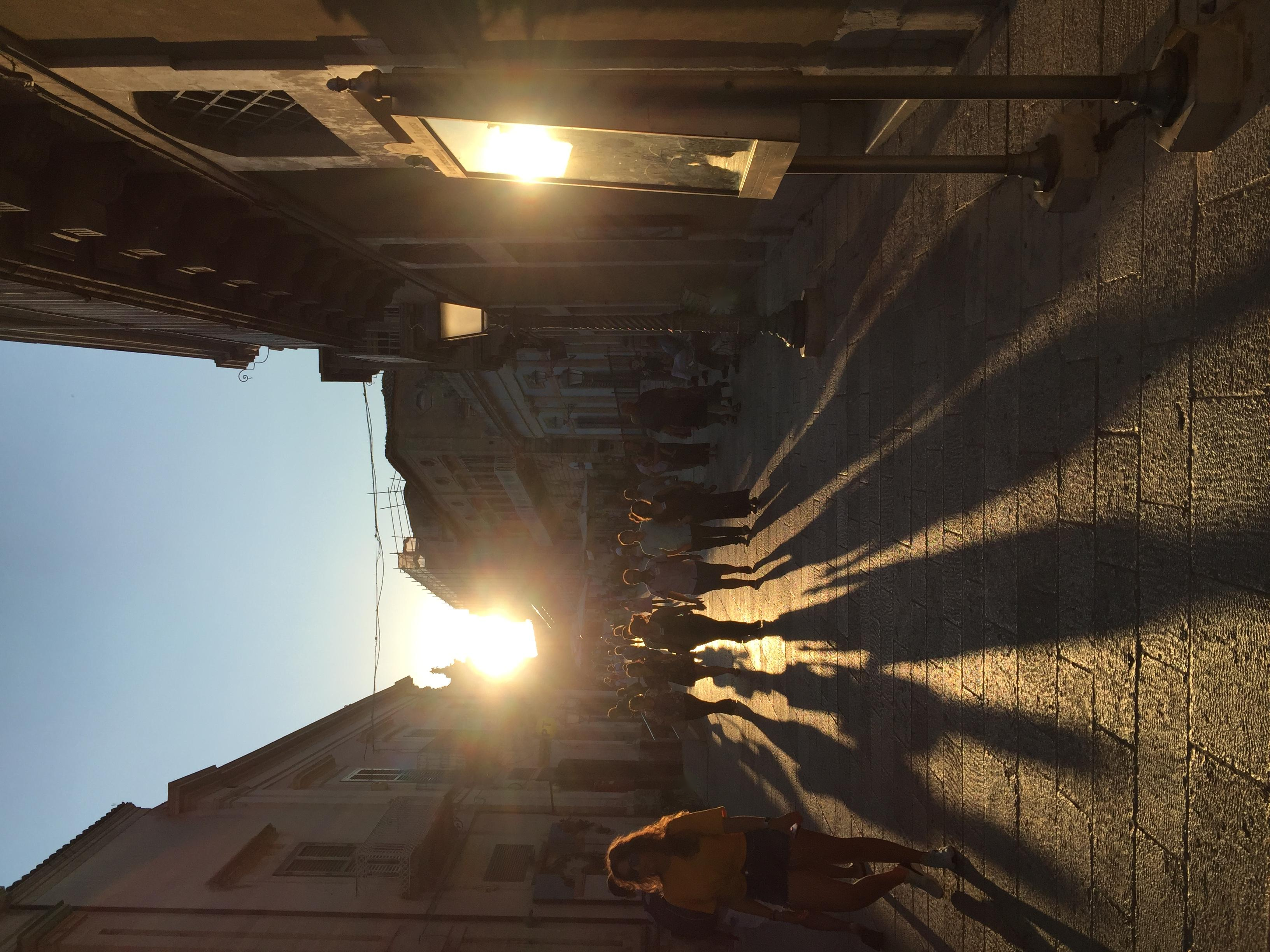}
    \includegraphics[height=3.6cm, angle=180]{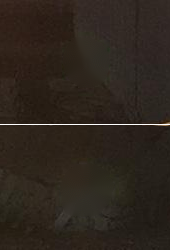}
         };
        \draw[red] (0.40,1.58) rectangle (0.57,1.705);
        \draw[red] (5.74,1.58) rectangle (5.91,1.705);
        \draw[red] (11.08,1.58) rectangle (11.25,1.705);
        
        \draw[red] (1.77,1.49) rectangle (1.94,1.615);
        \draw[red] (7.11,1.49) rectangle (7.28,1.615);
        \draw[red] (12.45,1.49) rectangle (12.62,1.615);

    \end{tikzpicture}
    
          \vspace{0.2cm}
     
   \begin{tikzpicture}
        \node[anchor=south west,inner sep=0] at (0,0) {
        \includegraphics[angle=270,width=2.7cm]{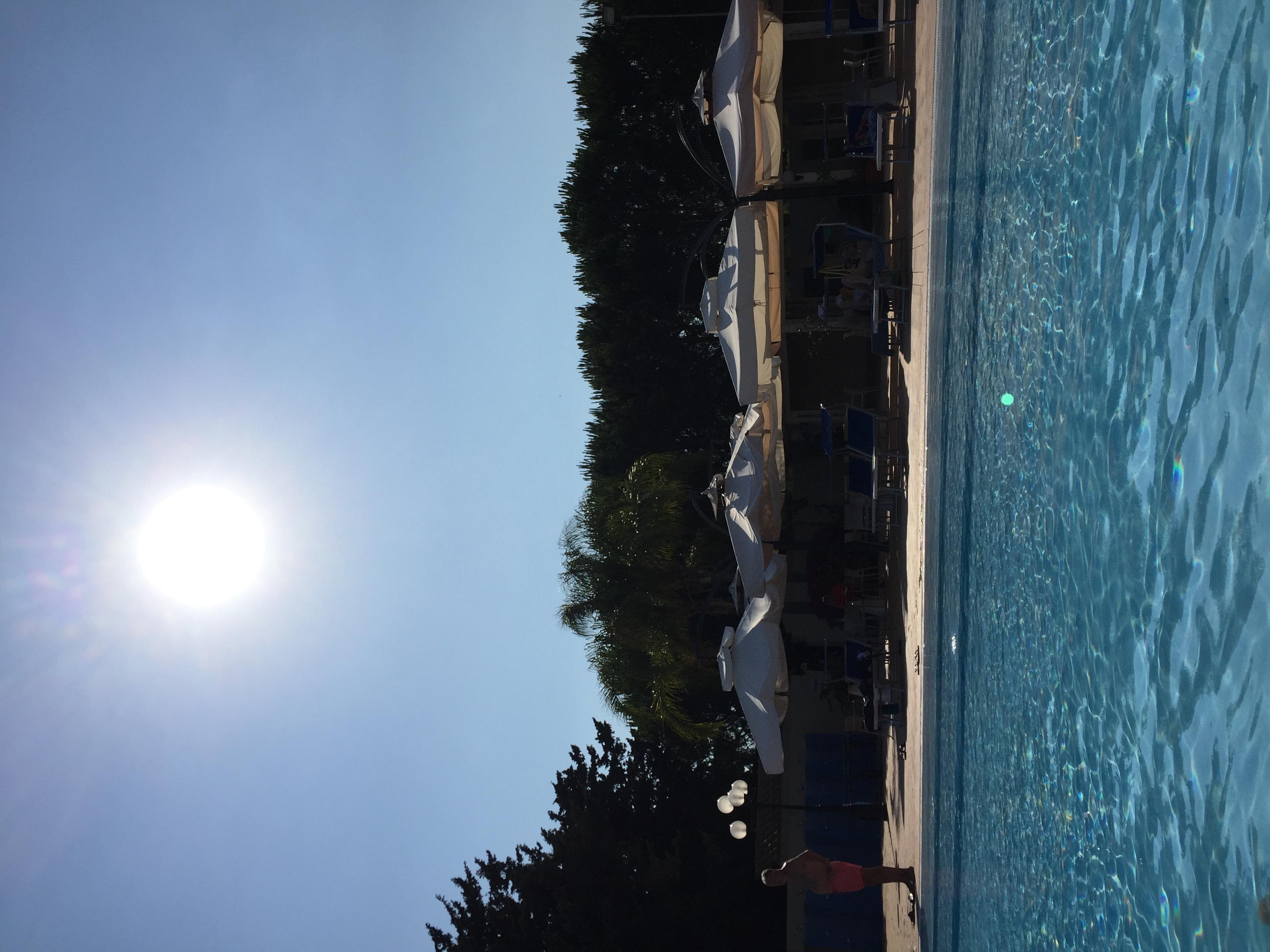}
         \includegraphics[width=3.6cm, angle=270]{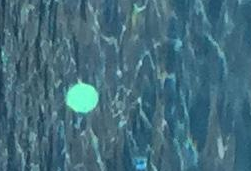}
         \includegraphics[ angle=270,width=2.7cm]{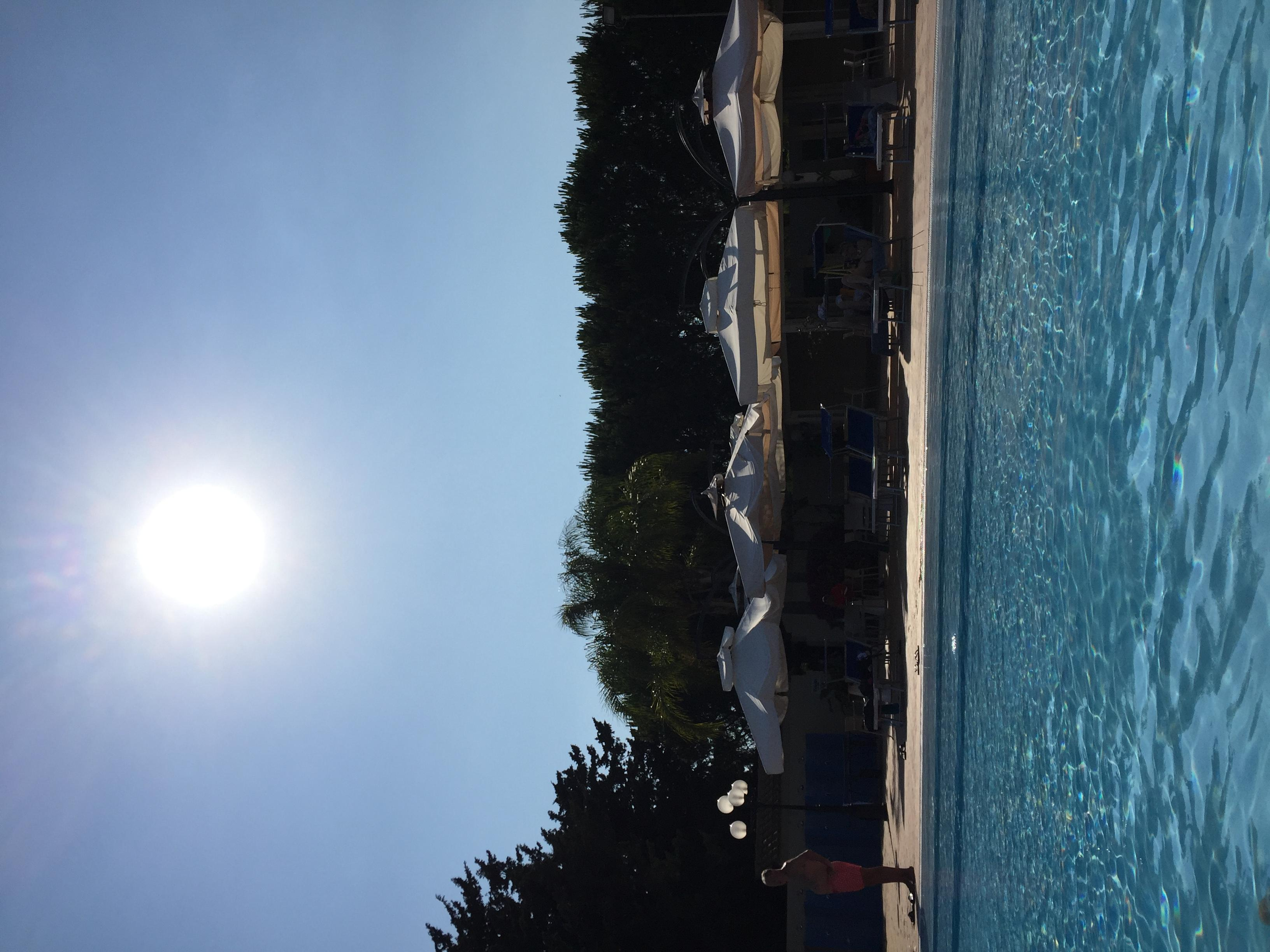}
    \includegraphics[width=3.6cm, angle=270]{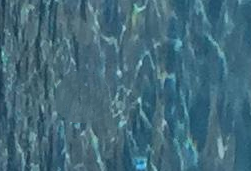}
    \includegraphics[ angle=270,width=2.7cm]{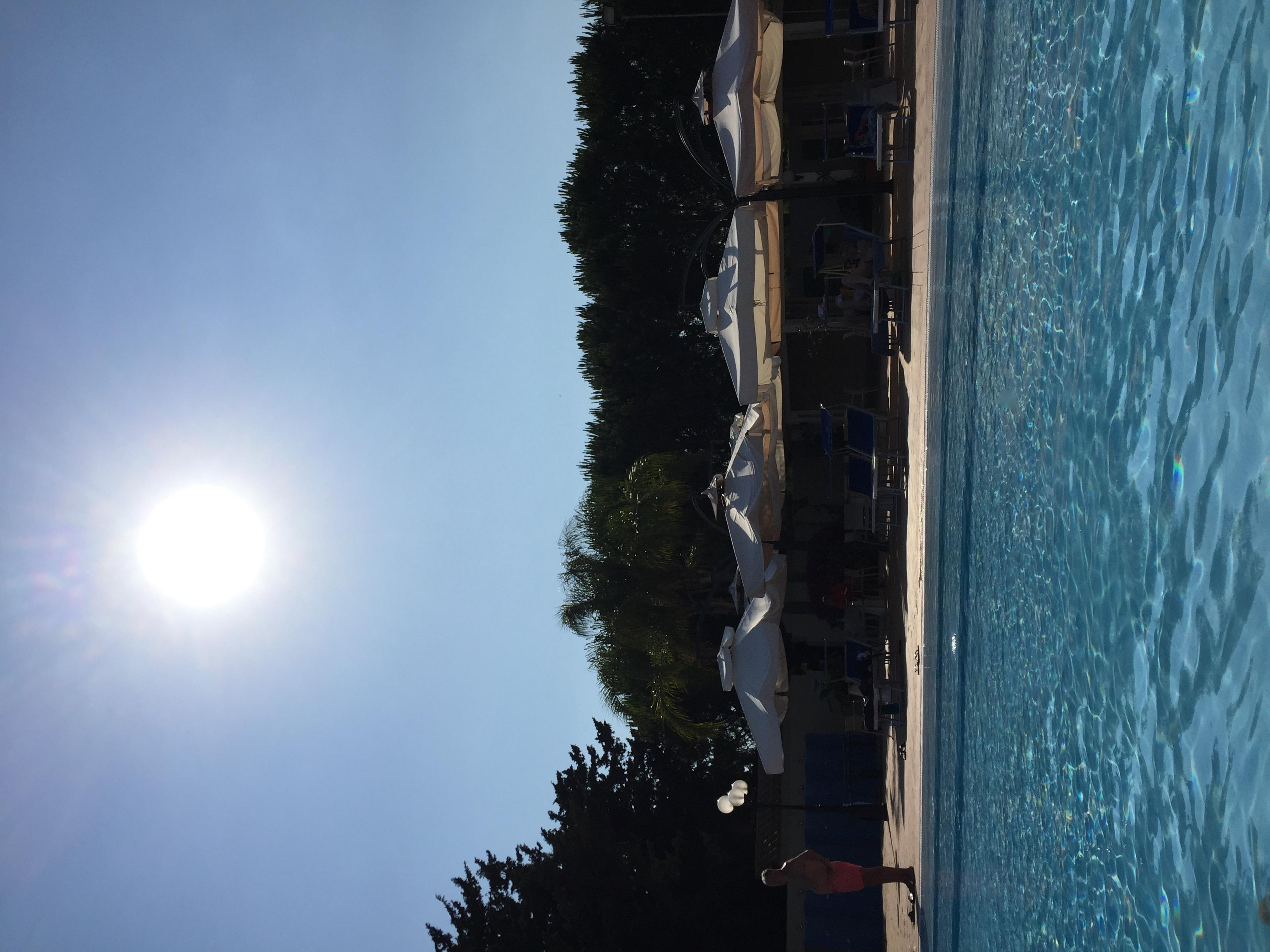}
    \includegraphics[width=3.6cm, angle=270]{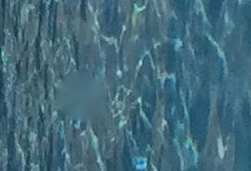}
         };
        \draw[red] (1.5,0.6) rectangle (1.67,0.85);
        \draw[red] (6.84,0.6) rectangle (7.01,0.85);
        \draw[red] (12.18,0.6) rectangle (12.35,0.85);
    \end{tikzpicture}
    
           \vspace{0.2cm}
     
   \begin{tikzpicture}
        \node[anchor=south west,inner sep=0] at (0,0) {
        \includegraphics[width=2.7cm]{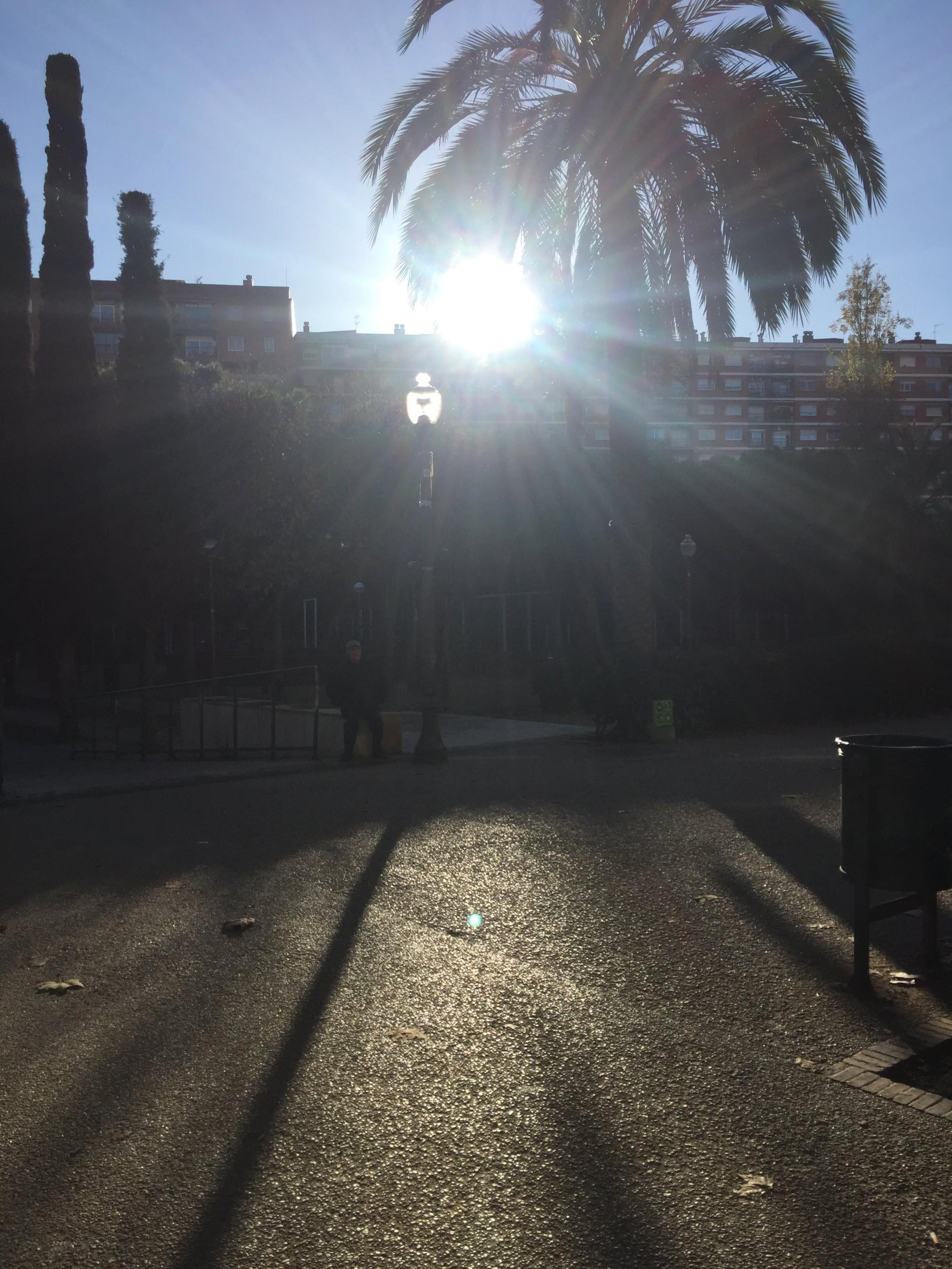}
         \includegraphics[height=3.6cm]{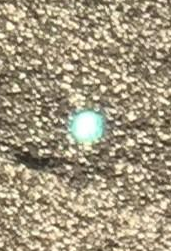}
         \includegraphics[ width=2.7cm]{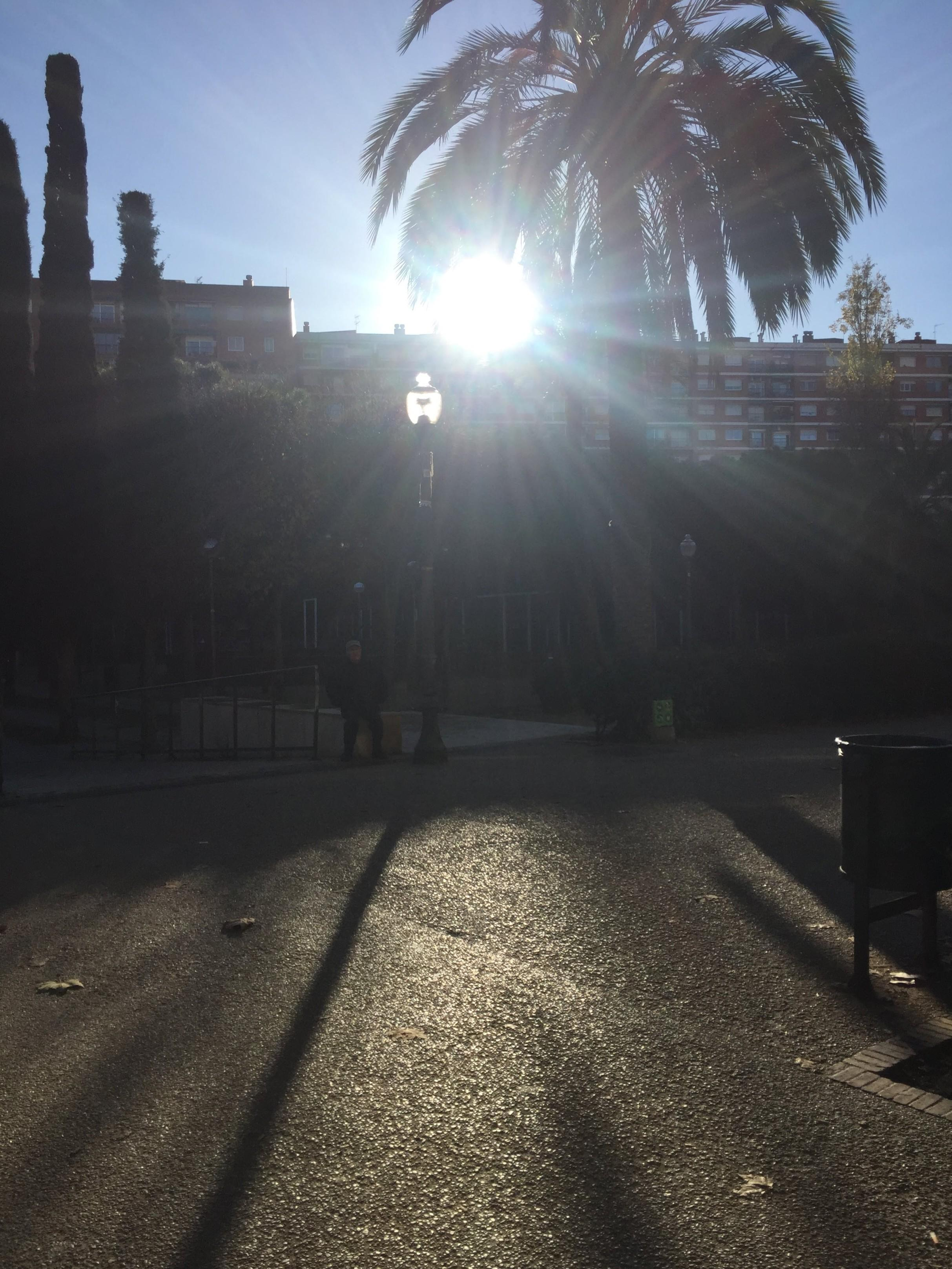}
    \includegraphics[height=3.6cm]{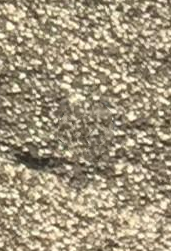}
    \includegraphics[ width=2.7cm]{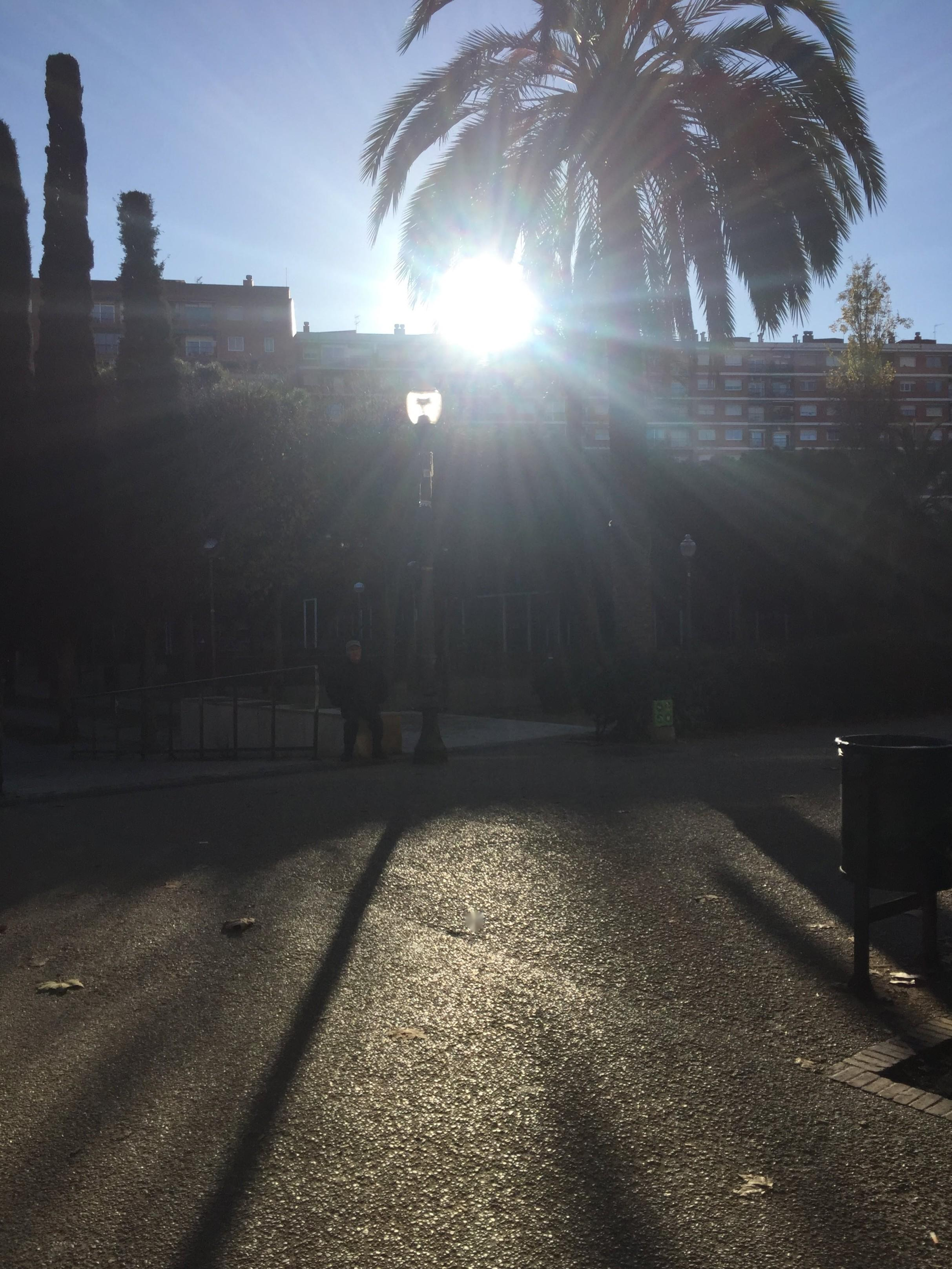}
    \includegraphics[height=3.6cm]{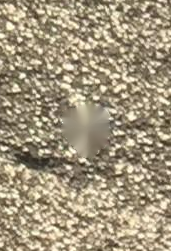}
         };
        \draw[red] (1.27,0.85) rectangle (1.44,1.1);
        \draw[red] (6.61,0.85) rectangle (6.78,1.1);
        \draw[red] (11.95,0.85) rectangle (12.12,1.1);
    \end{tikzpicture}

    \caption{Resulting images after flare spot reconstruction. Each row corresponds to one example. First and second column: original image and zoom around the flare spot artifact (pointed out with a red box in the original image). Third and fourth column: resulting image using our algorithm and zoom around the reconstructed flare spot. Fifth and sixth column: resulting image using ALFR algorithm \citep{chabertautomated} and zoom around the reconstructed flare spot. In the example on the first row, two flare spot artifacts appear which corresponding zooms are shown, one in top of the other, in the second, fourth and sixth colums}
    \label{fig:InpaintingResultschabertOurs2}
    
\end{figure*}

\begin{figure*}
    \centering
           \vspace{0.2cm}
           \begin{tikzpicture}
        \node[anchor=south west,inner sep=0] at (0,0) {
    \includegraphics[angle=270,width=2.6cm]{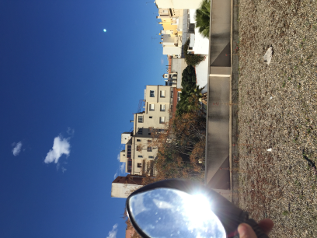}
    \includegraphics[angle=270,width=2.6cm]{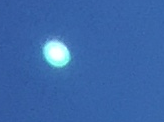}
    \includegraphics[angle=270, width=2.6cm]{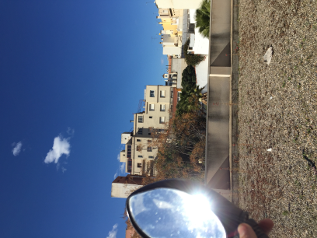}
    \includegraphics[angle=270,width=2.6cm]{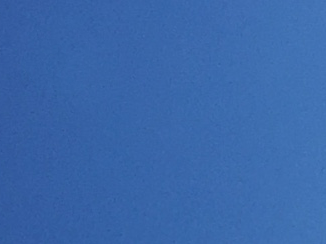}
         };
        \draw[red] (2.12,2.15) rectangle (2.37,2.45);
        \draw[red] (7.53,2.15) rectangle (7.78,2.45);
    \end{tikzpicture}
  \vspace{0.2cm}
 \begin{tikzpicture}
        \node[anchor=south west,inner sep=0] at (0,0) {
 \includegraphics[angle=270,width=2.6cm]{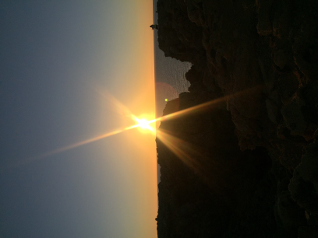}
    \includegraphics[width=3.45cm, angle=270]{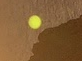}
    \includegraphics[angle=270, width=2.6cm]{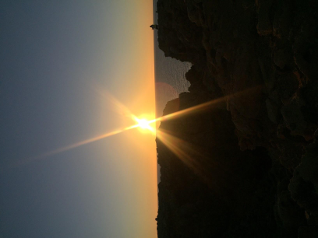}
   \includegraphics[width=3.45cm, angle=270]{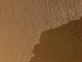}
         };
        \draw[red] (1.42,1.55) rectangle (1.57,1.75);
        \draw[red] (6.83,1.55) rectangle (6.98,1.75);
    \end{tikzpicture}
          \vspace{0.2cm}
               \begin{tikzpicture}
        \node[anchor=south west,inner sep=0] at (0,0) {
    \includegraphics[angle=270,width=2.6cm]{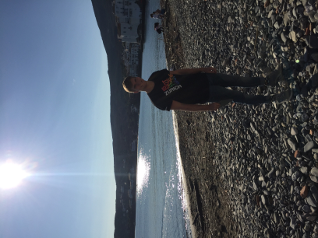}
    \includegraphics[width=3.45cm, angle=270]{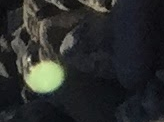}
    \includegraphics[angle=270, width=2.6cm]{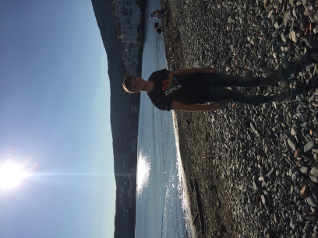}
    \includegraphics[width=3.45cm, angle=270]{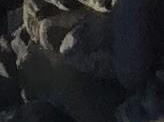}
         };
        \draw[red] (1.88,0.1) rectangle (2.03,0.3);
        \draw[red] (7.29,0.1) rectangle (7.44,0.3);
    \end{tikzpicture}
          \vspace{0.2cm}
                  \begin{tikzpicture}
        \node[anchor=south west,inner sep=0] at (0,0) {
 \includegraphics[angle=270,width=2.6cm]{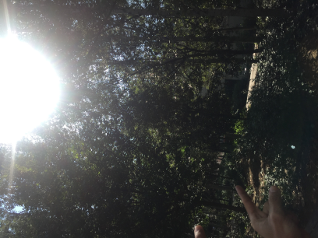}
    \includegraphics[width=3.45cm, angle=270]{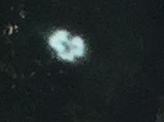}
    \includegraphics[angle=270, width=2.6cm]{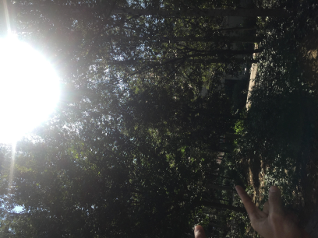}
    \includegraphics[width=3.45cm, angle=270]{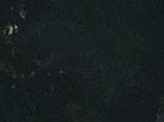}
         };
        \draw[red] (0.92,0.15) rectangle (1.07,0.35);
        \draw[red] (6.33,0.15) rectangle (6.48,0.35);
    \end{tikzpicture}
    \begin{tikzpicture}
        \node[anchor=south west,inner sep=0] at (0,0) {
 \includegraphics[angle=270,width=2.6cm]{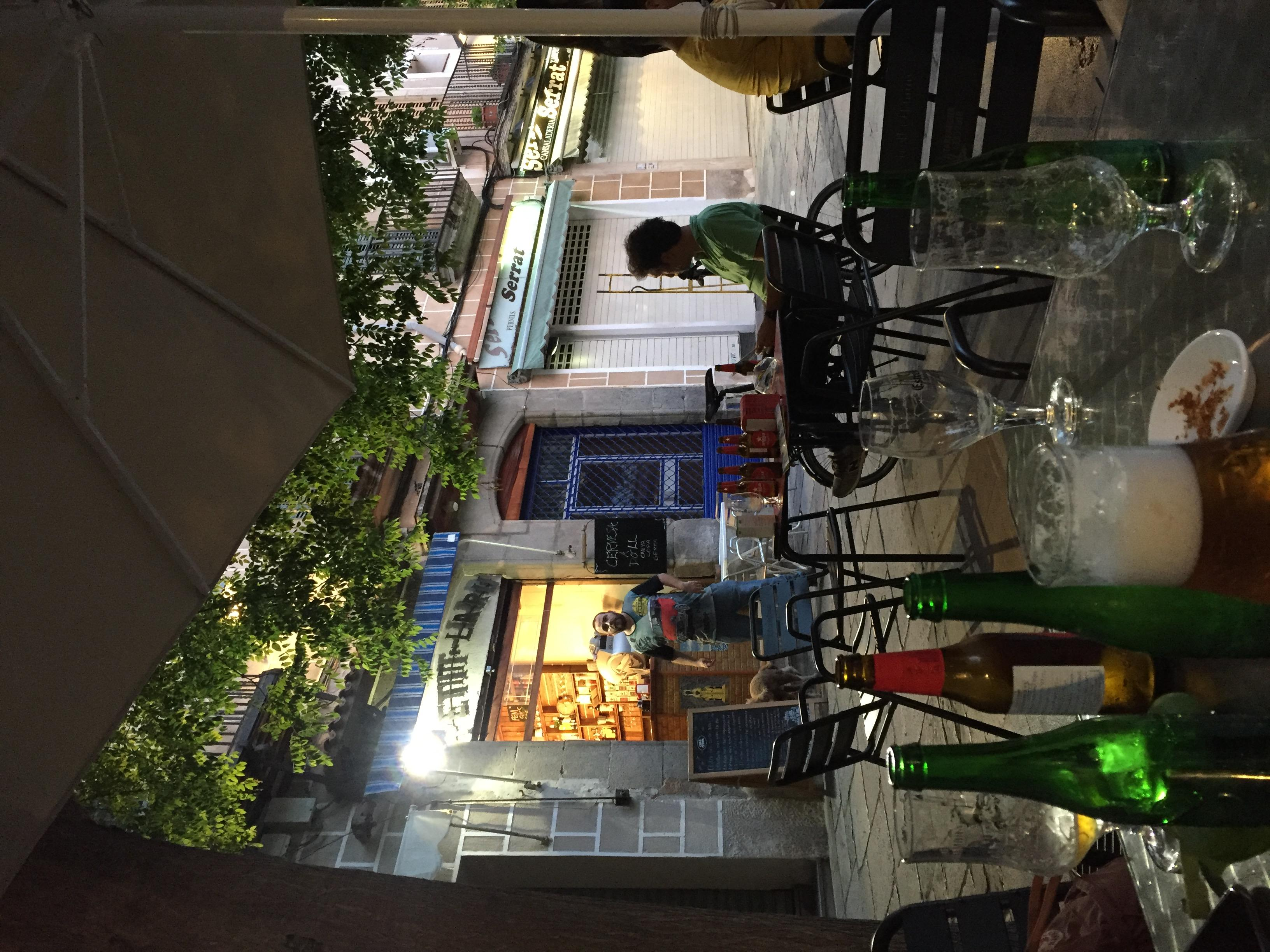}
    \includegraphics[angle=270,width=2.6cm]{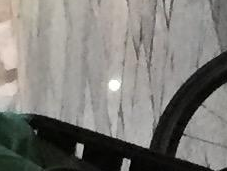}
    \includegraphics[angle=270, width=2.6cm]{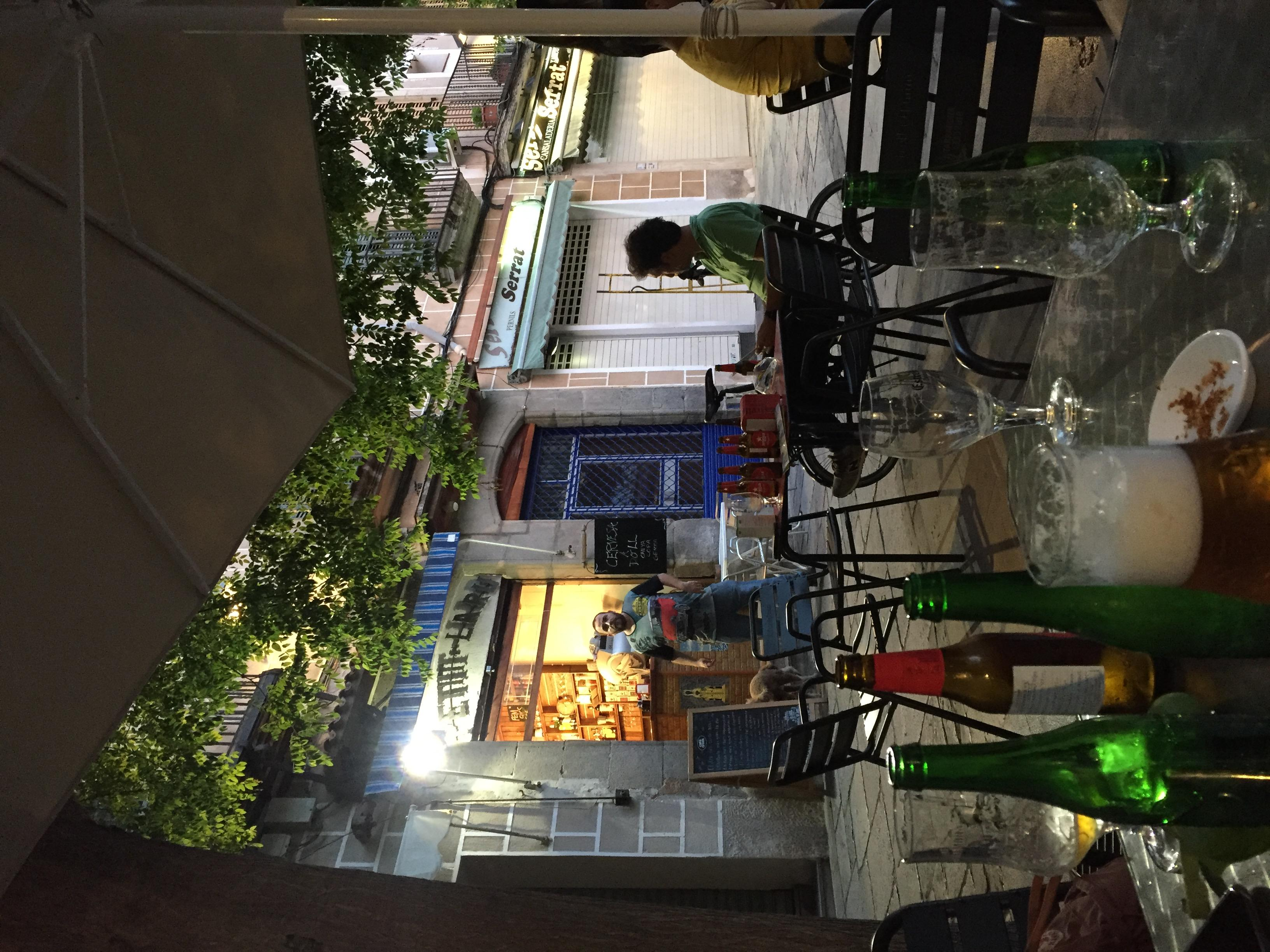}
    \includegraphics[angle=270,width=2.6cm]{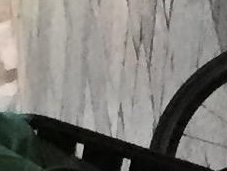}
         };
        \draw[red] (2.0,1.2) rectangle (2.15,1.45);
        \draw[red] (7.41,1.2) rectangle (7.56,1.45);
    \end{tikzpicture}
    
          \vspace{0.2cm}
                            \begin{tikzpicture}
        \node[anchor=south west,inner sep=0] at (0,0) {
           \includegraphics[width=2.6cm]{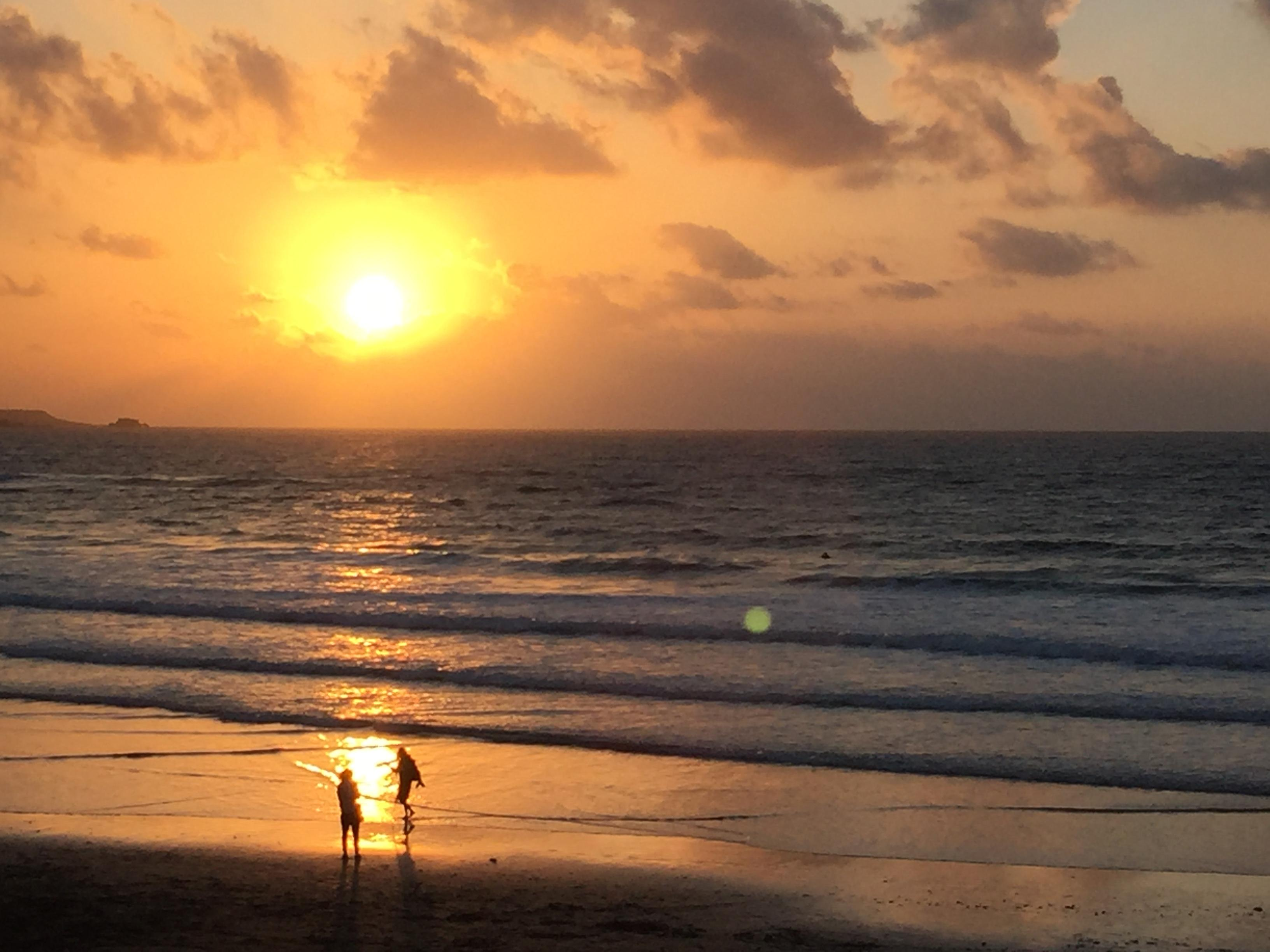}
    \adjincludegraphics[width=2.6cm]{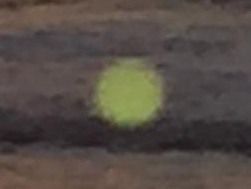}
    \includegraphics[width=2.6cm]{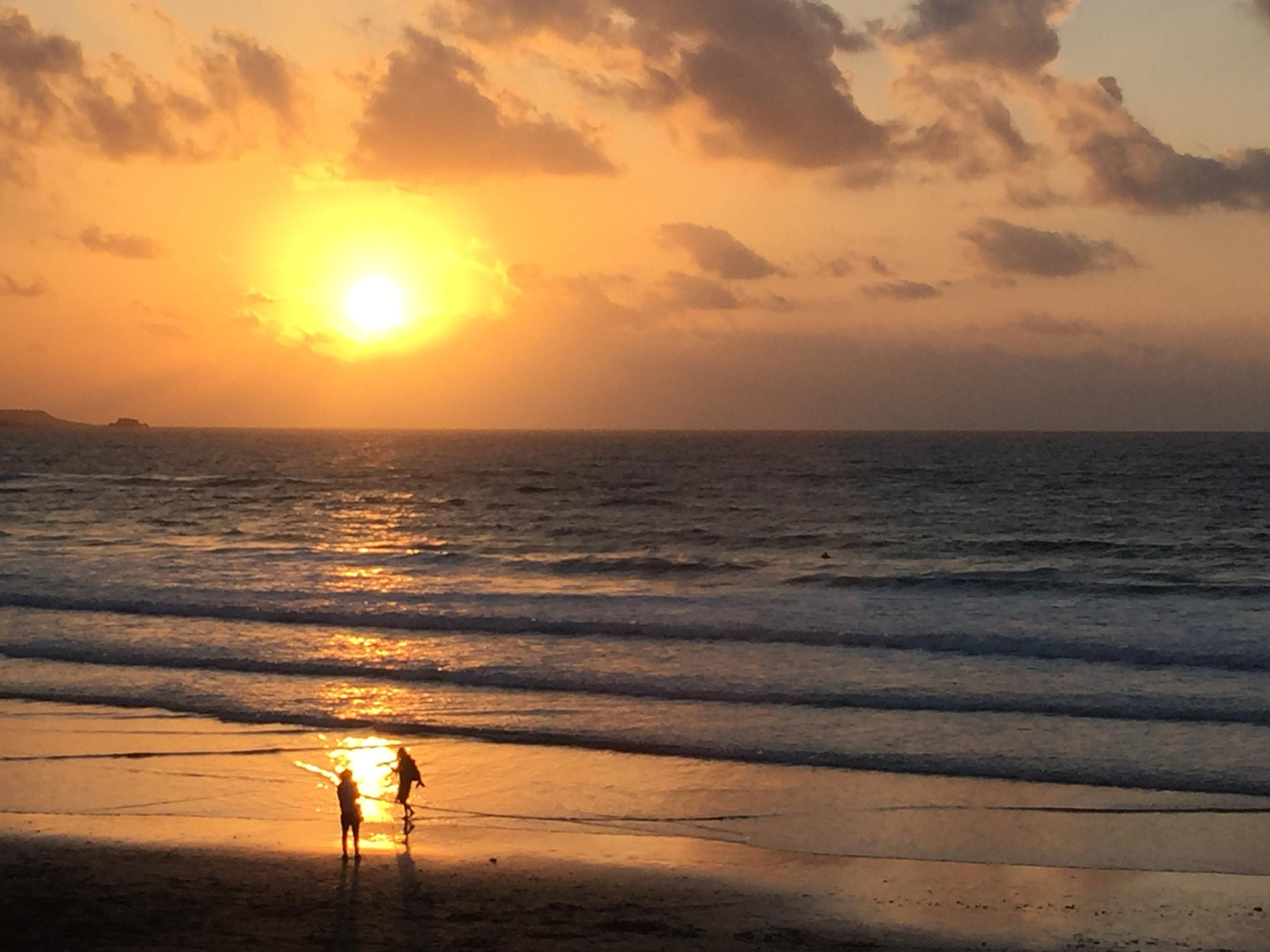}
    \adjincludegraphics[width=2.6cm]{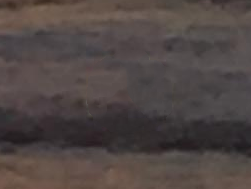}
         };
        \draw[red] (1.5,0.62) rectangle (1.65,0.75);
        \draw[red] (6.91,0.62) rectangle (7.06,0.75);
    \end{tikzpicture}
          \vspace{0.2cm}
          
                  \begin{tikzpicture}
        \node[anchor=south west,inner sep=0] at (0,0) {
           \includegraphics[width=2.6cm]{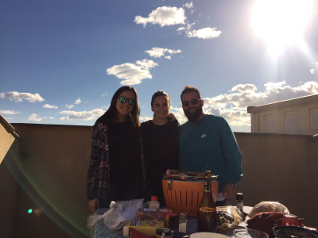}
    \adjincludegraphics[width=2.6cm]{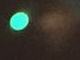}
    \includegraphics[width=2.6cm]{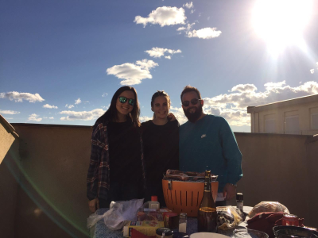}
    \adjincludegraphics[width=2.6cm]{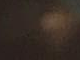}
         };
        \draw[red] (0.2,0.15) rectangle (0.35,0.28);
        \draw[red] (5.61,0.15) rectangle (5.76,0.28);
    \end{tikzpicture}
          \vspace{0.2cm}

    \caption{Resulting images after flare spot reconstruction, for some examples where our algorithm successfully detects the flare spot and ALFR does not. Each row corresponds to one example. First and second column: original image and zoom around the flare spot artifact (indicated with a red box in the original image). Third and fourth column: resulting image using our algorithm and zoom around the reconstructed flare spot}
    \label{fig:InpaintingResultsOurs1}
\end{figure*}

\begin{figure*}
    \centering
    \begin{tikzpicture}
        \node[anchor=south west,inner sep=0] at (0,0) {
    \includegraphics[width=2.6cm]{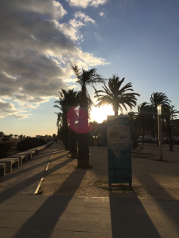}
    \includegraphics[width=2.6cm]{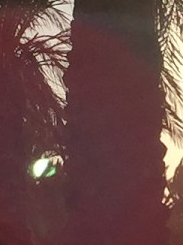}
    \includegraphics[ width=2.6cm]{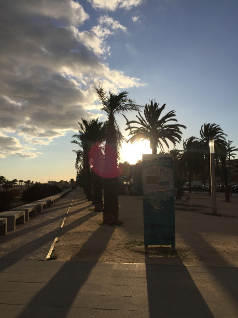}
    \includegraphics[width=2.6cm]{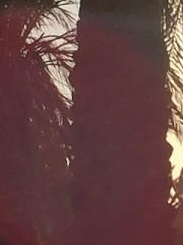}
         };
        \draw[red] (1.05,1.6) rectangle (1.3,1.9);
        \draw[red] (6.46,1.6) rectangle (6.71,1.9);
    \end{tikzpicture}
          \vspace{0.2cm}
                      \begin{tikzpicture}
        \node[anchor=south west,inner sep=0] at (0,0) {
    \includegraphics[width=2.6cm]{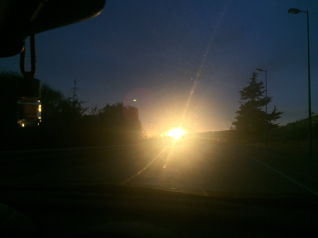}
    \includegraphics[width=2.6cm]{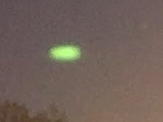}
    \includegraphics[width=2.6cm]{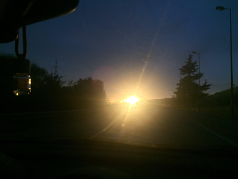}
   \includegraphics[width=2.6cm]{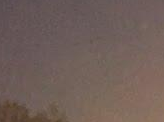}
         };
        \draw[red] (1.05,1.05) rectangle (1.20,1.19);
        \draw[red] (6.46,1.05) rectangle (6.61,1.19);
    \end{tikzpicture}
          \vspace{0.2cm}
          \begin{tikzpicture}
        \node[anchor=south west,inner sep=0] at (0,0) {
     \includegraphics[width=2.6cm]{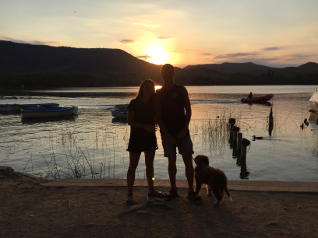}
    \includegraphics[width=2.6cm]{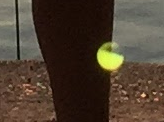}
    \includegraphics[width=2.6cm]{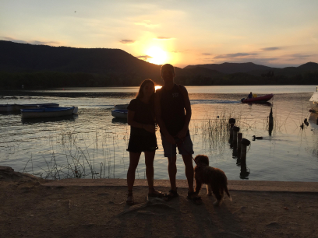}
   \includegraphics[width=2.6cm]{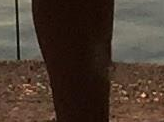}
         };
        \draw[red] (1.17,0.43) rectangle (1.32,0.57);
        \draw[red] (6.58,0.43) rectangle (6.73,0.57);
    \end{tikzpicture}
          \vspace{0.2cm}
        \begin{tikzpicture}
        \node[anchor=south west,inner sep=0] at (0,0) {
         \includegraphics[width=2.6cm]{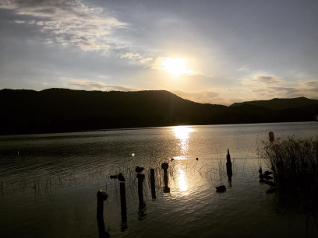}
    \includegraphics[width=2.6cm]{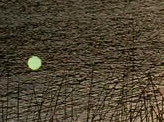}
    \includegraphics[width=2.6cm]{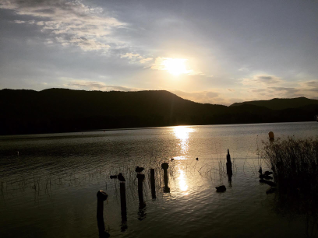}
   \includegraphics[width=2.6cm]{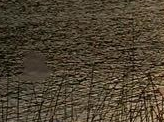}
         };
        \draw[red] (1.05,0.61) rectangle (1.2,0.75);
        \draw[red] (6.46,0.61) rectangle (6.61,0.75);
    \end{tikzpicture}
          \vspace{0.2cm}

    \caption{Resulting images after flare spot reconstruction, for some examples where our algorithm successfully detects the flare spot and ALFR does not. Each row corresponds to one example. First and second column: original image and zoom around the flare spot artifact (pointed out with a red box in the original image). Third and fourth column: resulting image using our algorithm and zoom around the reconstructed flare spot}
    \label{fig:InpaintingResultsOurs2}
\end{figure*}

Since there is not public dataset for flare spot detection we have 
built a dataset with $405$ natural images in which a minimum of one flare spot artifact appears. The sources of light can be the sun, light bulbs or specular surfaces, among others. The images have been captured by different cameras with different technical specifications. Thus, in particular, the size of the images is not the same, nor the technical and natural conditions. 
Some of them are shown in  Figures \ref{fig:InpaintingResultschabertOurs}, \ref{fig:InpaintingResultschabertOurs2},  \ref{fig:InpaintingResultsOurs1} and \ref{fig:InpaintingResultsOurs2}.
Notice that the flare spot has been created in a natural manner, in others words, it is not added by a computer after the acquisition. Generally our algorithm would not be able to remove artifacts added manually in post-processing since, for instance, the position of a real flare spot is dependent on the position of the root cause while artificially added flare spots may well not meet, e.g., spatial properties.  Thus, the dataset used in the work of \cite{chabertautomated} where the flare spot is created synthetically and without any pretext is not used in this work. 
 
As mentioned above, we divide this section in three subsections and provide quantitative assessment for each of them.


\subsection{Flare Spot Detection}

We start analyzing the performance of the proposed method for flare spot detection in comparison with ALFR~\citep{chabertautomated}. We evaluate its robustness by testing both algorithms on our dataset and computing its precision, recall, F-measure as well as the average number of false positives which have been detected per image.  A false positive is a region detected as a flare spot when actually it is not a flare spot. Thus, the optimum average false positive rate is zero. Recall refers to the true positive rate or sensitivity and precision to the positive predictive value. Their formulas are
\begin{equation*}
        \text{recall}=\frac{\text{number of true positive}}{\text{number of real positive}}
\end{equation*}
\begin{equation*}
\centering
    \text{precision}=\frac{\text{number of true positive}}{\text{number of detections as positive}}
\end{equation*}
The close are the values of recall and precision to $1$ the better. 
Finally, we also provide the F-measure (also called F score or F1 score). Let us recall that the F-measure is the harmonic mean of the precision and recall. It is computed as
\begin{equation*}
\text{F-measure} = 2\frac{\text{precision}\cdot \text{recall}}{\text{precision} + \text{recall}}
\end{equation*}
F-measure reaches its best value at 1 and worst at 0.
Table \ref{tab:SummaryTable} shows the corresponding quantitative results on the dataset of $405$ images. In particular, our method obtains a better detection performance than the one of ALFR. The first four rows of Table \ref{tab:SummaryTable} display the average detection quantitative measures on that dataset.  

\begin{table}[ht!]
        \caption{Flare spot detection performance, shown in the first four rows of quantitative results, and flare mask computation accuracy, in the last two rows, for ALFR method \citep{chabertautomated} and the proposed method}
  \begin{center}
      \label{tab:SummaryTable}
\begin{tabular}{lll}
\hline\noalign{\smallskip}
& ALFR  & Ours \\
\noalign{\smallskip}\hline\noalign{\smallskip}

Precision & 0.2813 & \textbf{0.8092} \\
\noalign{\smallskip}\noalign{\smallskip}
Recall & 0.6769  & \textbf{0.7862} \\
\noalign{\smallskip}\noalign{\smallskip}
F-measure  & 0.3974 & \textbf{0,7975} \\
\noalign{\smallskip}\noalign{\smallskip}
Average of false positive per image & 1.7975 & \textbf{0.1925} \\
\noalign{\smallskip}\noalign{\smallskip}
Average Mask Accuracy  & 0.4947 & \textbf{0.7231} \\
\noalign{\smallskip}\noalign{\smallskip}
Median Mask Accuracy  & 0.5000 & \textbf{0.7408} \\

\noalign{\smallskip}\hline\noalign{\smallskip}
    \end{tabular}
  \end{center}
\end{table}

Regarding recall, our higher value results (0.7862 in comparison to the 0.6769 from ALFR) could be understood as that our detection method is based on a more complete characterization of flare spot artifacts. In the case of ALFR, it restricts the search of flare spot into the most common case of circular, convex, bright blobs. Our algorithm is able to detect a flare spot which, for example, is not circular. 

Our precision is again closer to one (0.8092) and much higher than the precision obtained by ALFR  (0.2813). ALFR is prone to detect as flare spot all the regions of the image with similar geometric properties leading to a low precision.  Our algorithm, on the other hand, besides the geometric properties also has into account other properties, such as chromatic and location properties, resulting in a more accurate detection.

\begin{figure*}
    \centering
        \begin{tikzpicture}
            \begin{axis}[
                ybar,
                bar width=.4cm,
                width=0.95\textwidth,
                height=.5\textwidth,
                xtick=data,
                ymin=0,ymax=100,
            	ylabel= \% of the experiments,
            	xlabel= \# false detections per images
            ]
            \addplot 
            	coordinates {(0,42.96) (1, 20.74) (2,10.86) (3,7.4) (4,5.92) (5,3.45) (6,1.72) (7,1.97) (8,1.97) (9,0.49) (10,0.49) (11,0.24) (12,1.23) (13,0) (14,0.24) (15,0.24)};
            \addplot 
            	coordinates {(0,81.48) (1,17.53) (2,0.98) (3,0) (4,0) (5,0) (6,0) (7,0) (8,0) (9,0) (10,0) (11,0) (12,0) (13,0) (14,0) (15,0) };
            \legend{ALFR \citep{chabertautomated}, Ours}
            \end{axis}
        \end{tikzpicture}
    \caption{Graph showing the percentage of experiments (vertical axis) in relation to a number of false positive flare spots detected per image (horizontal axis) using ALFR \citep{chabertautomated} (in blue) and our algorithm (in red). A false positive flare spot is a position identified as a flare spot without being a real flare spot }
        \label{fig:falsepositiveChart}
\end{figure*}
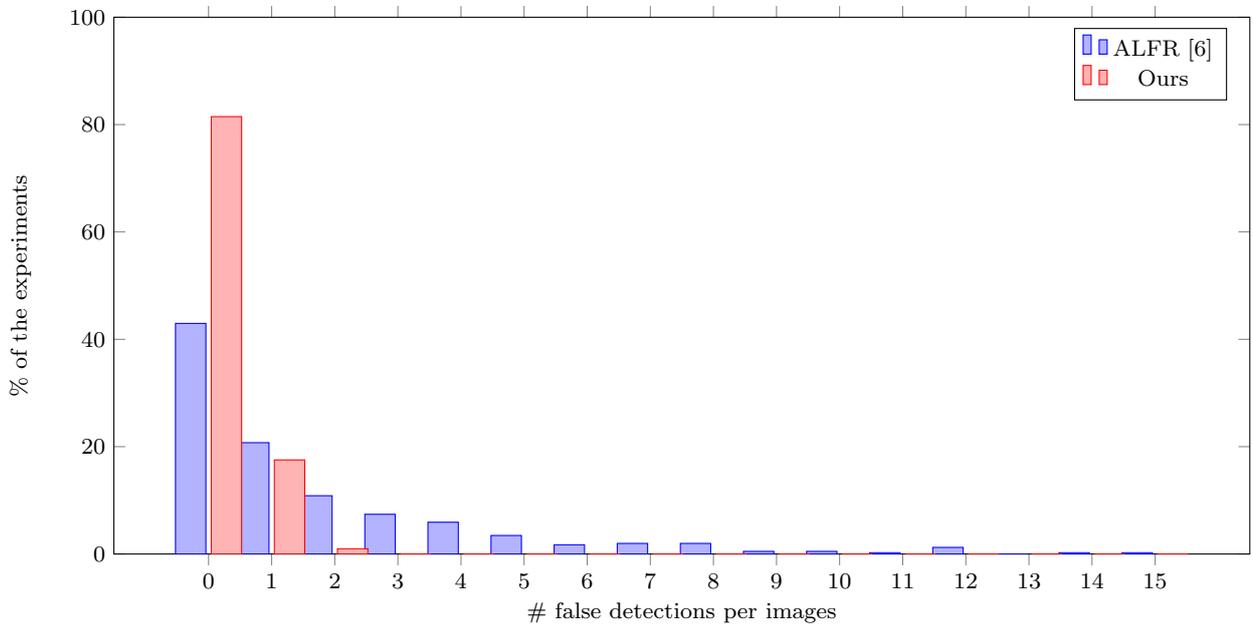

To further analyze it, we include the average number of false positives over all results. Our result is almost ten times smaller than ALFR: It obtains an average of 1.7975 false positives per image compared to the 0.1925 obtained by our method. Let us notice that, for an automatic detection and restoration method, it is really important to have a low false positive rate in order to do not lose information while applying inpainting (thus modifying the image) in regions where no flare spot is present. For a more detailed analysis, Figure \ref{fig:falsepositiveChart} shows the percentage of images from the dataset (vertical axis) with a certain number of false positives (horizontal axis).  Let us notice that our algorithm generally (in 81,5\% of the cases) does not have any false positive and, in the worst case, only has two. In the remaining intervals of the horizontal axis, our algorithm has zero or one false positives. On the other hand,  in 43 \% of the cases identifies at least one false positive per image and in the worst case could go up to a maximum of fifteen false positives in a single image.

\begin{figure}
    \centering
    \includegraphics[angle=270, width=4cm]{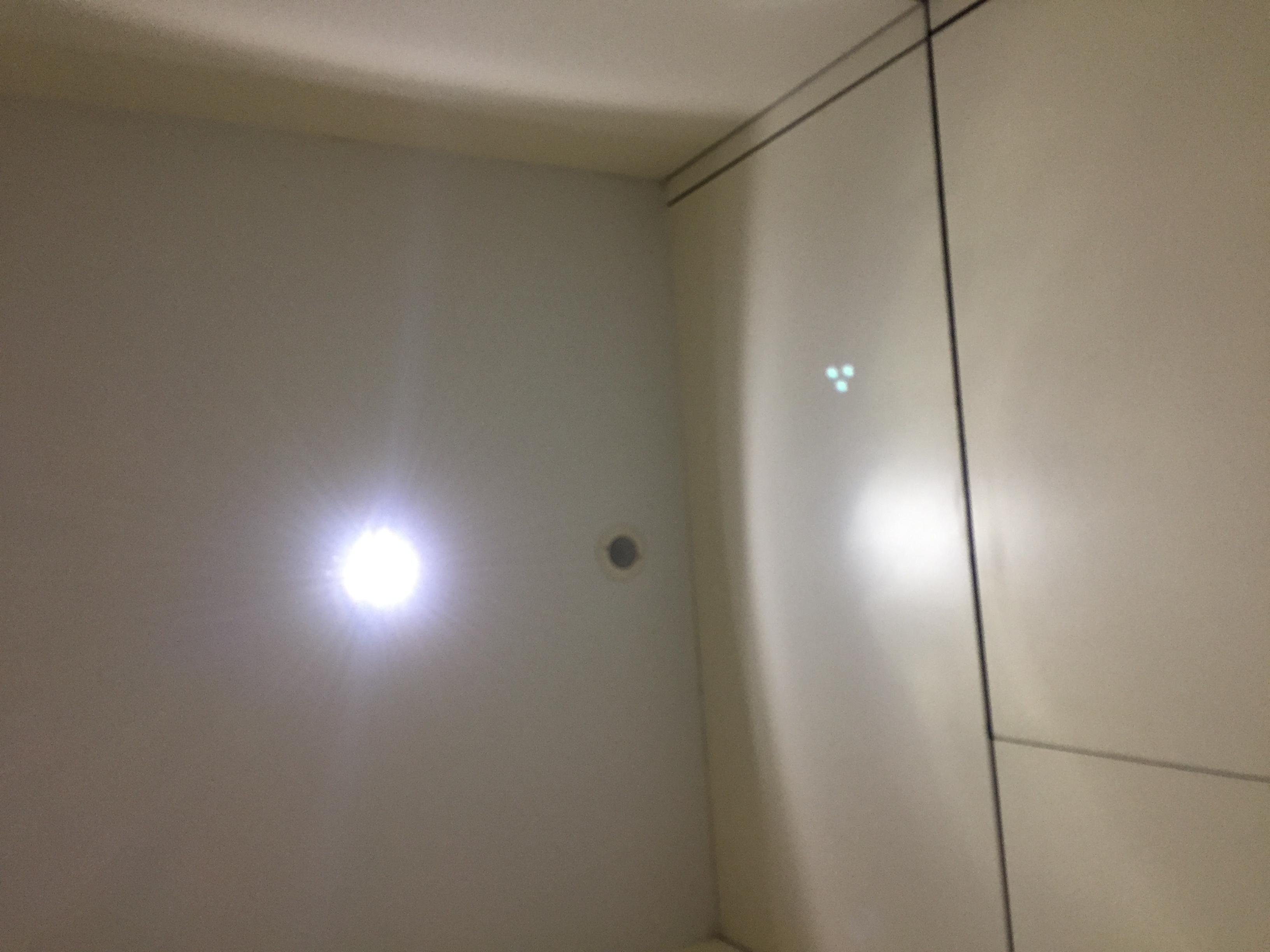}
    \includegraphics[angle=270, width=4cm]{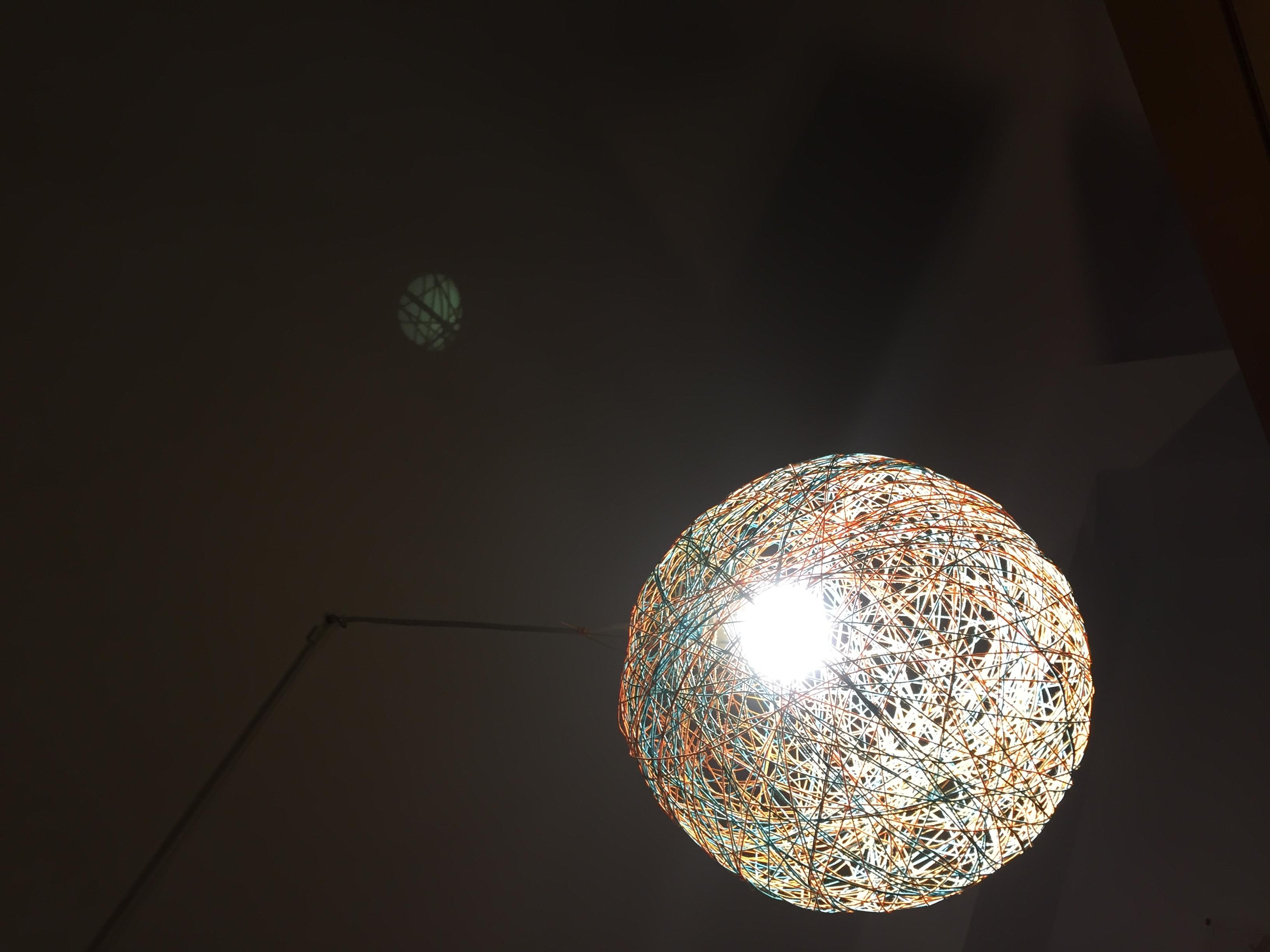}
        \includegraphics[ width=8.2cm]{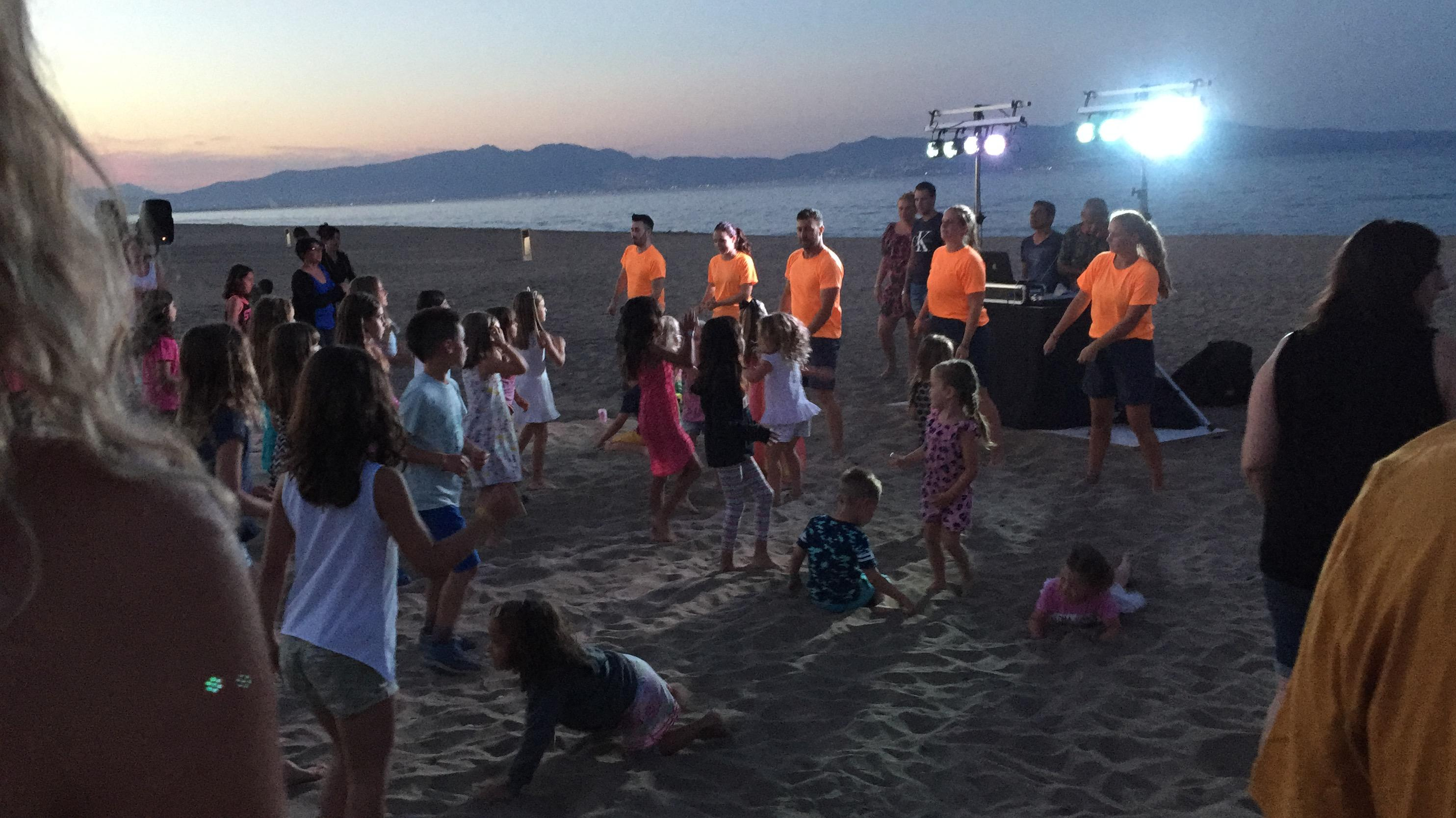}
    \caption{Examples of failure cases where the flare spot was not fully detected}
    \label{fig:typicalErrorFlare}
\end{figure}

We can see that our method for flare detection works correctly even in hard conditions: for example when the flare is close to the sun (second row in Figure \ref{fig:InpaintingResultsOurs1}), when several bright circular shape textures are present in the image (third and fourth rows in Figure \ref{fig:InpaintingResultsOurs1}), when the flare has a complex shape ( fourth row in Figure \ref{fig:InpaintingResultsOurs1} and first, second and third row in Figure \ref{fig:InpaintingResultsOurs2}), when the bright light source is an specular reflection surface (first row in Figure  \ref{fig:InpaintingResultsOurs1}) or in a highly complex scene where the scene is composed by reflective products such as glasses  (fifth row in Figure \ref{fig:InpaintingResultsOurs1}).

\paragraph{Discussion on failure cases}
The obtained recall and precision are not equal to one and neither the average false positive equal to zero which translate that some of the artifacts are not detected and a false positive has been detected for some cases. These failure cases are mainly due to the light source detector. When there exists a combination of light bulbs (one next to each other) present in the scene, the light source detector detects them as a single light source. Thus, if the algorithm finds only one light source, it looks only for a single flare spot artifacts, while indeed each light bulb is generating an independent flare spot. For that reason, our algorithm only finds a part of each flare spot artifact, instead of finding the combination of each of them. This problem could be solved by a more sophisticated light source detector. In our case, we lose precision on light source detector by saving computational time.  Figure \ref{fig:typicalErrorFlare} shows three examples of these situations.

\subsection{Flare Mask estimation}

\begin{figure*}
    \centering
    \includegraphics[angle=270,width=2.8cm]{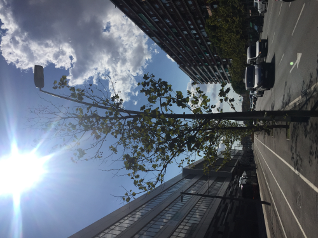}
    \includegraphics[angle=270,width=2.8cm]{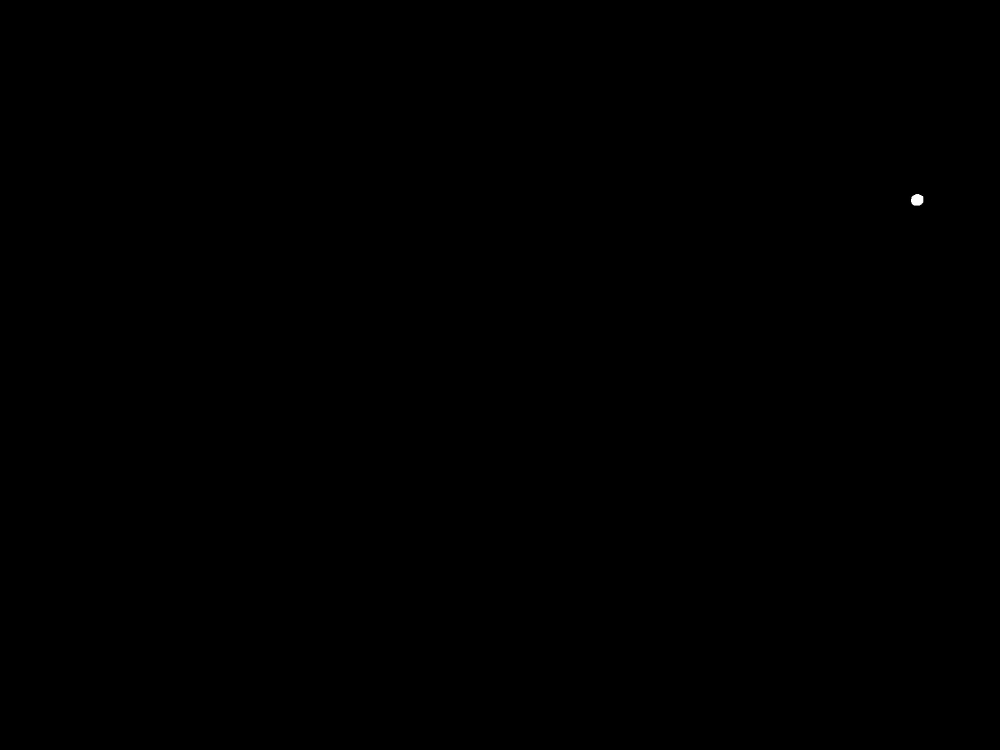}
    \includegraphics[ angle=270,width=2.8cm]{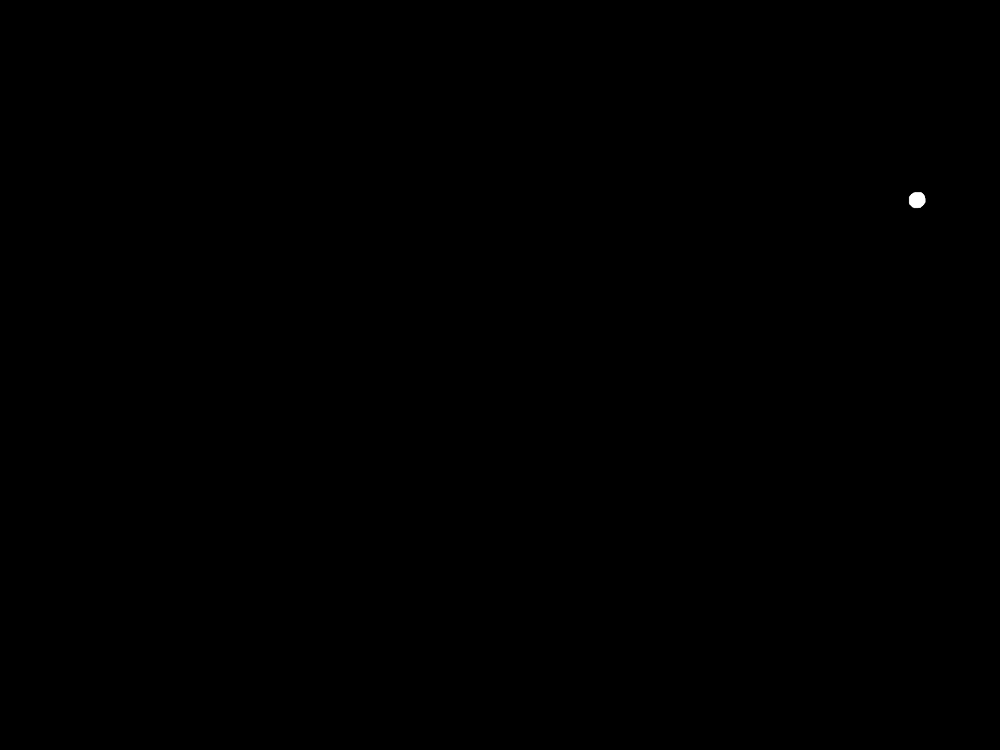}
    \includegraphics[ angle=270,width=2.8cm]{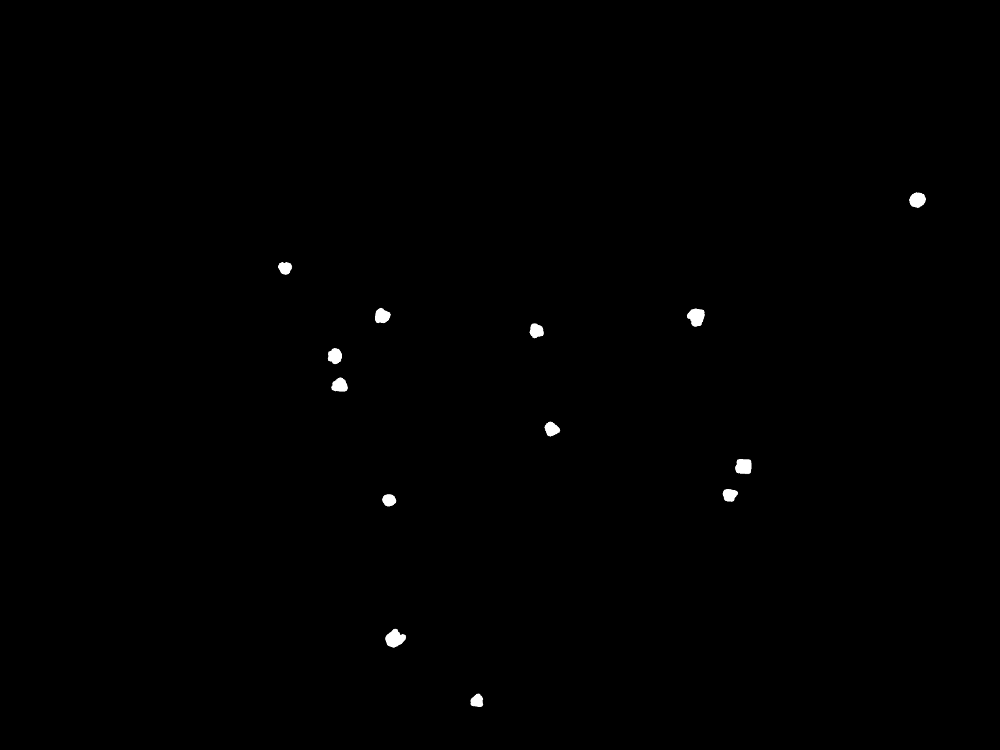}
     
    \includegraphics[angle=270,width=2.8cm]{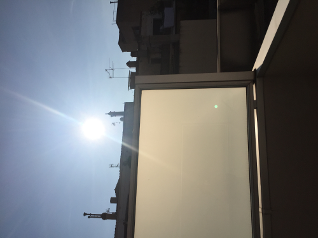}
    \includegraphics[angle=270,width=2.8cm]{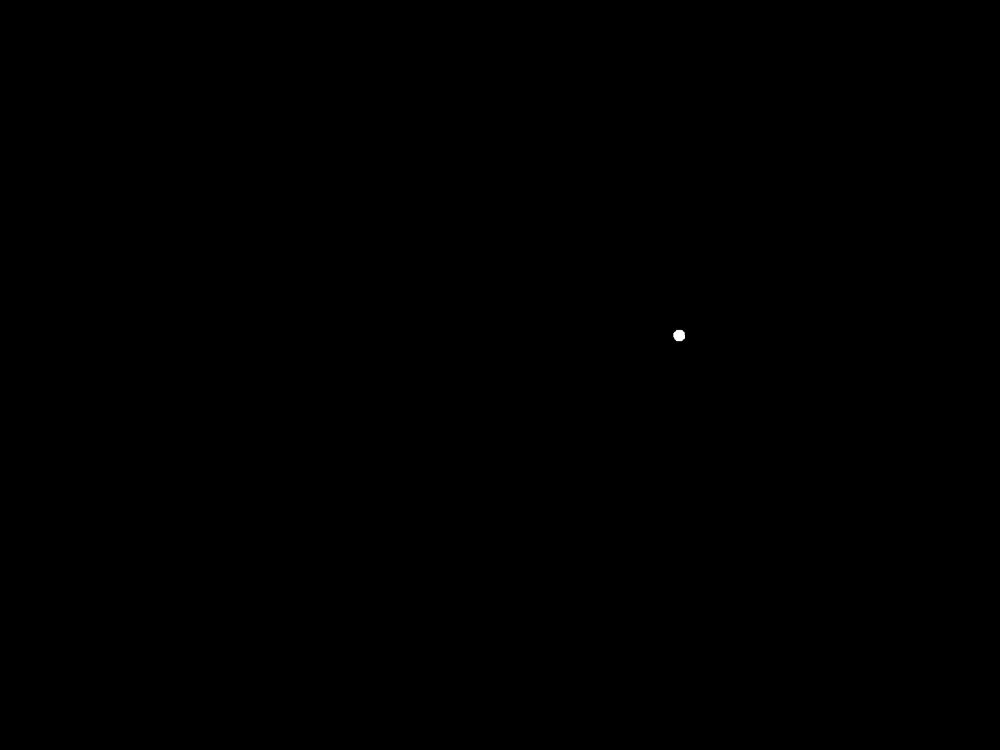}
    \includegraphics[ angle=270,width=2.8cm]{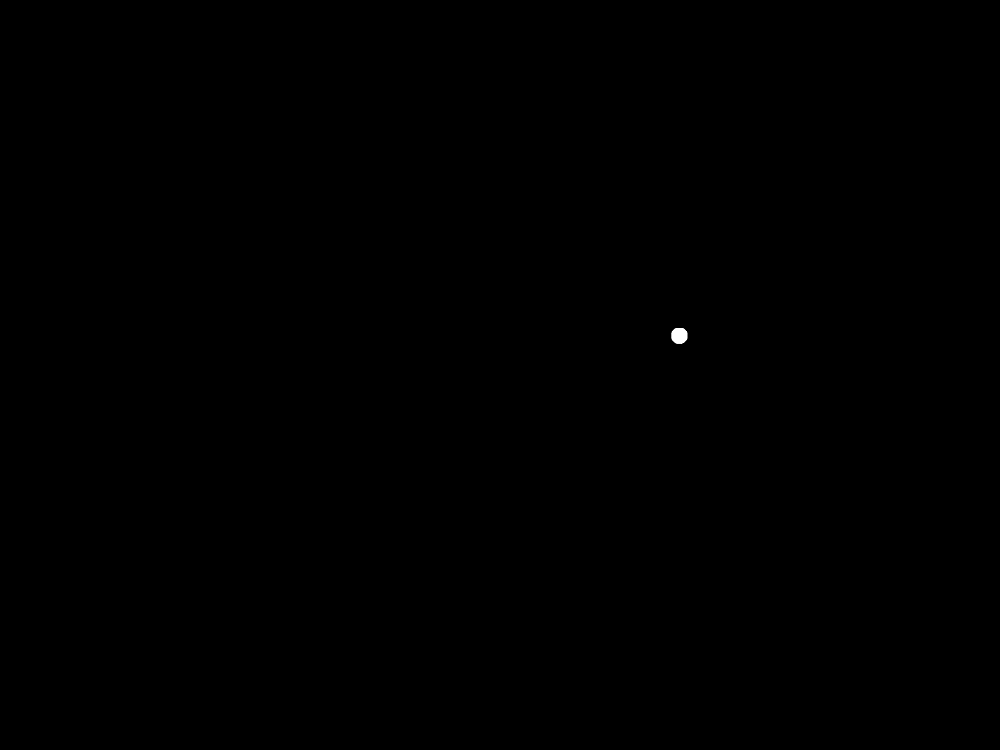}
    \includegraphics[ angle=270,width=2.8cm]{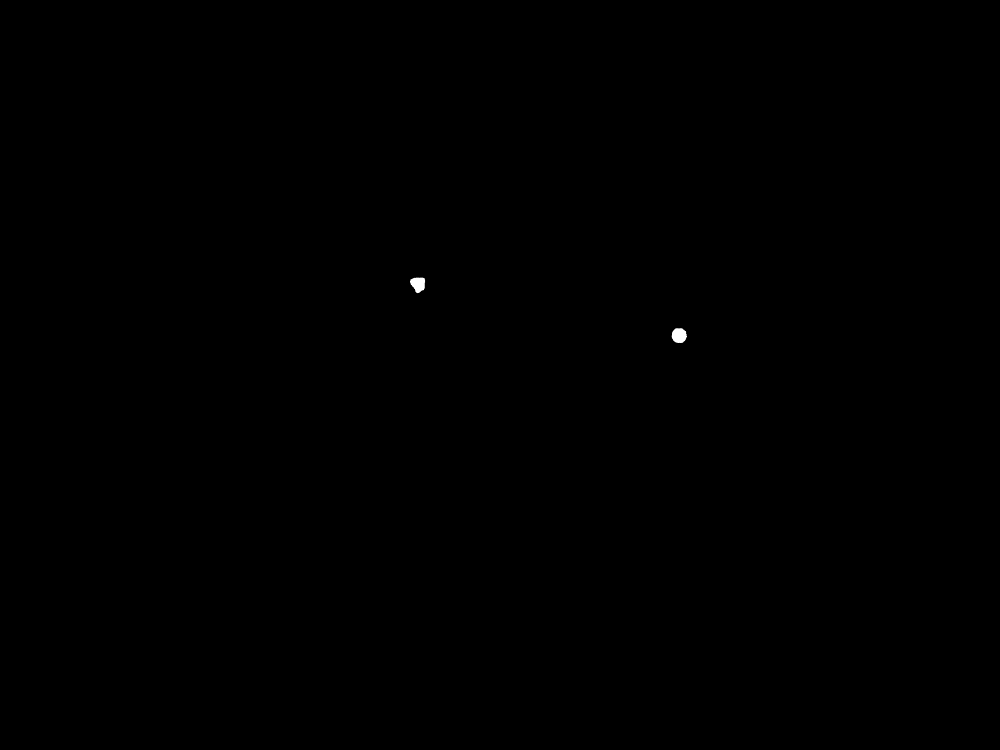}
      
    \includegraphics[angle=270,width=2.8cm]{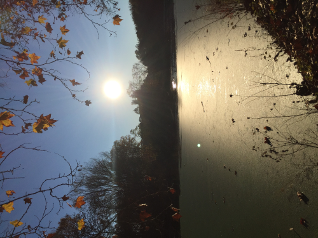}
    \includegraphics[angle=270,width=2.8cm]{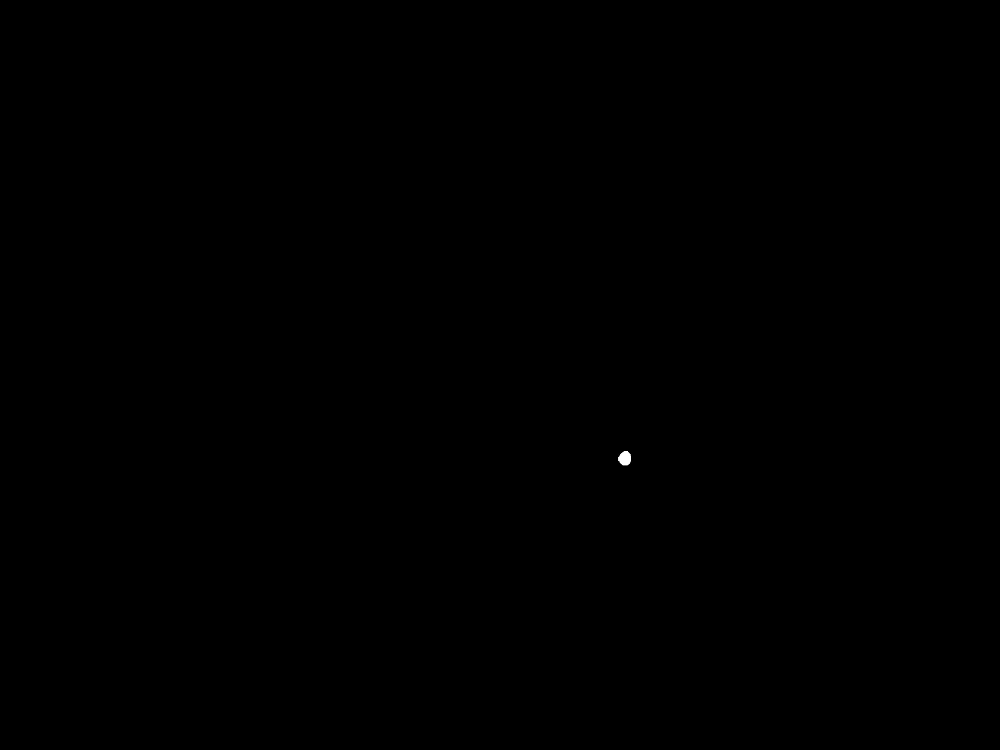}
    \includegraphics[ angle=270,width=2.8cm]{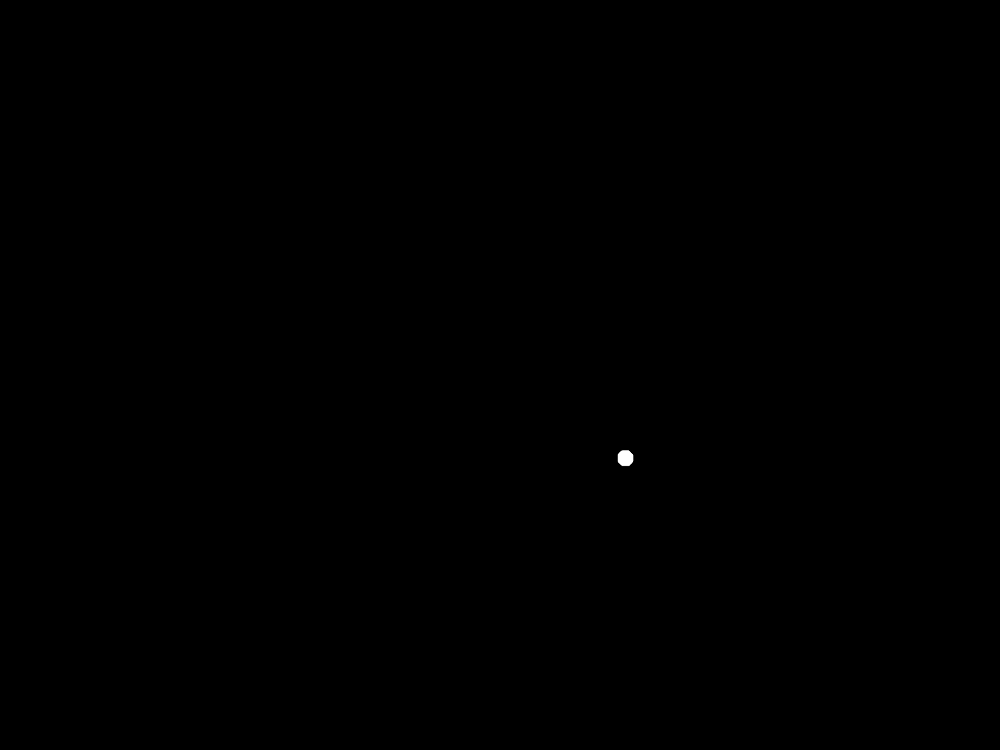}
    \includegraphics[ angle=270,width=2.8cm]{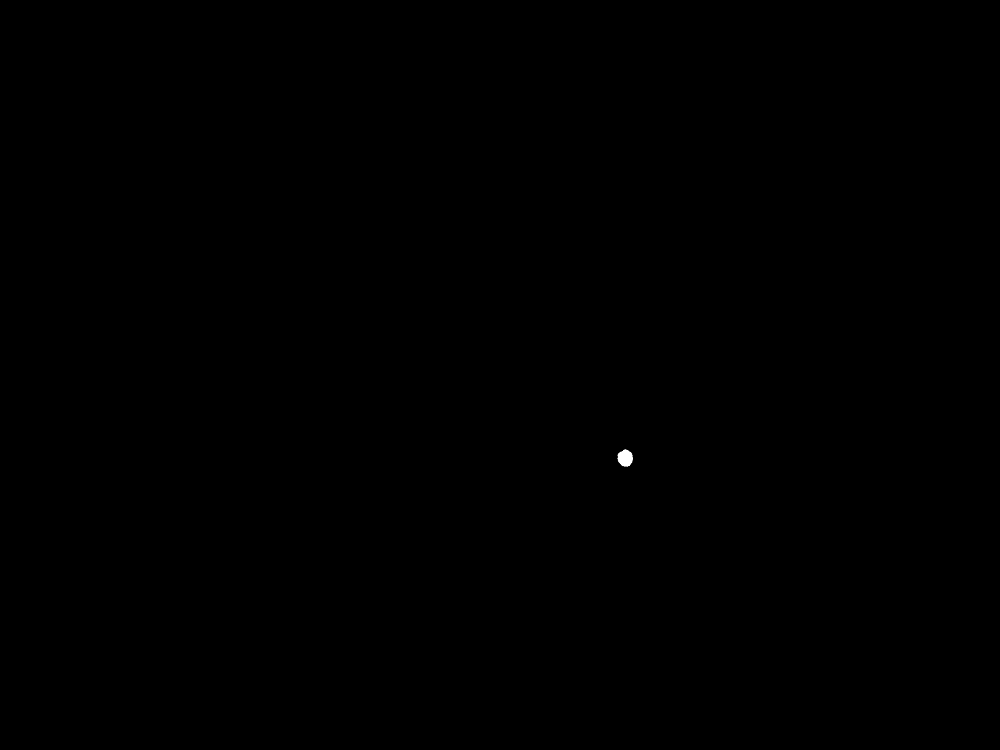}
      
       \includegraphics[angle=270, width=2.8cm]{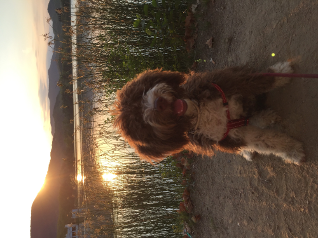}
    \includegraphics[angle=270, width=2.8cm]{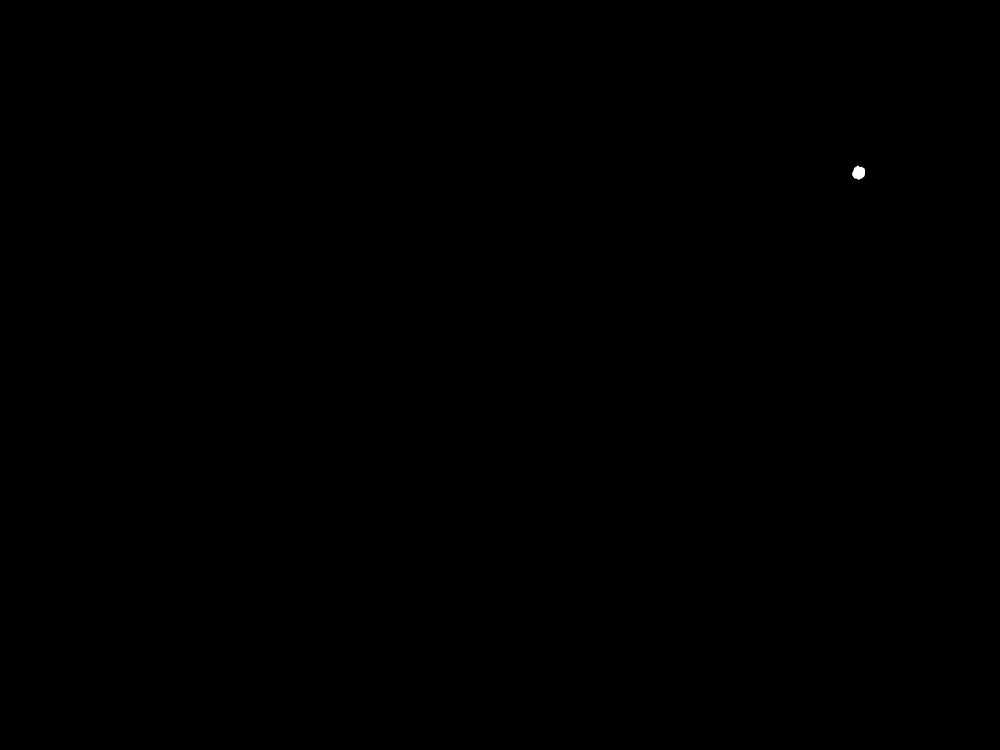}
    \includegraphics[angle=270, width=2.8cm]{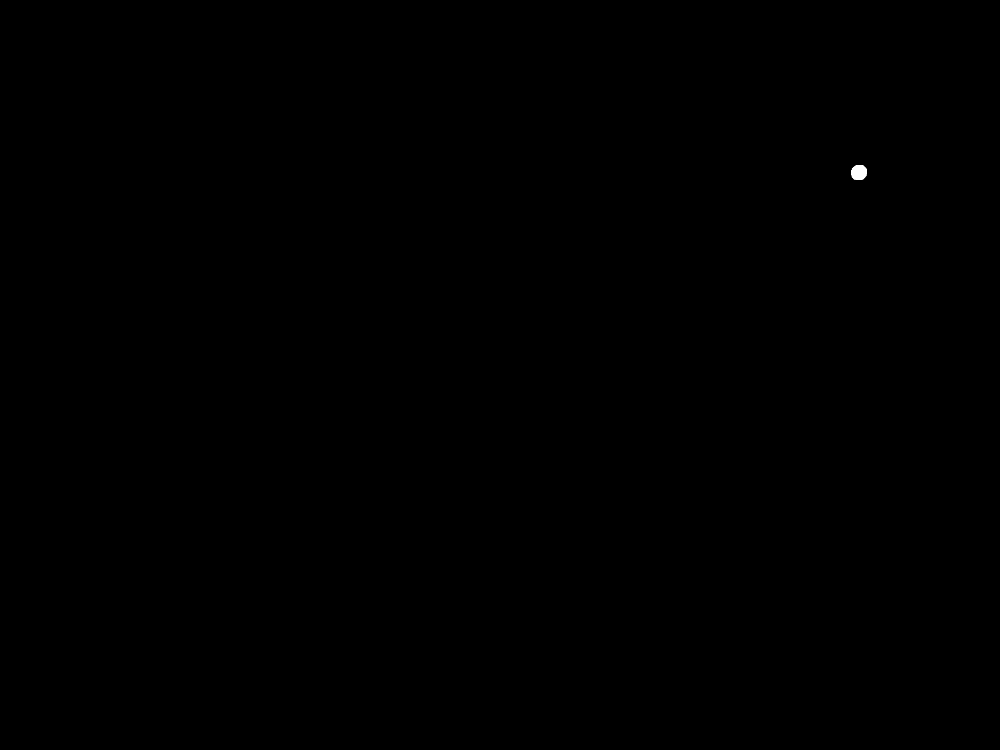}
    \includegraphics[ angle=270, width=2.8cm]{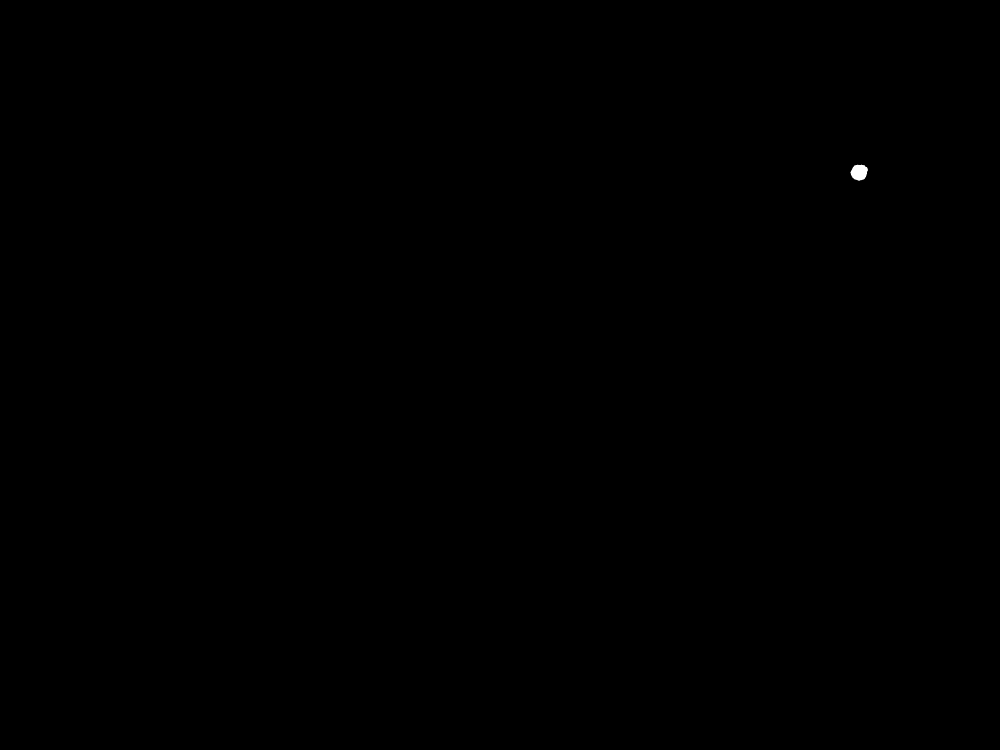}
          
    \includegraphics[width=2.8cm]{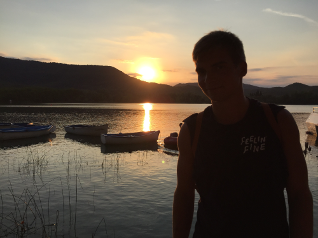}
    \includegraphics[width=2.8cm]{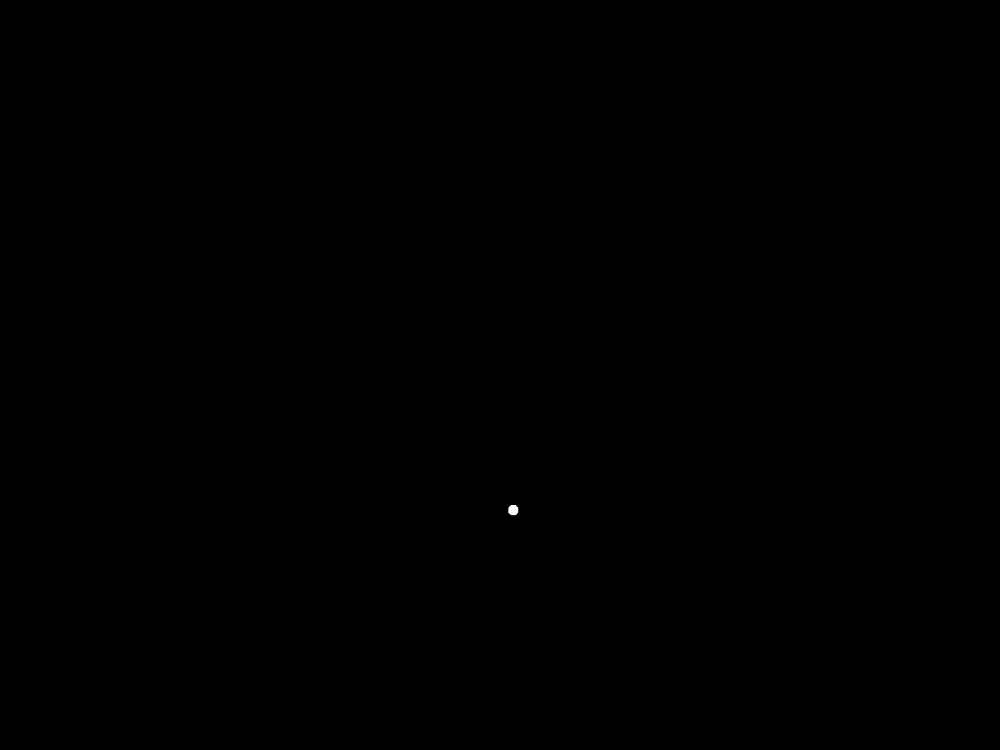}
    \includegraphics[ width=2.8cm]{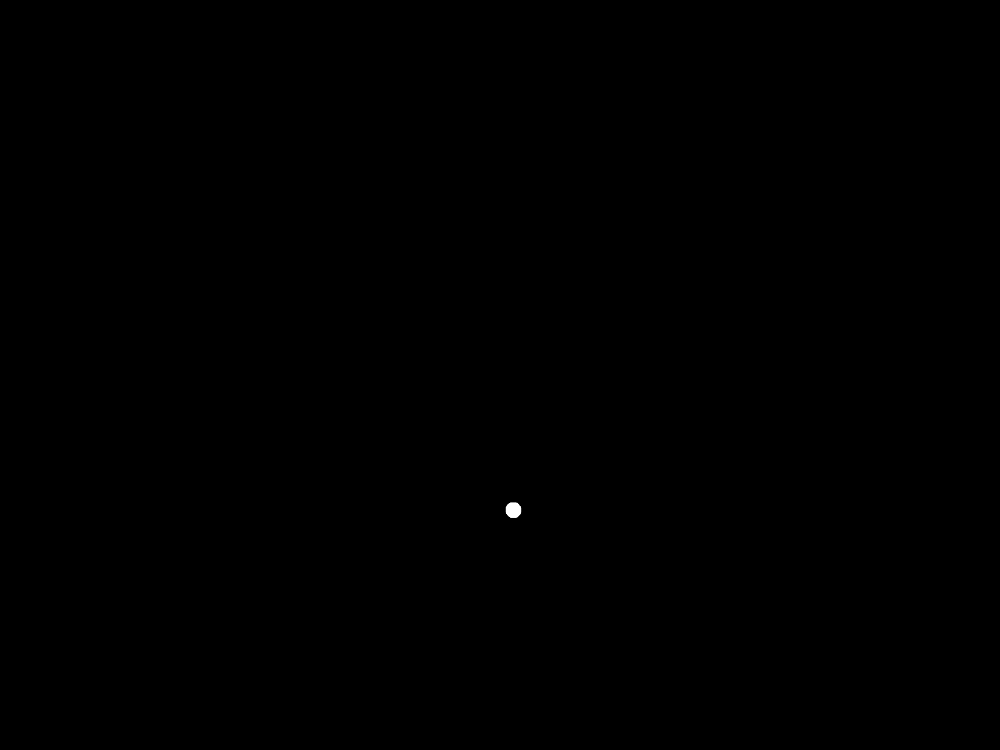}
    \includegraphics[ width=2.8cm]{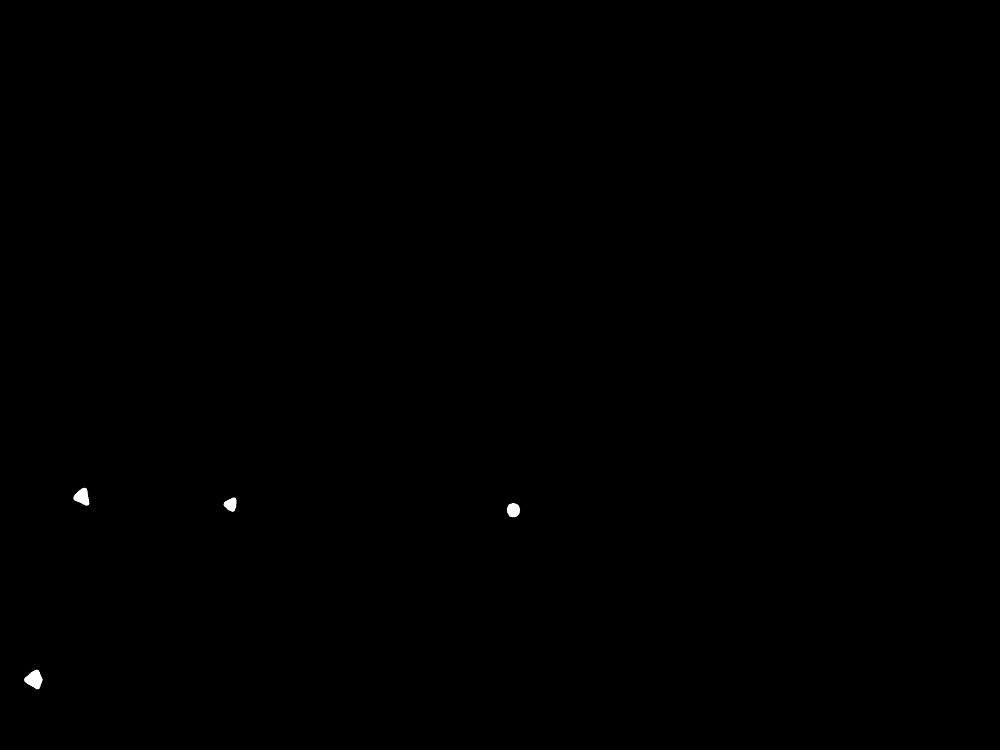}
      
    \includegraphics[width=2.8cm]{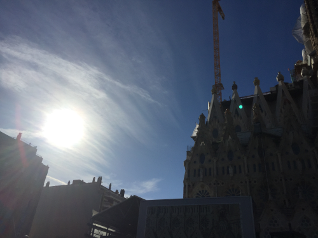}
    \includegraphics[width=2.8cm]{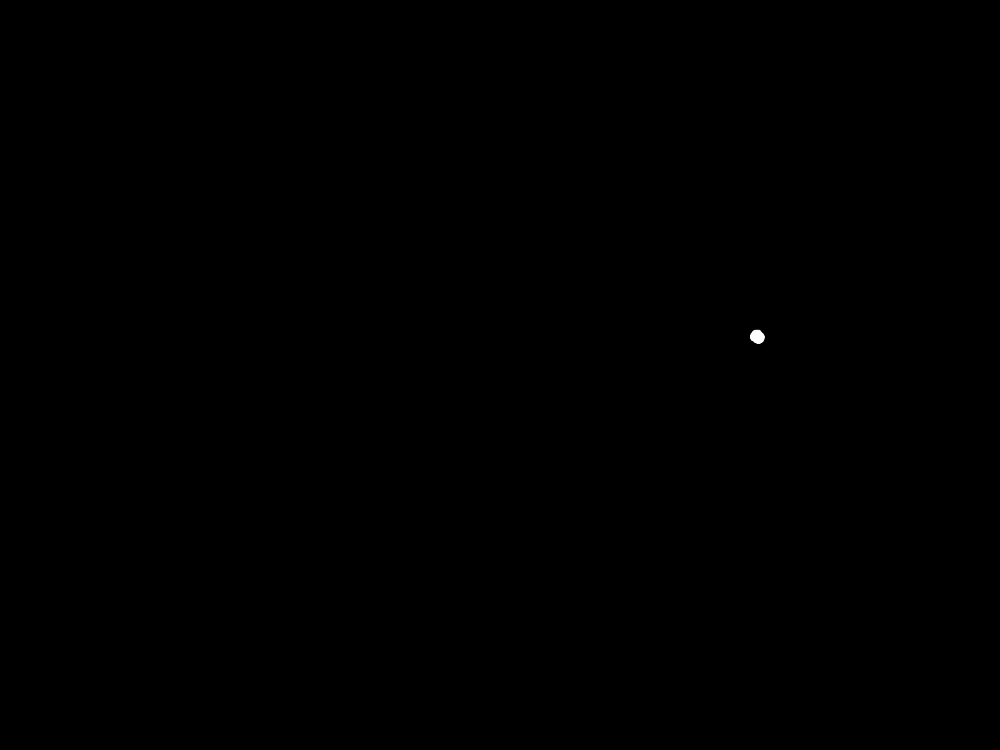}
    \includegraphics[width=2.8cm]{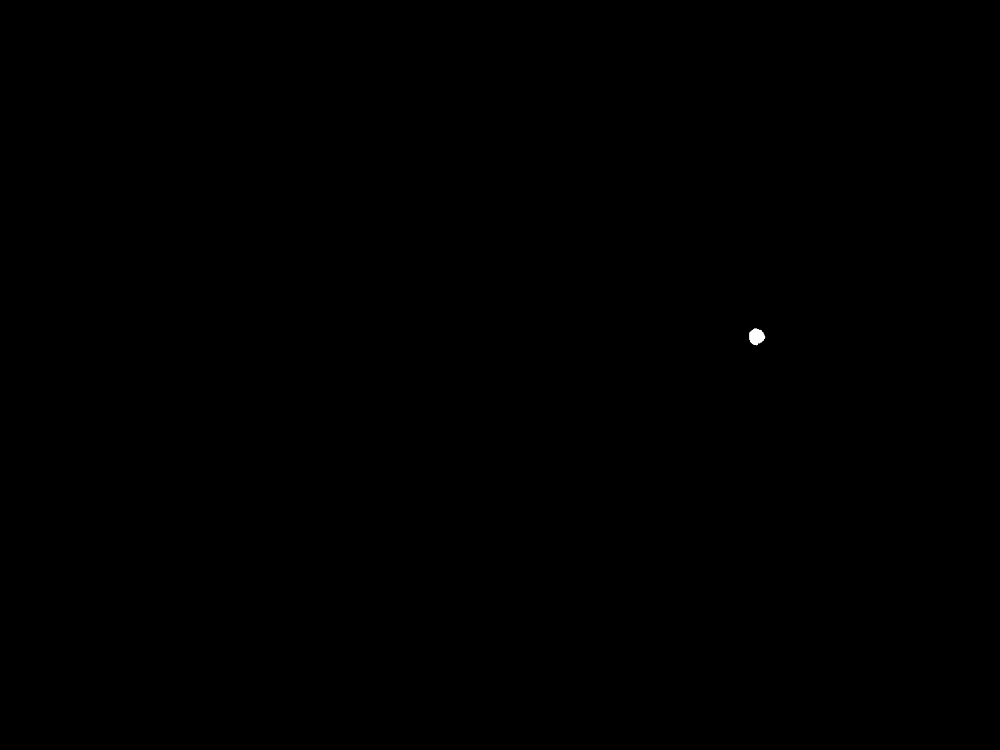}
    \includegraphics[width=2.8cm]{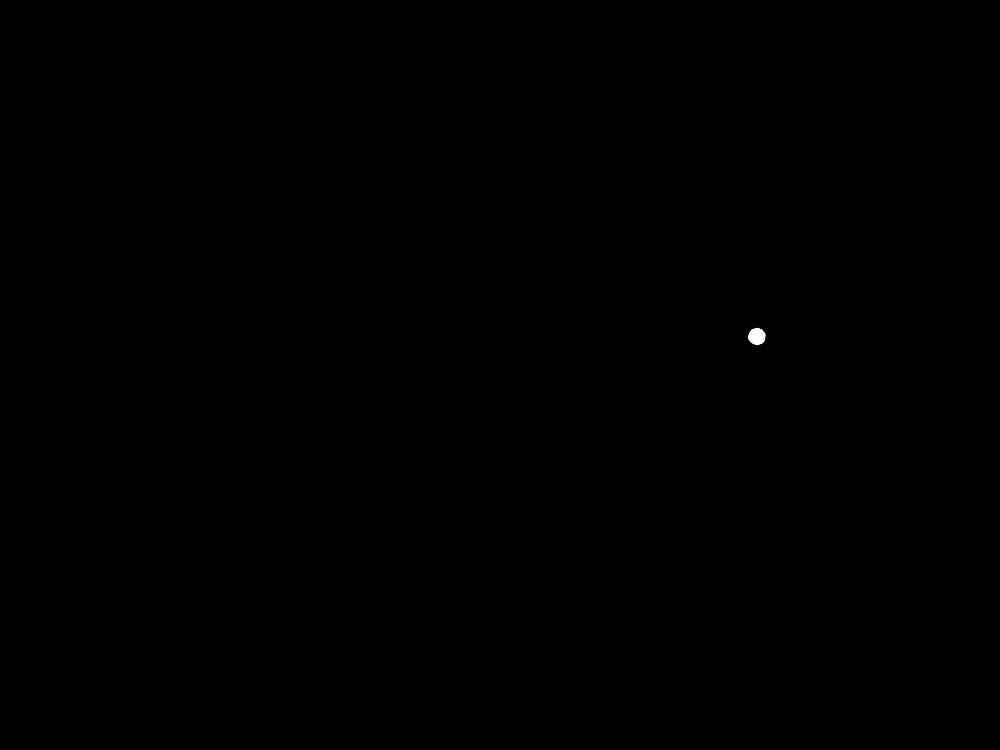}

    \caption{Resulting flare mask using our algorithm and ALFR \citep{chabertautomated}. First column: original image. Second column: ground truth flare mask. Third column: our results. 
    Fourth column: results with ALFR}
    \label{fig:MaskResultschabertOurs}
\end{figure*}

\begin{figure*}
    \centering
    \includegraphics[angle=270,width=2.8cm]{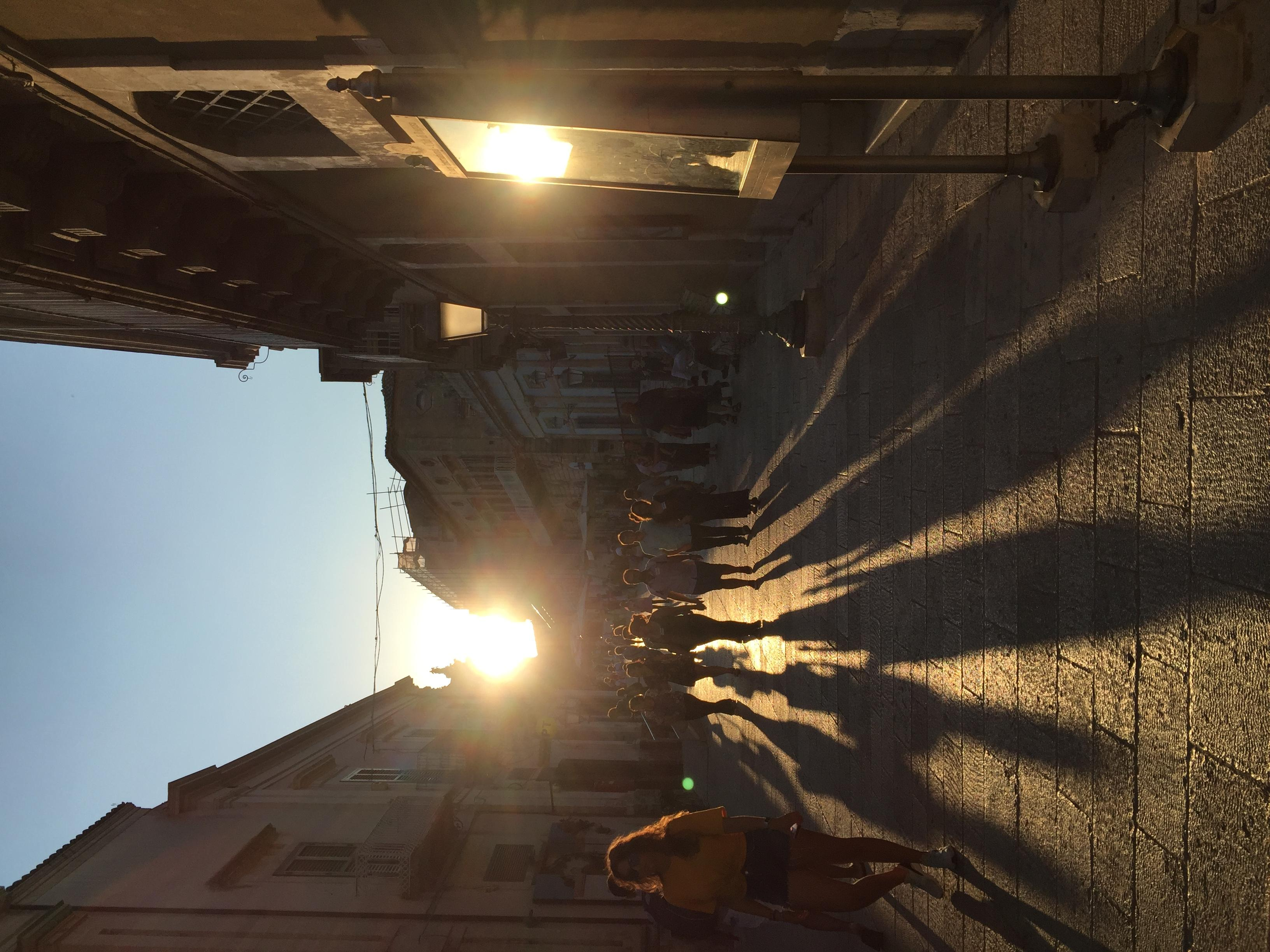}
    \includegraphics[angle=270,width=2.8cm]{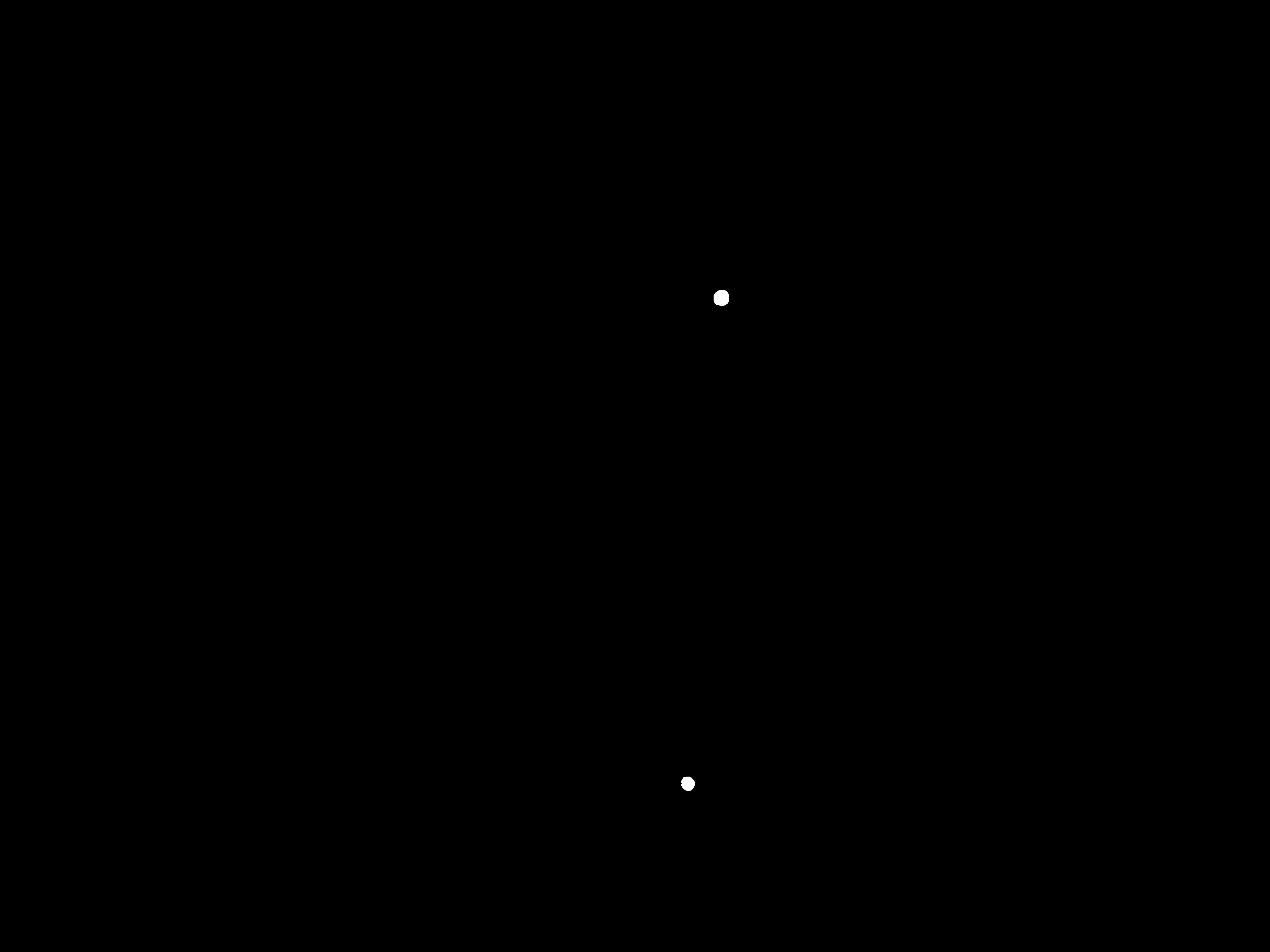}
    \includegraphics[ angle=270,width=2.8cm]{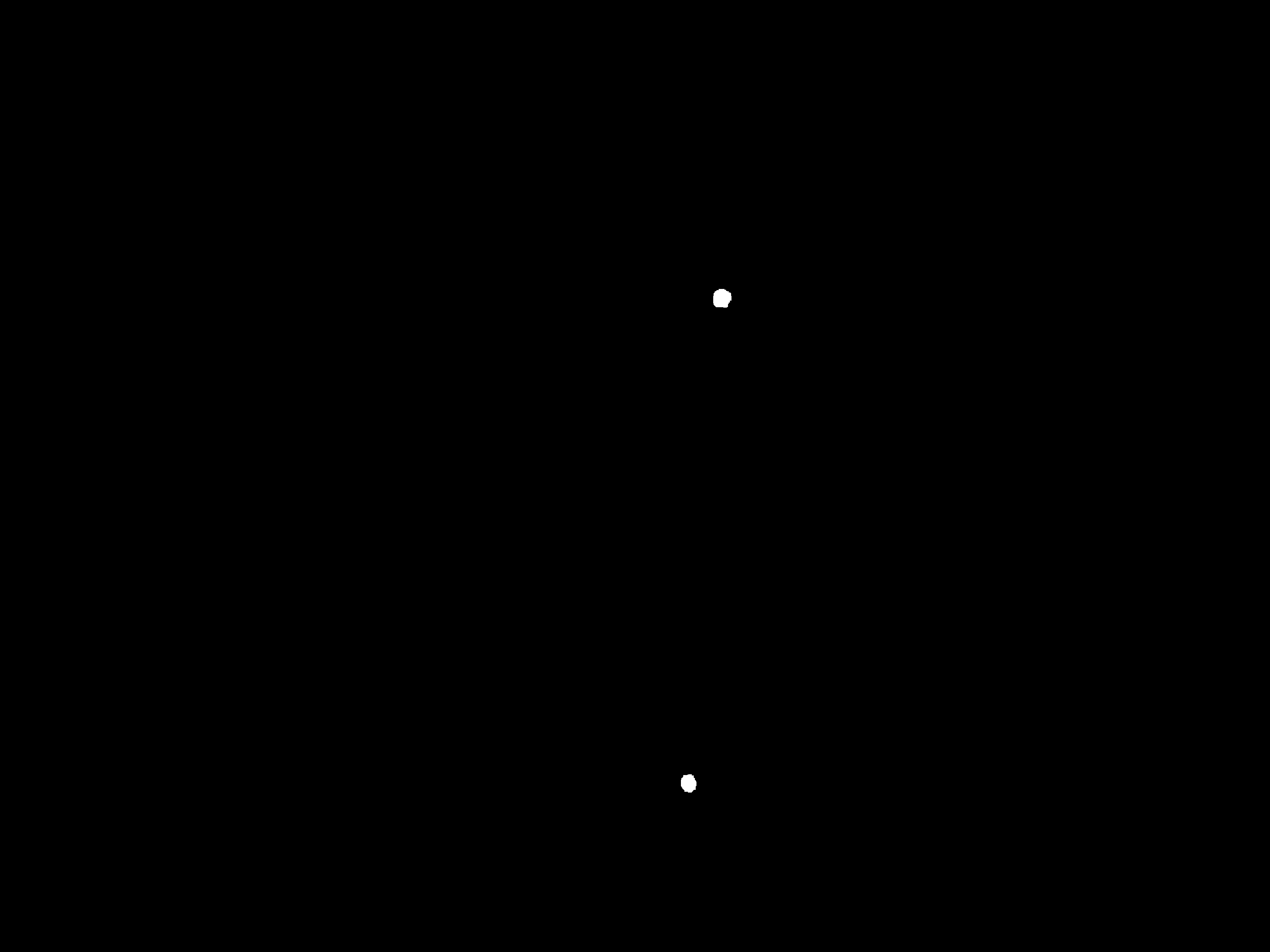}
    \includegraphics[ angle=270,width=2.8cm]{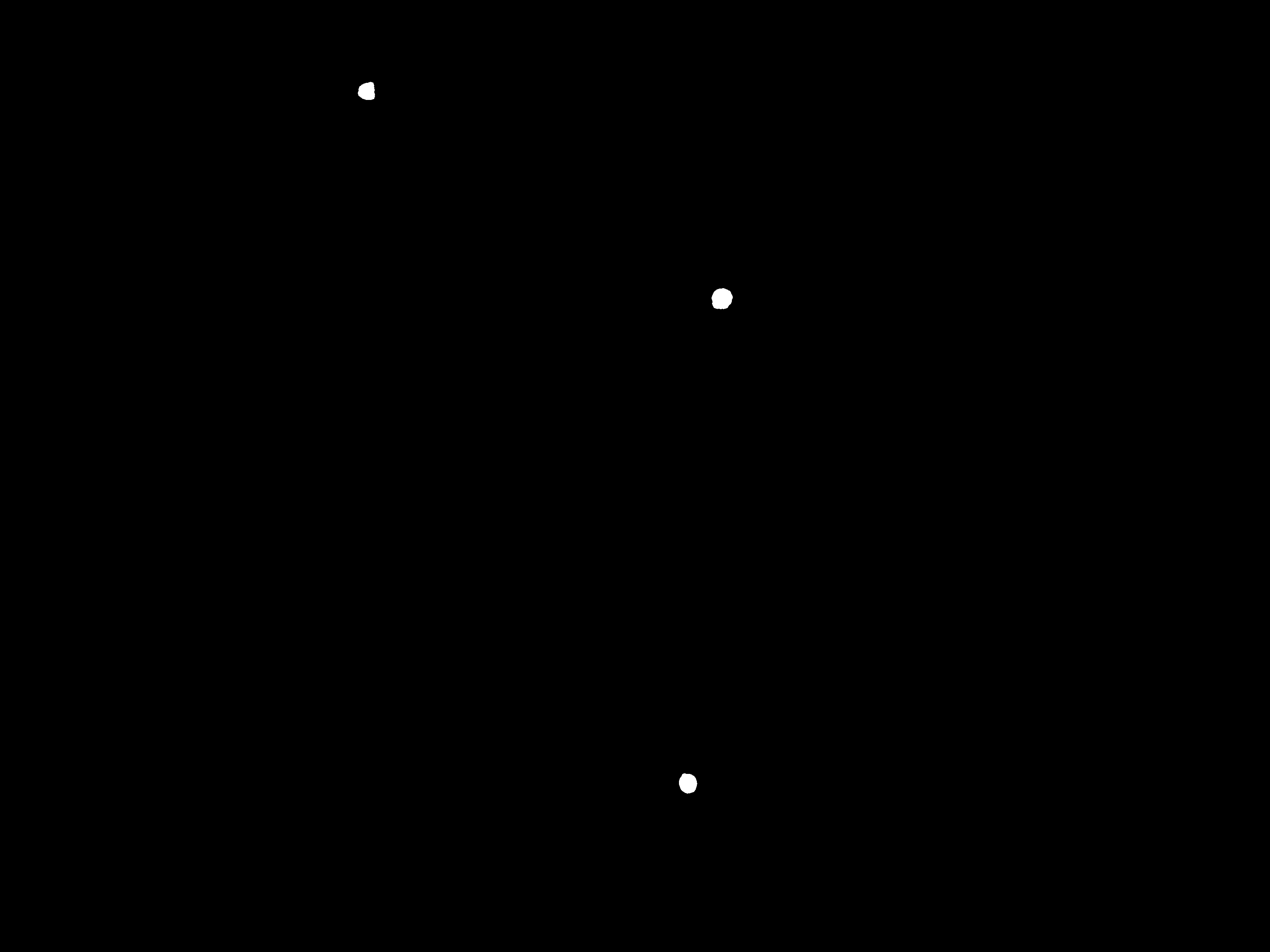}
     
    \includegraphics[angle=270,width=2.8cm]{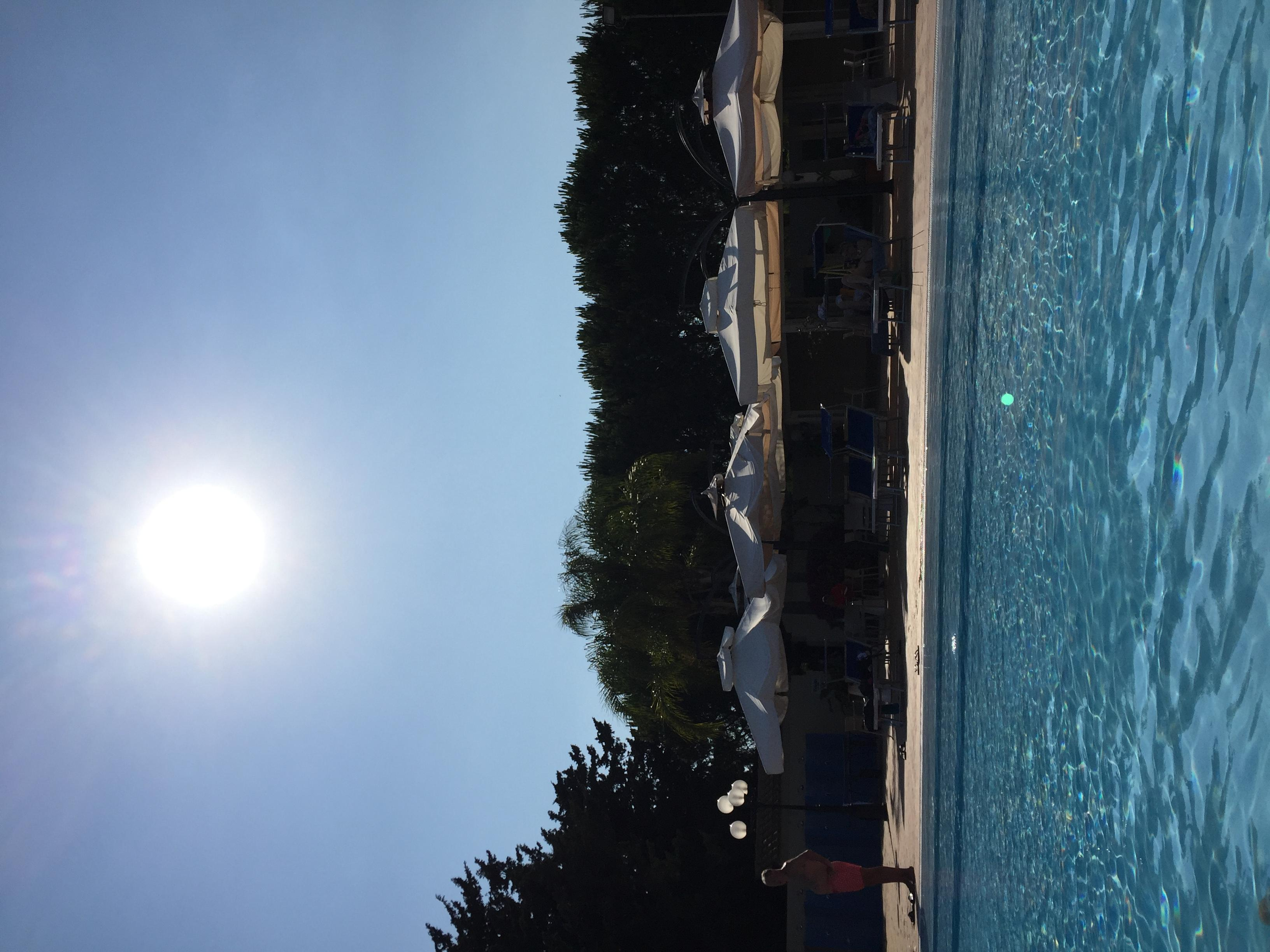}
    \includegraphics[angle=270,width=2.8cm]{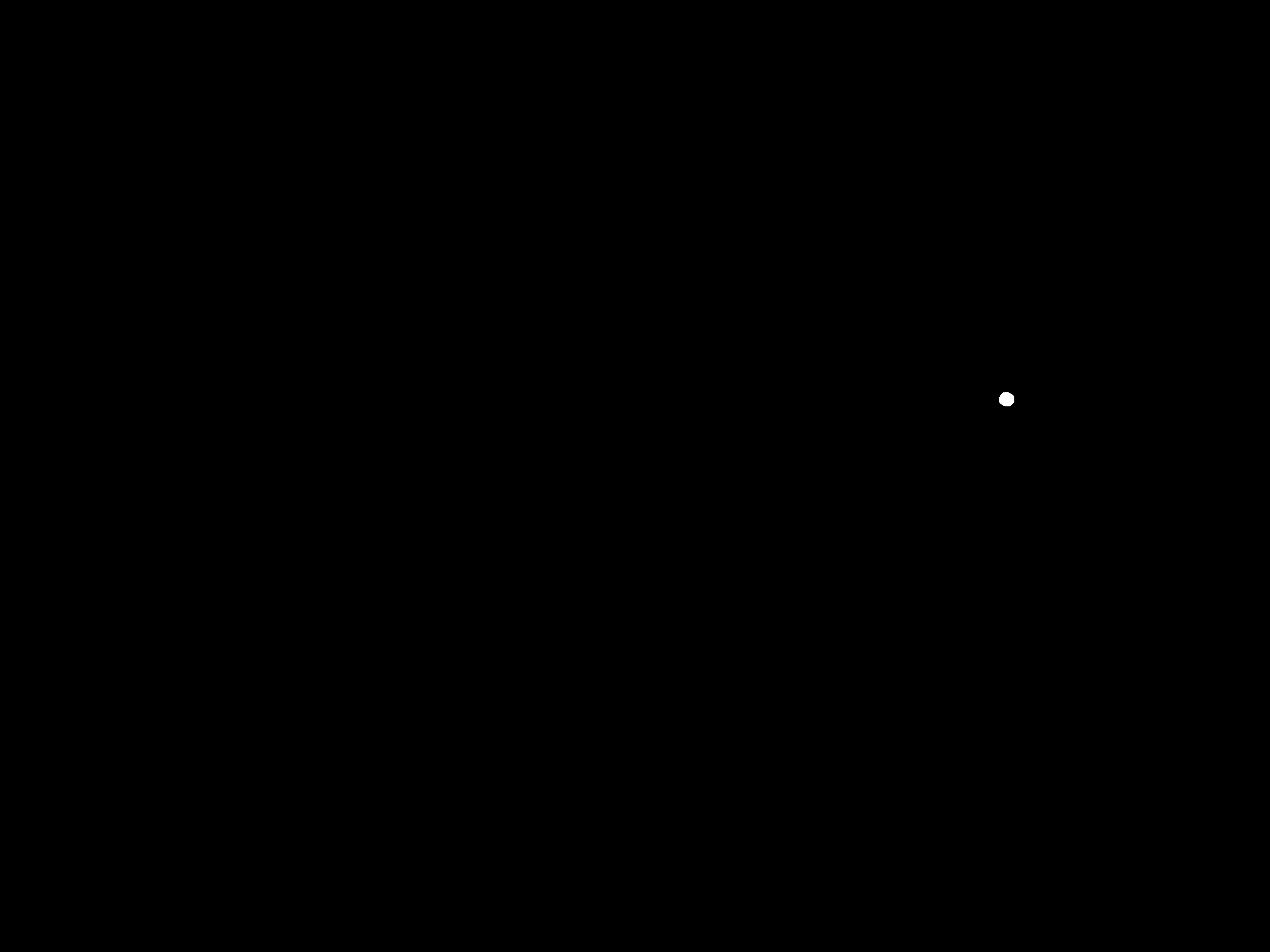}
    \includegraphics[ angle=270,width=2.8cm]{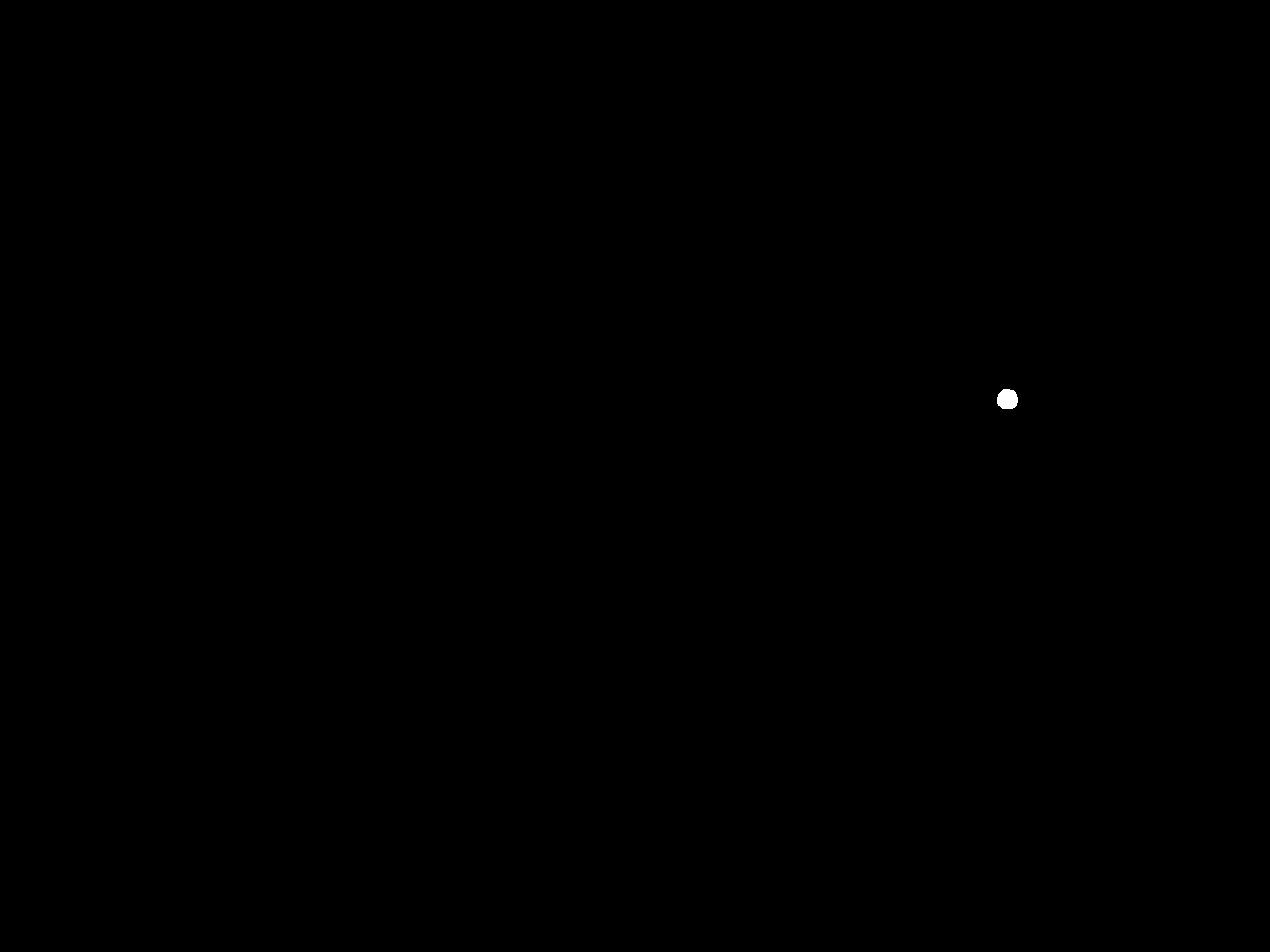}
    \includegraphics[ angle=270,width=2.8cm]{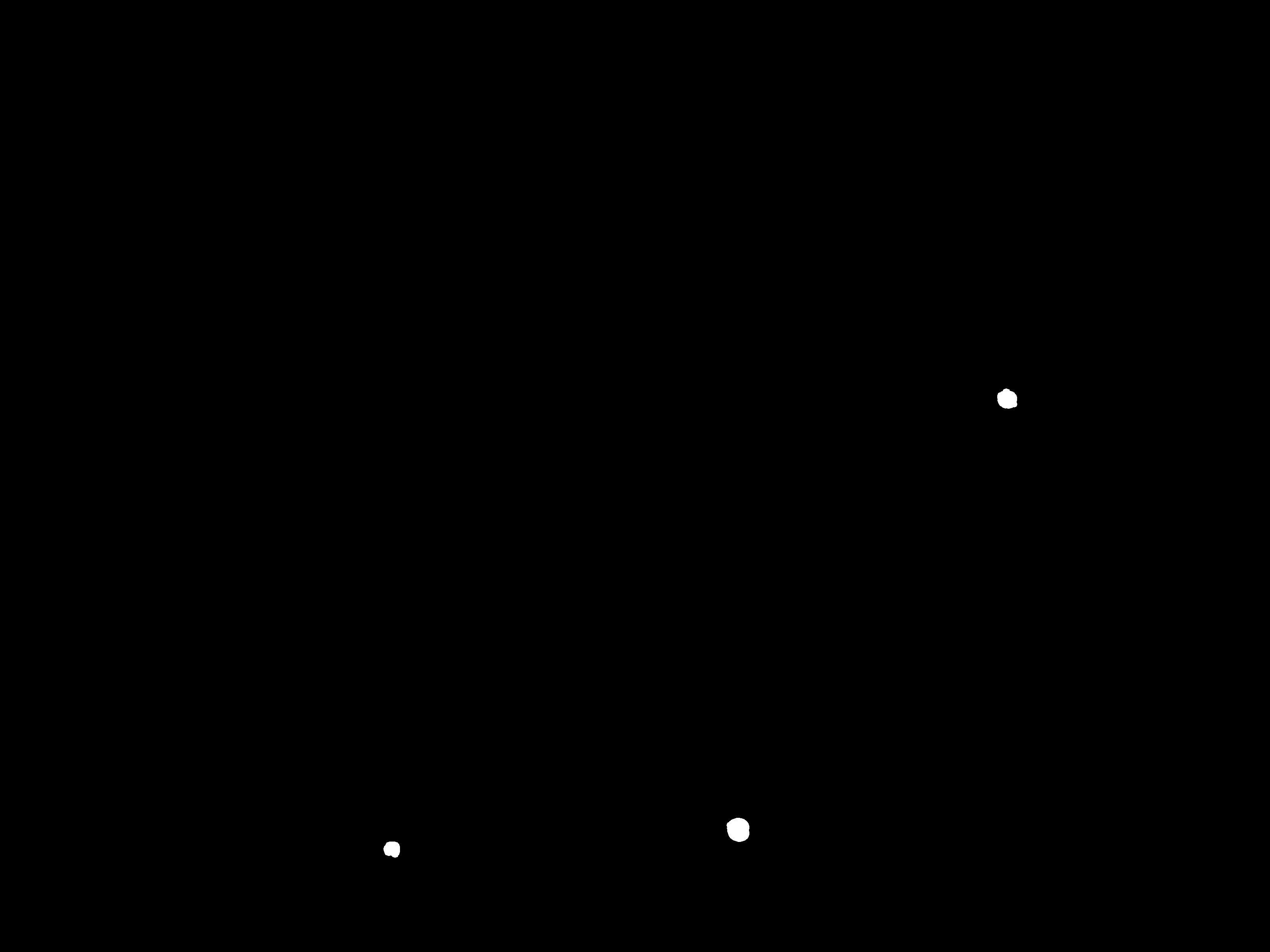}
     
       \includegraphics[width=2.8cm]{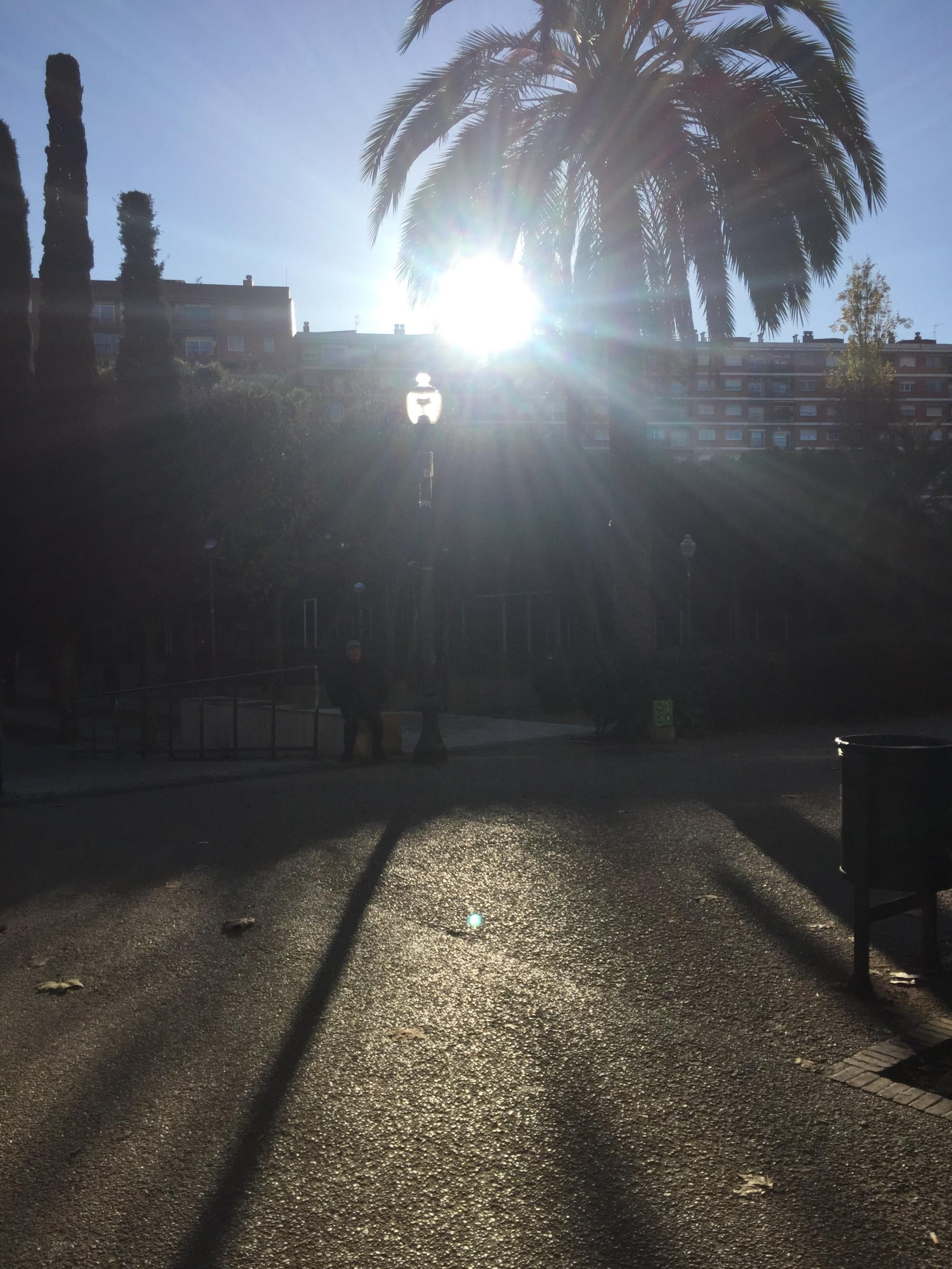}
    \includegraphics[width=2.8cm]{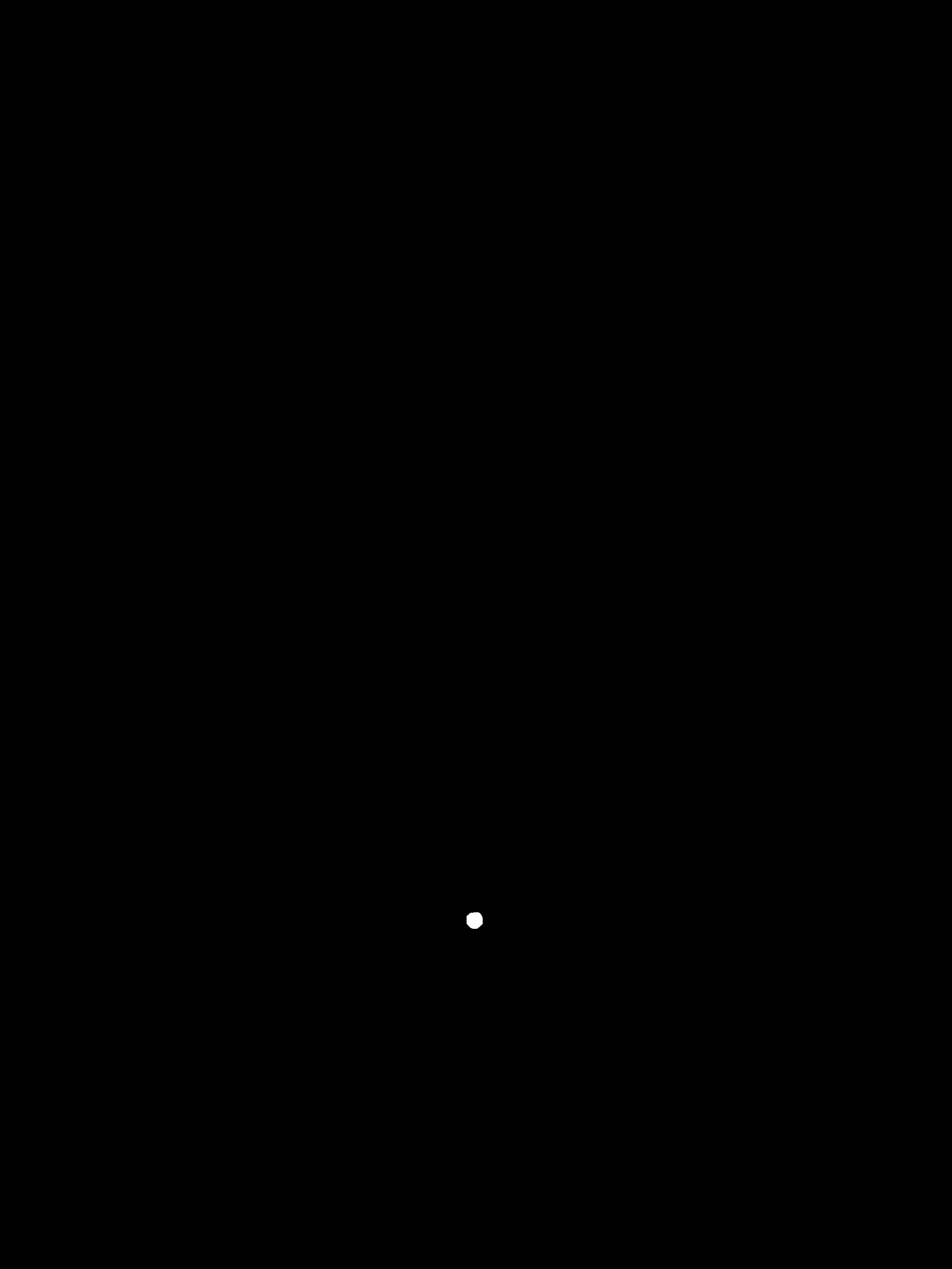}
    \includegraphics[ width=2.8cm]{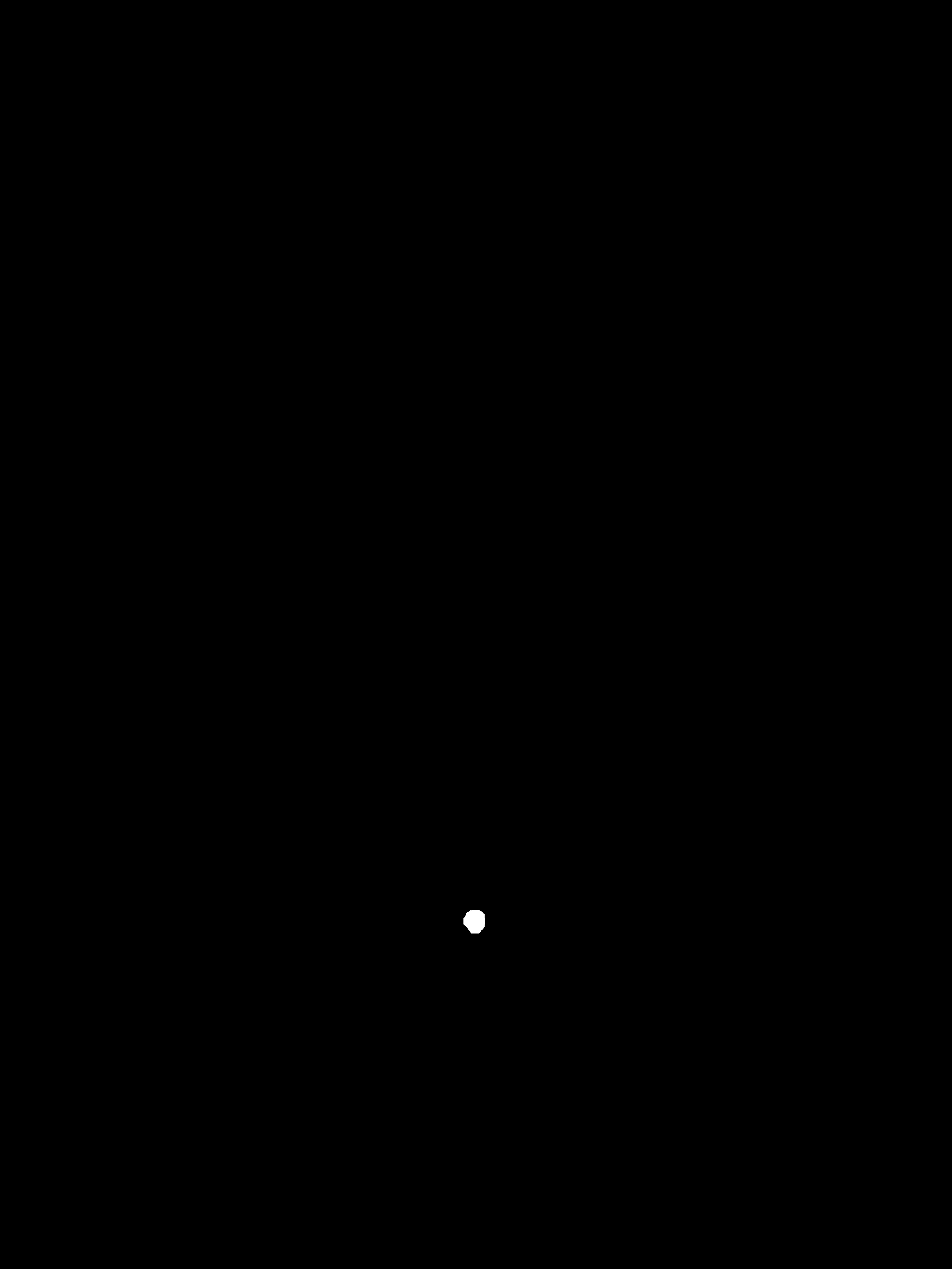}
    \includegraphics[ width=2.8cm]{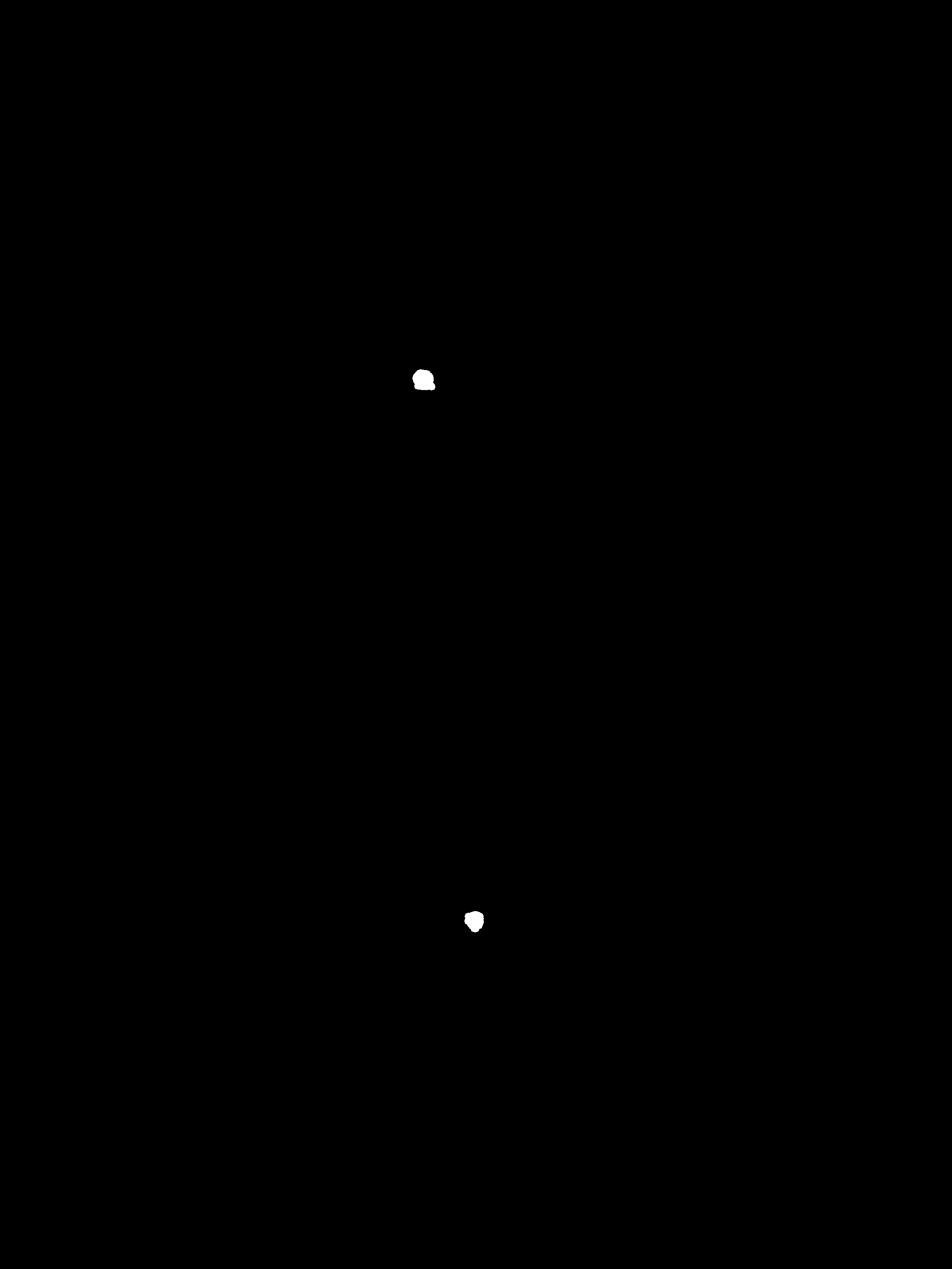}

    \caption{Resulting flare mask using our algorithm and ALFR \citep{chabertautomated}. First column: original image. Second column: ground truth flare mask. Third column: our results. 
    Fourth column: results with ALFR}
    \label{fig:MaskResultschabertOurs2}
\end{figure*}

To evaluate the quality of the computed mask $M$ (which represents the characteristic function of $O=\cup_{i=1}^s F(\mathbf{x}^i_{fs})$), we have manually drawn a ground truth flare mask $M_{gt}$, that delimits the affected area by flare spot artifacts, for each image of our dataset of $405$  images. Some of this ground truth dataset is displayed in the second column of Figure \ref{fig:MaskResultschabertOurs} and \ref{fig:MaskResultschabertOurs2}. In order to quantify the quality of the resulting mask with respect to the ground truth, we consider the \textit{Dice similarity coefficient} \cite{dice1945measures}. The Dice coefficient is a well-known quantitative measure used in the literature for evaluating region-based results such as, e.g., in image segmentation \cite{aljabar2009multi} or for the evaluation of the pixel-wise classification  task  in  recognition  \cite{wels2009fast}, to mention but a few. The Dice similarity coefficient is given by: 
\begin{equation*}
\mathcal{D}(M,M_{gt}) = \frac{2|M \cap M_{gt}|}{|M|+|M_{gt}|}    
\end{equation*}
Notice that $\mathcal{D}(M,M_{gt})$  is equal to 1 when the mask $M$ is equal to $M_{gt}$, and 0 when they do not have any pixel in common. The Dice similarity coefficient is not very different in form from the \emph{Intersection over Union} or \emph{Jaccard index} and the latter can be deduced from the former (and viceversa). 

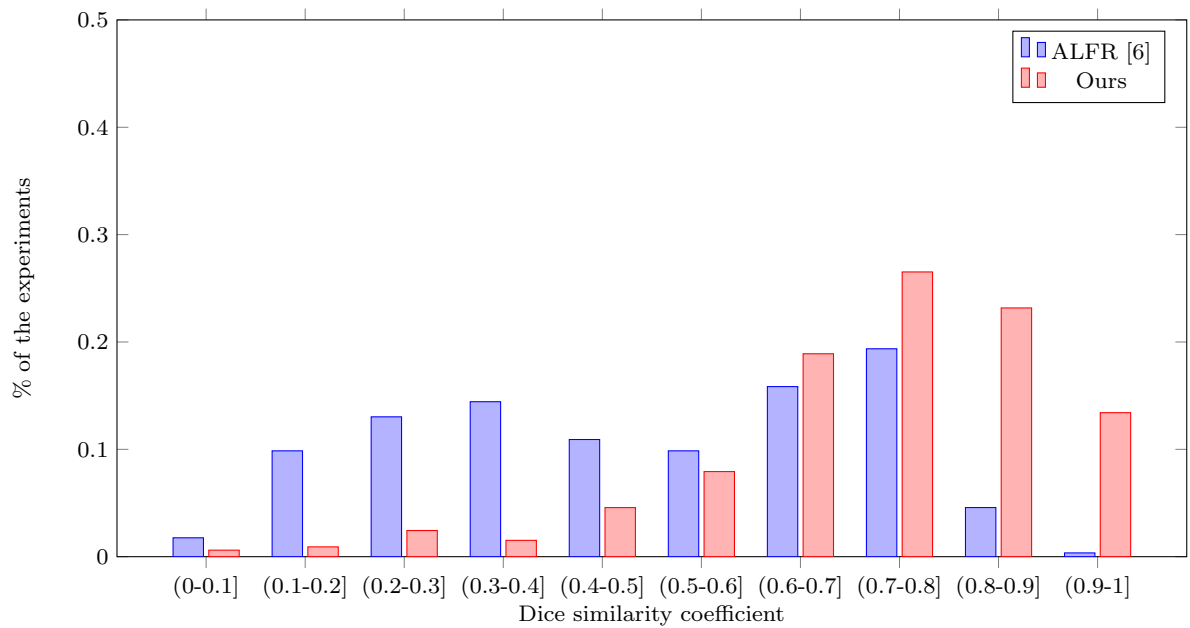
\begin{figure*}
    \centering
        \begin{tikzpicture}
            \begin{axis}[
                ybar,
                bar width=.4cm,
                width=0.9\textwidth,
                height=.5\textwidth,
                symbolic x coords={0,(0-0.1],(0.1-0.2],(0.2-0.3],(0.3-0.4],(0.4-0.5],(0.5-0.6],(0.6-0.7],(0.7-0.8],(0.8-0.9],(0.9-1]},
                xtick=data,
                ymin=0,ymax=0.5,
            	ylabel= \% of the experiments,
            	xlabel= Dice similarity coefficient 
            ]
            \addplot 
            	coordinates { ((0-0.1],0.017605) ((0.1-0.2],0.098591)((0.2-0.3],0.130281) ((0.3-0.4],0.144366) ((0.4-0.5], 0.109154) ((0.5-0.6],0.098591) ((0.6-0.7],0.158450) ((0.7-0.8], 0.193661) ((0.8-0.9], 0.045774) ((0.9-1], 0.003521) };
            \addplot 
            	coordinates { ((0-0.1],0.006097) ((0.1-0.2],0.009146) ((0.2-0.3],0.024390) ((0.3-0.4],0.015243) ((0.4-0.5], 0.045731) ((0.5-0.6], 0.079268) ((0.6-0.7], 0.189024) ((0.7-0.8], 0.265243) ((0.8-0.9], 0.231707) ((0.9-1], 0.134146)};
            \legend{ALFR \citep{chabertautomated}, Ours}
            \end{axis}
        \end{tikzpicture}
    \caption{Percentage of experiments (vertical axis) related to a certain range of values of the dice similarity coefficient (horizontal axis)  obtained using ALFR \citep{chabertautomated} (in blue) and ours (in red). Notice that a dice similarity coefficient equal to one means that the flare mask is equal to the ground truth, and when it is equal to zero that it  does not have any pixel in common }
        \label{fig:maskAcurracyGraph}
\end{figure*}

For each of the images from the dataset we have computed the flare mask $M$ using both algorithms: ALFR and ours. The distribution of the Dice similarity coefficient  values can be seen in Figure \ref{fig:maskAcurracyGraph}. We would like to remark that the Dice similarity coefficient has been computed only over the images where the flare spot was successfully detected, in order to avoid hindering the results of ALFR, due to its lower recall. For each range of values in the horizontal axis, the related percentage of cases with a Dice coefficient between these values is displayed (vertical axis). Thus, a good result is one that has the maximum percentage of experiments on the right hand side of the diagram (i.e., Dice coefficient close to 1).  Let us notice that in the half of the cases, the mask obtained by ALFR has a dice coefficient lower than 0.5, and on the other half, the results are closer to 0.5 than to 1. In our case, more than half of experiments have a dice coefficient over 0.7. The average accuracy value obtained using both algorithms can be seen in the last two rows of Table \ref{tab:SummaryTable}.  By comparing the distribution and the average of the mask accuracy, we can infer that the obtained resulting mask using our algorithm is, in average, much closer to the ground truth than the one obtained by ALFR.

In Figure \ref{fig:MaskResultschabertOurs} and \ref{fig:MaskResultschabertOurs2} we show some resulting flare masks obtained using ALFR and our algorithm. As we can see our algorithm tends to be closer to the ground truth than ALFR. Notice also that, in some  cases, ALFR identifies multiple false positive flare spot artifacts (for example, in the  first, second and fifth row of Figure \ref{fig:MaskResultschabertOurs} or in all the examples of Figure \ref{fig:MaskResultschabertOurs2}), which translates in his high false positive rate (see Table \ref{tab:SummaryTable} and Figure \ref{fig:falsepositiveChart}) and, moreover, creating additional regions to be inpainted. Also, in the  first row of Figure \ref{fig:MaskResultschabertOurs2} we can see one example where more than one flare spot is present in the image. Both algorithms detect both flares. While our algorithm  restricts in detecting only real flare spots, in the case of ALFR, it goes beyond detecting false positive artifacts. 

In terms of computational time, our whole flare spot detection method, including the flare spot mask computation, takes on average 39.77 seconds. We do not include the computation time for the last exemplar-based inpainting step which uses algorithm by \cite{fedorov2015variational} and which is discussed in the next section.

\subsection{Flare Spot Region Recovery}

Finally, we evaluate the complete flare removal method and present comparison results where the flare spot regions are reconstructed. As explained above, to reconstruct the image in the flare spot regions, we use the exemplar-based inpainting algorithm implementation by Fedorov et al. \cite{fedorov2015variational} with the default parameters proposed by the authors. As an input for the inpainting algorithm we employ the mask obtained in Section \ref{sec:MaskCreation} together with the original image. 

For the sake of comparison, Figure \ref{fig:InpaintingResultschabertOurs} and \ref{fig:InpaintingResultschabertOurs2} include some reconstructed images where the flare spot artifacts were correctly identified both by ALFR and our algorithm. Let us remark that, for instance, in the first row of Figure \ref{fig:InpaintingResultschabertOurs}, while our reconstruction have into account the edge between the white arrow and the asphalt, ALFR creates a diffusion of the white arrow without having into account the boundaries between both regions. Also, our algorithm is able to reconstruct textures; on the contrary, ALFR does not tend to have into account them and creates an over-smoothed area (see, e.g., the examples in the third and fourth rows of Figure \ref{fig:InpaintingResultschabertOurs} and second and third rows of Figure \ref{fig:InpaintingResultschabertOurs2}). 
First row in Figure \ref{fig:InpaintingResultschabertOurs2} shows an example with more than one flare spot. See more reconstructed examples, that only our algorithm detects the artifacts, in Figures \ref{fig:InpaintingResultsOurs1} and \ref{fig:InpaintingResultsOurs2}.

On the other hand, the used exemplar-based method produces a plausible completion inside the flare spot area that preserves the textures of the surrounding region and preserves edges (this can be seen, for example, in the third row in Figure \ref{fig:InpaintingResultsOurs2}), leading to a plausible flare spot area's reconstruction.

\section{Conclusions}\label{sec:conclusions}

We have proposed an automatic method to detect and remove flare spot artifacts from a photograph.  Our method consists in three main blocks.  First we detect the presence and location of light sources in the image which in turn can be used as a location hint for a faster flare spot detection. Then, we propose a characterization of the main properties of flare spot artifacts and a computational method which identifies a small set of candidates based on such properties. A confidence measure is then proposed that is able to select the correct flare spot among them. Finally, as a last step, we remove the flare spot artifacts by means of exemplar-based inpainting.

The numerical results show that the proposed algorithm outperforms related works in flare spot detection not only by having a higher recall and precision (0.7862 and 0.8092, respectively, over a maximum of 1), but also having an average number of false positive close to zero. Furthermore, the obtained flare mask also gives better accuracy than in ALFR \citep{chabertautomated}.
The visual results of the reconstructed regions also outperform the ones obtained by applying ALFR both by better preserving textures and geometric edges. Summarizing, our whole flare spot detection and removal method obtains, to the best of our knowledge, quantitative state-of-the-art results and eventually outputs a qualitative image close to what we would expect in the real world.

\begin{acknowledgements}
The authors acknowledge partial support by MINECO/FEDER UE project, with reference TIN2015-70410-C2-1-R, and by H2020-MSCA-RISE-2017 project with reference 777826 NoMADS.
\end{acknowledgements}

\bibliography{references}

\end{document}